\documentclass[10pt]{iopart}
\usepackage{iopams}
\usepackage{bbm}
\usepackage{color}
\usepackage{graphicx} \graphicspath{{./Images/}}
\usepackage{float}
\usepackage{etoolbox}

\makeatletter
\def\@mkboth#1#2{}
\newlength\appendixwidth
\preto\appendix{\addtocontents{toc}{\protect\patchl@section}}
\newcommand{\patchl@section}{%
  \settowidth{\appendixwidth}{\textbf{Appendix }}%
  \addtolength{\appendixwidth}{1.5em}%
  \patchcmd{\l@section}{1.5em}{\appendixwidth}{}{\ddt}%
}
\renewcommand\@makefnmark{\textsuperscript{\@thefnmark}}
\renewcommand\@makefntext[1]%
    {\noindent\makebox[0pt][r]{\textsuperscript{\@thefnmark}\,}#1}
\makeatother

\begin{document}

\renewcommand{\fnsymbol}{\alph}
\newcommand{\bx}{\mathbf{x}}
\newcommand{\bbr}{\mathbf{r}}
\newcommand{\be}{\begin{equation}}
\newcommand{\ee}{\end{equation}}
\newcommand{\bea}{\begin{eqnarray}}
\newcommand{\eea}{\end{eqnarray}}
\newcommand{\KS}{\mathrm{KS}}
\newcommand{\kin}{\mathrm{kin}}

\def\ket#1{\vert #1\rangle}
\def\bra#1{\langle#1\vert}
\def\braket#1#2{\langle #1 \vert #2 \rangle}

\def\diff{\mathrm{d}}
\def\imagi{\mathrm{i}}

\newcommand{\ux}{\underline{x}}
\newcommand{\ur}{\underline{r}}
\newcommand{\us}{\underline{\sigma}}
\newcommand{\hH}{\hat{H}}
\newcommand{\hA}{\hat{A}}
\newcommand{\hT}{\hat{T}}
\newcommand{\hV}{\hat{V}}
\newcommand{\hW}{\hat{W}}
\newcommand{\ext}{\mathrm{ext}}
\newcommand{\Ha}{\mathrm{H}}
\newcommand{\xc}{\mathrm{xc}}
\newcommand{\Hxc}{\mathrm{Hxc}}
\newcommand{\prt}{\partial_t}

\topical[Density-potential mapping in TDDFT]{Existence, Uniqueness, and Construction of the Density-Potential Mapping in Time-Dependent Density-Functional Theory}

\author{Michael Ruggenthaler}
\address{Institut f\"ur Theoretische Physik, Universit\"at Innsbruck, Technikerstra{\ss}e 21a, A-6020 Innsbruck, Austria}

\author{Markus Penz}
\address{Institut f\"ur Theoretische Physik, Universit\"at Innsbruck, Technikerstra{\ss}e 21a, A-6020 Innsbruck, Austria}

\author{Robert van Leeuwen}
\address{Department of Physics, Nanoscience Center, University of Jyv\"askyl\"a, 40014 Jyv\"askyl\"a, Finland}

\date{\today}

\begin{abstract}
In this work we review the mapping from densities to potentials in quantum mechanics, which is the basic building block of time-dependent density-functional theory and the Kohn-Sham construction. We first present detailed conditions such that a mapping from potentials to densities is defined by solving the time-dependent Schr\"odinger equation. We specifically discuss intricacies connected with the unboundedness of the Hamiltonian and derive the local-force equation. This equation is then used to set up an iterative sequence that determines a potential that generates a specified density via time propagation of an initial state. This fixed-point procedure needs the invertibility of a certain Sturm-Liouville problem, which we discuss for different situations. Based on these considerations we then present a discussion of the famous Runge-Gross theorem which provides a density-potential mapping for time-analytic potentials. Further we give conditions such that the general fixed-point approach is well-defined and converges under certain assumptions. Then the application of such a fixed-point procedure to lattice Hamiltonians is discussed and the numerical realization of the density-potential mapping is shown. We conclude by presenting an extension of the density-potential mapping to include vector-potentials and photons.
\end{abstract}
\submitto{\JPCM}
\maketitle

\tableofcontents

\section{Introduction}

\subsection{General overview}

The Schr\"{o}dinger equation \cite{schroedinger1926} (together with its variants, e.g., \cite{pauli1927, hubbard1963}) is ubiquitous in physics, chemistry, material science, and biology as it describes in detail the interactions between
electrons and atomic nuclei which are the building blocks of atoms, molecules, and solids.
These interactions completely determine the physical and chemical properties of atomic, molecular, and
condensed matter systems and a deep understanding of these properties therefore requires the solution of the Schr\"odinger equation. This is, however, a very difficult problem for realistic systems due to the Coulomb interaction between the electrons which prohibits the decoupling of the many-electron Schr\"odinger equation \cite{fetter2003, stefanucci2013} into single-particle problems, which can be solved efficiently on modern computers (see e.g.~\cite{andrade2012, leforestier1991}). 
Therefore, in principle, we have to treat the huge number of degrees of freedom of an interacting many-body system explicitly, which can only be done for simple, i.e., small systems. This exponential increase of complexity with the number of interacting particles is known as the exponential wall \cite{kohn1999}. Several approaches (see e.g.~\cite{fetter2003, stefanucci2013, bonitz1998}) have been developed that try to avoid this exponential scaling by considering reduced quantities instead of the full many-body wave function. 

For the description of electronic systems in their ground state one of the most successful of these approaches \cite{burke2012} is density-functional theory (DFT) \cite{gross1995, engel2011}, which allows to determine the exact ground-state observables by only knowing the one-particle density. The foundation of ground-state DFT is the (time-independent) density-potential mapping that was first established in the seminal paper of Hohenberg and Kohn \cite{hohenberg1964}. By applying the Rayleigh-Ritz minimal principle of quantum mechanics, they could show that there exists a one-to-one correspondence, i.e., a bijective mapping, between the set of ground-state densities and their respective external scalar potentials. Densities which are connected via the solution of a Schr\"odinger equation to an external potential are termed $v$-representable.
The second cornerstone of DFT is the Kohn-Sham construction that allows to determine the density of an interacting quantum system by considering an auxiliary non-interacting system \cite{kohn1965} (and thus decoupling the problem into single-particle problems). The Kohn-Sham construction actually employs a composition of the Hohenberg-Kohn mapping of an interacting and a non-interacting system. To be able to do so, one has to assume that the set of interacting and non-interacting ground-state densities is the same, i.e.~that every interacting density is also non-interacting $v$-representable \cite{gross1995, engel2011}. That this is true has been shown under certain restrictions \cite{chayes1985,eschrig1996,lammert2010,kvaal2014}, but in the most general situation \cite{lieb1983} this is still an open issue\footnote{The main reason is that the Lieb functional (a generalization of the Hohenberg-Kohn functional to arbitrary densities with finite kinetic energies) is not functionally differentiable at the $v$-representable densities in the usual Banach-norm sense.}.
However, what has been proven so far \cite{chayes1985,eschrig1996,lammert2010,kvaal2014}
already provides a sound theoretical basis for numerical implementations of the Kohn-Sham method
which is gratifying since 
it is currently the most widely used electronic structure method in solid state
physics and quantum chemistry \cite{burke2012}.

The discussion so far was on ground-state properties. However, 
for the description of dynamical properties of electronic systems an extension of the ground-state
formalism of DFT is required. The first such extension of DFT to dynamical systems was discussed by Peuckert \cite{peuckert1978}.
This work assumed the existence of a time-dependent density-potential mapping, but did not present a formal proof of a bijective mapping between a set of time-dependent external scalar potentials and their respective time-dependent one-particle densities.
The formal justification for time-dependent density-functional theory (TDDFT) \cite{marques2012,ullrich2012} was presented in \cite{runge1984}, where Runge and Gross showed a one-to-one correspondence between time-analytic potentials and their respective time-dependent densities based on the (divergence of the) local-force equation of quantum mechanics \cite{martin1959}. 
Similar to the ground-state case the construction of a corresponding Kohn-Sham scheme requires that the set of interacting $v$-representable densities is the same as the set of non-interacting $v$-representable densities.
A first proof of the existence of a time-dependent Kohn-Sham scheme \cite{vanleeuwen1999} has also been based on the local-force equation (and the assumption of time-analytic potentials as well as densities). Again, the mapping from a time-dependent interacting to an auxiliary non-interacting system is the composition of two density-potential mappings and demands that the set of densities does not depend on the specific interaction potential.
We further like to mention that an alternative proof of the Runge-Gross theorem exists in the linear response regime. One can prove the invertibility of the density response function for perturbations from a non-degenerate
ground state \cite{vanleeuwen2001,ruggenthaler2012b}. This relaxes the constraint of having Taylor expandable external potentials and only requires that the Laplace transform in time of the potentials exist. However, since in this review we want to deal with the density-potential mapping in
its most general context we will not consider this more specific case.

Since TDDFT is a younger field of research its mathematical foundations are not as established as that of ground-state DFT. Also the required mathematical proofs are of a different nature.  Ground-state DFT singles out a specific state, namely the ground state, which can be obtained from a minimum principle.
Many of the proofs are therefore based on certain properties (such as convexity) of energy functionals
which are peculiar to the ground-state problem. The basis of TDDFT on the other hand is formed
by the time-dependent Schr\"{o}dinger equation which describes an initial-value problem, i.e.~for a
given initial state its time-evolution is to be sought. Mathematical proofs must therefore be based
on evolution equations. The ground state does not play any special role except that it can be chosen
as the initial state of a time-evolution.

In the following we will give an extensive review of the density-potential mapping in TDDFT, which, as explained in the preceding short historical introduction, forms the basis of TDDFT and the time-dependent Kohn-Sham approach. We will not discuss any approximations that are needed to perform TDDFT in practice, but will rather concentrate on what is known about the exact properties of the one-to-one correspondence between densities and potentials. In the following we will first set the stage for the closer investigation of the density-potential mapping. We highlight how the density appears as a fundamental and useful quantity in time-dependent quantum mechanics in Sec.~\ref{dens_pot_map}. Assuming the existence of a density-potential mapping, in Sec.~\ref{sec:KohnSham} we demonstrate how Kohn-Sham TDDFT can help to avoid the exponential wall and approximately determines properties of large quantum systems. Before we consider the direct mapping of potentials to densities we first illustrate in several examples the intricacies of the time-dependent Schr\"odinger equation in Sec.~\ref{sec:wavepacket}. This section will illustrate that we carefully need to take into account the unboundedness of the Schr\"odinger Hamiltonian if we want to discuss the density-potential mapping. In Sec.~\ref{sec:history} we give a short overview of the history of the mathematical treatment of explicitly time-dependent initial-value problems, before we define exactly what we mean by a solution of the time-dependent Schr\"odinger equation in Sec.~\ref{sec:Cauchy}. Next we give precise conditions for the existence of such solutions in Sec.~\ref{sec:existence}. We will then properly define the potential-density mapping in Sec.~\ref{sec:observables}. 

After the definition of $v \mapsto n$ we will give an analytical example of the inverse mapping $n \mapsto v$ in Sec.~\ref{sec:Example} and discuss certain implications. In Sec.~\ref{sec:IterativeScheme} we present an iterative procedure to construct the density-potential mapping in the most general case. For this iteration to be possible we need that a certain operator is invertible, which will be discussed in Sec.~\ref{sec:SturmLiouville}. Based on these considerations we will present the famous Runge-Gross theorem that establishes the existence of a density-potential mapping for analytic potentials in Sec.~\ref{sec:Uniqueness}. Then in Sec.~\ref{sec:FixedPoint} we extend these results to more general potentials and give conditions such that the iterative construction of the density-potential mapping converges.

How one can apply the iterative procedure in the case of lattice systems (and hence circumvent some of the mathematical problems of the continuum case) is shown in Sec.~\ref{sec:Lattice}. Further, in Sec.~\ref{sec:numerics} we present a numerical scheme that is able to construct the potential for a given density and initial state for interacting and non-interacting systems. Next we show how the ideas developed for the Schr\"odinger equation can be extended to systems with vector potentials and photons in Sec.~\ref{sec:extension}. Finally we conclude with a summary and outlook in Sec.~\ref{sec:summary}.

\subsection{The time-dependent many-particle problem and the density-potential mapping}

\label{dens_pot_map}

Before we go in much more detail into the properties of the density-potential mapping we want to describe
here in general terms the main ideas and motivations for the consideration of such a mapping.
We start with a discussion of the  time-dependent Schr\"odinger equation (TDSE).
The TDSE of an $N$-electron system is a partial differential equation of the form
\bea
\imagi \partial_t \Psi (\ux, t) &=& \hH (t) \Psi (\ux,t) \label{eq:TDSE_1}\\
\Psi (\ux,t_0) &=& \Psi_0 (\ux) \nonumber
\eea
where $\ux$ is the collection of spatial and spin variables of the system, i.e.~$\ux=(\bx_1, \ldots,\bx_N )$ 
where $\bx_j=\bbr_j \sigma_j$ is a space-spin variable in which each spin variable $\sigma_j$ can have two discrete values. The wave function $\Psi_0 (\ux)$ is the prescribed
state at the initial time $t_0$, which is often taken to be an eigenstate of a time-independent Hamiltonian.
The Hamilton operator $\hH (t)$ contains all the information on the system and has the structure 
\be
\label{TDSEHamiltonian}
\hH (t) = \hT + \hW + \hV (t),
\ee
where $\hT$ represents the kinetic energy operator, $\hV (t)$ the time-dependent external potential and $\hW$ the
two-body interactions. Their explicit form in first quantization is (in atomic units)
\bea
\hT &= - \frac{1}{2} \sum_{j=1}^N \nabla_j^2, \nonumber \\
\hW &= \sum_{i > j}^N w ( \bbr_i - \bbr_j ), \\
\hV (t) &= \sum_{j=1}^N v (\bbr_j,t). \label{potential} \nonumber  
\eea
The two-body interaction $w$ is in realistic applications almost always the Coulomb potential $w(\bbr)=1/|\bbr |$ 
(for the mathematical considerations in this review it should at least have the property that $w(\bbr)=w(-\bbr)$). 
The external potential $v(\bbr,t)$ is typically the sum of
a static and a time-dependent part and depends on the physical system of interest. For instance, for a general molecule with $M$ fixed atomic nuclei it is given by
\be
\label{MolecularPotential}
v (\bbr,t) = - \sum_{\alpha=1}^M \frac{Z_\alpha}{|\bbr- \mathbf{R}_\alpha |} + v_{\ext} (\bbr,t),
\ee 
where $Z_\alpha$ and $\mathbf{R}_\alpha$ are the charge and position of atomic nucleus $\alpha$ and $v_{\ext} (\bbr,t)$ is an externally applied field, for example a laser pulse (which can be approximately described by a scalar
potential in some suitable approximation).
The many-body problem is now completely defined. The task is to solve the TDSE and once the wave function is obtained we can calculate all physical observables of interest. The central problem which remains in practice, is to do this in a preferably efficient manner for realistic systems of
physical interest. This is also often referred to as the quantum many-body problem.\\
An important observation is now that when, for instance, studying different molecules with the same number of electrons we only change the
potential $v$ but always keep the functional form of the two-body interaction and the kinetic energy operator the same.
 Any system of interest in this setting is fully specified by the external potential and the initial state, i.e., the wave function that we calculate from the 
TDSE is a functional of $v$ for a given initial state $\Psi_0$. We can therefore talk about a mapping 
$v \mapsto \Psi [v] $ from potentials to wave functions for a given initial state. 
This idea immediately raises the question for which class of potentials the initial-value problem of the TDSE has
an acceptable solution. One can, for instance, imagine that for singular enough potentials (for instance making the Hamilton operator unbounded from below) a solution 
may not exist or develop undesirable properties such as infinite expectation values for certain physical observables such as the kinetic energy. 
This is a question about the proper domain of the mapping  $v \mapsto \Psi [v] $ (see Sec.~\ref{sec:existence}). 
For the moment it will suffice that we consider a domain of physically reasonable potentials and refer for a more detailed discussion to later sections of this review.
If for a given set of potentials we can construct the many-body state $\Psi [v]$ then we can also construct any observable of interest as a functional of $v$. In particular if $\hA$ is any operator representing a physical quantity then its expectation value
\[
A ([v ],t) = \langle \Psi  ([v],t) | \hA  \Psi ([v], t) \rangle
\]
is a functional of the external potential $v$. It is clear that $A([v],t)$ is a functional of the potential $v$ but
the converse is, in general, not true as $A([v],t)$ could be an object of lower dimensionality than the potential $v(\bbr,t)$. For instance, there could be many different potentials $v(\bbr,t)$ that all generate the same time-dependent dipole moment $d(t)$ of some molecule.
For a given initial state $\Psi_0$ the knowledge of the function $A([v],t)$ is, in general, not sufficient to determine the potential $v$. There is, however, a natural variable
for which such an inversion is possible, namely the time-dependent density. The reason is that this observable is related in a special way to the potential. Let us explain this in more detail.
We can rewrite (\ref{potential}) as
\be
\hV (t) = \int \diff \bbr \, \hat{n} (\bbr) v (\bbr,t),
\label{V_expec}
\ee
where 
\[
\hat{n} (\bbr) = \sum_{j=1}^N \delta (\bbr - \bbr_j)   \nonumber
\]
is called the density operator. Its expectation value
\[
n([v],\bbr,t) = \langle \Psi  ([v],t) | \hat{n} (\bbr) \Psi  ([v],t) \rangle   \nonumber
\]
is called the particle density or simply density. This is a physical quantity for which
$n(\bbr,t) d \bbr$ gives the probability to find a particle in a small
volume $d \bbr$ around point $\bbr$ \cite{stefanucci2013}.
More explicitly this can be written as
\be
n (\bbr, t) = N \sum_{\sigma} \int \diff\ux^{N-1} \, |\Psi (\ux,t)|^2,
\label{density_def}
\ee
where the volume element $\int \diff\ux^{N-1}$ refers to a integration of the spatial coordinates and summation over
the spin coordinates of $N-1$ particles. Which of the coordinates these are is not relevant if we take $|\Psi (\ux,t)|^2$ properly symmetrized under interchange of particles. The coordinates that we integrate over we can label
by $\bx_2,\ldots, \bx_N$ and the space-spin-coordinate not included in the volume element $\diff\ux^{N-1}$ 
we will call $\bx=(\bbr,\sigma)$.
Due to special form of (\ref{V_expec}) an equation of motion, known as the local-force equation,
can be derived that directly relates the density $n(\bbr,t)$ and the potential $v(\bbr,t)$. Given this equation it can be 
proven that under certain conditions there exists a one-to-one mapping between densities and potentials. 
This equation will be discussed in detail in Sec.~\ref{sec:observables} and therefore we restrict ourselves at this point to an intuitive argument.
We can imagine, at least for slow enough temporal variations of the potential, that by making the potential more attractive in some region of space we will increase the probability of finding particles in this region and therefore its particle density.
From this intuitive picture we may conjecture that to a given density profile $n(\bbr,t)$ there corresponds a unique potential $v(\bbr,t)$ that produces it for a given initial state. It is one of the main topics of this review to specify in which sense this statement is true. As a first step we note that if we change the potential by a purely
time-dependent function $v(\bbr,t) \mapsto v(\bbr,t) + C(t)$ then the wave function solving the TDSE will only change by a time-dependent phase factor, i.e.~$\Psi \mapsto \e^{\imagi \alpha (t)} \Psi$ where $\alpha (t)=-\int_{t_0}^t \diff t' C(t')$,
and therefore will not change any of the physical observables. Such a change of the potential is just a gauge and we will regard potentials that differ only in a
gauge as physically equivalent. Therefore the more correct conjecture is that to a given density profile $n(\bbr,t)$ there corresponds a unique equivalence class of
potentials that produces it for a given initial state. In order to specify a unique potential within this class we can always choose a particular gauge. For instance, if the class of potentials that we consider remains spatially constant at 
$|\bbr| \rightarrow \infty$ we can choose a gauge such that $v(\bbr,t) \rightarrow 0\, (|\bbr| \rightarrow \infty)$. 
Assume that we have made a particular gauge choice, then our conjecture implies that there exists a
mapping $n \mapsto v$ which, for a given initial state $\Psi_0$, maps the density to the potential
that generates it. 
This means that, rather than parametrizing the wave function by the potential, we could parametrize it by the
density and write $\Psi [n]$. This implies that, in particular, the expectation value of an observable $\hA$ will be a
functional of the density and we can write
\[
A ([n],t) = \langle \Psi ([n],t) | \hA  \Psi ([n],t) \rangle .
\]
These considerations form the basis of TDDFT as we will summarize in the next section
and in which it will be shown that the existence of a density to potential mapping is used to derive the Kohn-Sham
equations.
From a more mathematical point of view the possible existence of a density to potential mapping immediately raises the new question for which prescribed density profiles such a potential can be found. In other words, what is the domain of the density-potential mapping? It would, for instance, be very useful to know if for every density profile $n$ with certain continuity or differentiability properties there exists a potential $v$ with certain other specified properties that generates it.
Densities which are produced by a potential $v$ by solving the TDSE are called $v$-representable and the problem of the characterization of the set of $v$-representable densities is often referred to as the $v$-representability problem. The elucidation of this problem is one of the main topics of this review.
 
\subsection{Summary of time-dependent Kohn-Sham density-functional theory} 

\label{sec:KohnSham} 

This section gives a summary of the Kohn-Sham (KS) approach to TDDFT with a focus on conceptual issues. The KS equations are derived under the assumption of the existence of the density-potential mapping which will be discussed in detail later in this review.
The KS formalism is heavily used in applications and the practitioner of TDDFT is in this way already introduced to a familiar framework before going deeper in the more fundamental issues of existence and uniqueness of the density functionals.

The central idea of KS DFT is to associate with an interacting
system an effective noninteracting system, known as the KS system, which has the same time-dependent density as the true system. Since the KS system is a noninteracting system it is much easier to treat in numerical applications than the fully interacting system that we started out with. 
The price we pay for this simplification is that the one-body potential of the KS system, known as the KS potential, is an in general unknown functional of the density.
Despite this difficulty it has turned out that practically useful approximations for this potential can be devised.
 \\
Let us describe the KS approach in more detail.
The existence of a density-potential mapping $n \mapsto v$ for a fixed initial state $\Psi_0$, which we also denote by $v[\Psi_0,n]$, is assumed not to depend on the chosen two-body interaction for a physically reasonable class of interactions
(we will be more specific later). Specifically this means that we have a density-potential mapping for interacting as well as noninteracting systems. For the case of a noninteracting system this mapping is called $v_s[\Phi_0,n]$ 
(the subscript $s$ is usually not explained in the density-functional literature but we can assume that it refers to ``single particle'' as the potential often appears in effective single-particle equations as explained below).
Since in this case we have no two-body interactions the Hamiltonian is then simply given by
\be
 \hH_s (t) = \sum_{j=1}^N \left[ - \frac{1}{2} \nabla_j^2 + v_s (\bbr_j, t) \right].
\ee
Let us assume that this is a closed-shell system of $2N$ electrons and that the initial state $\Phi_0$ of the noninteracting system is chosen to be a single Slater determinant of orbitals $\varphi_{j,0} (\bbr)$. 
This allows us to reduce the TDSE for the noninteracting system to single-orbital equations of the form
\bea
\imagi \partial_t \varphi_j (\bbr,t) &= \left[  - \frac{1}{2}  \nabla^2 + v_s ([\Phi_0,n],\bbr ,t)  \right] \varphi_j (\bbr,t)  \label{KS_1} \\
n(\bbr,t) &= 2 \sum_{j=1}^N |\varphi_j (\bbr,t)|^2, \label{KS_2}
\eea
where the initial conditions are given by $\varphi_j (\bbr,t_0) = \varphi_{j,0} (\bbr)$.
For a given density profile $n(\bbr,t)$ we have to construct a proper initial state $\Phi_0$ producing the initial density
$n(\bbr,t_0)$ (this can be done in practice, for example, using the so-called Harriman construction \cite{harriman1981}) and the initial current \cite{lieb2013, tellgren2014}. Once this is done
our assumption guarantees the existence of the potential $v_s [\Phi_0,n]$.
If the density that we prescribe happens to be the density of an interacting system then we have succeeded
in reproducing this density within a noninteracting framework. However, this is not relevant in practice, since
the (\ref{KS_1}) and  (\ref{KS_2}) do not allow us to predict the density of the interacting system as they
obviously do not have any information on which interacting system we would like to solve. To make a
predictive scheme we have to connect the true and the KS system.  To do this we introduce the
KS potential
\be
v_{\KS} [\Psi_0,\Phi_0,n, v_\ext] = v_\ext + v_s [\Phi_0,n] - v [\Psi_0,n],  \label{ks_pot}
\ee
where $v_\ext$ is the external potential of the interacting system of interest.
Note that here we make a careful distinction between $v_s$ and $v_{\KS}$ as they have different functional dependencies. This is usually not
done in the density-functional literature where both quantities are often denoted by the same symbol $v_s$ which can lead to misunderstandings.
If we assume the full knowledge
of the functionals $v [\Psi_0,n]$ and $v_s [\Phi_0,n]$ then the set of equations
\bea
\imagi \partial_t \varphi_j (\bbr,t) &= \left[  - \frac{1}{2}  \nabla^2 + v_\KS (\bbr ,t)  \right] \varphi_j (\bbr,t) \label{KS_3} \\
n(\bbr,t) &= 2 \sum_{j=1}^N |\varphi_j (\bbr,t)|^2 \label{KS_4}
\eea
does have a unique solution for a self-consistent density $n_\mathrm{sc}$. By definition of $v_s [\Phi_0,n]$ this
self-consistent density is exactly obtained whenever
\be
v_{\KS} [\Psi_0,\Phi_0,n_\mathrm{sc}, v_\ext] =  v_s [\Phi_0,n_\mathrm{sc}],
\ee
which according to (\ref{ks_pot}) is precisely satisfied when
\be
v_\ext = v[\Psi_0, n_\mathrm{sc}].
\ee
In turn, this is exactly true when $n_\mathrm{sc}$ is equal to the density produced by the potential $v_\ext$ in the
interacting system with initial state $\Psi_0$, which is precisely the density that we are interested in. 
The procedure that we outlined here therefore comprises a predictive computational scheme to calculate the
density of an interacting system of interest with the single-orbital equations (\ref{KS_3})  and (\ref{KS_4}). To make the scheme useful in practice we need an approximation to the functional $v_s [\Phi_0,n] - v [\Psi_0,n]$ appearing
in (\ref{ks_pot}). Usually this functional is split into two pieces as follows
\be 
\label{xcPotential}
v_s [\Phi_0,n] - v [\Psi_0,n] = v_\Ha [n] + v_\xc [\Psi_0,\Phi_0,n],
\ee
where $v_\Ha$ is the Hartree potential defined as
\be
v_\Ha ([n];\bbr , t) = \int \diff \bbr' \, w( \bbr-\bbr' ) \, n(\bbr',t) ,
\ee
which describes a mean-field classical electrostatic potential between the electrons, and $v_\xc$ is the so-called exchange-correlation (xc) potential.
By this redefinition we have shifted all difficult functional dependencies to the xc potential. Putting everything together we recover the standard KS equations as they appear 
in textbooks
\bea\label{KS_5}
\imagi \partial_t \varphi_j (\bbr,t) =& \left[-\frac{1}{2}  \nabla^2 + v_\ext (\bbr ,t) + v_\Ha ([n],\bbr,t) \right. \\
& + v_\xc ([\Psi_0,\Phi_0,n], \bbr,t) \bigg] \varphi_j (\bbr,t), \nonumber
\eea
\be
n(\bbr,t) = 2 \sum_{j=1}^N |\varphi_j (\bbr,t)|^2 ,
\ee
although most textbook discussions are less precise and do not indicate the initial state dependencies of the xc potential explicitly. The main obstacle for applying these equation to the calculation of electronic  properties is obtaining a good approximation for the xc potential. Several useful approximations for this quantity have been developed. 
The discussion of such approximations is, however, not the aim of this review. For a comprehensive overview of approximate xc potentials and their applications we refer to a recent textbook on TDDFT \cite{ullrich2012}.

\section{Overview of the time-dependent Schr{\"o}dinger equation}
 
\label{sec:TDSE}

The very foundation of the density-potential mapping is that for a given initial state the solution of the TDSE for two different external potentials (differing more than a gauge) 
leads to two different densities, i.e.~the mapping from potentials to densities is injective. However, to investigate the basic properties of this mapping, we first need to properly define it. 
To do so, we need to specify the domain of the mapping, i.e., the set of potentials, the codomain or range of the mapping, i.e., the set of densities, and the rule how the potentials are mapped to their respective densities.
It is obvious that we want to have a domain such that for every potential in this set we can actually solve the TDSE uniquely. Therefore we want to investigate under which conditions the TDSE has a unique solution. Although in general this is tacitly assumed, for a proper investigation of the density-potential mapping we need to know specifics. 
Thus, in Sec.~\ref{sec:Cauchy} we discuss what we mean by a solution to the TDSE and set the stage for a precise presentation of the density-potential mappings in TDDFT. We introduce the notion of classical and non-classical solutions to the initial-value problem. These non-classical solutions arise since the Hamiltonians in quantum mechanics are usually unbounded operators and thus can lead to infinities. To make sense of these generalizations we also need to present some (in physics often ignored) details about self-adjoint operators. We briefly discuss the idea of a self-adjoint domain and give conditions on the two-body interactions and the potentials such that the resulting Hamiltonians are self-adjoint on a common domain. We then present conditions (on the external potentials) for the existence of unique solutions to the TDSE. Regularity properties, e.g., under which conditions we have classical solutions, are discussed as well. In a next step we then investigate physical quantities derived from the wave functions. We give exact conditions such that the density obeys the continuity equation, and discuss which restrictions we need to impose on the potential in order to obey the fundamental equation of TDDFT. 
Since the rigorous discussion of the TDSE involves some abstract concepts we will first illustrate these concepts by discussing a simple but very common physical situation, namely the free propagation of a wave packet.

\subsection{Free propagation of a wave packet}

\label{sec:wavepacket}

\begin{figure} [H]
\centering
\includegraphics[width=9cm]{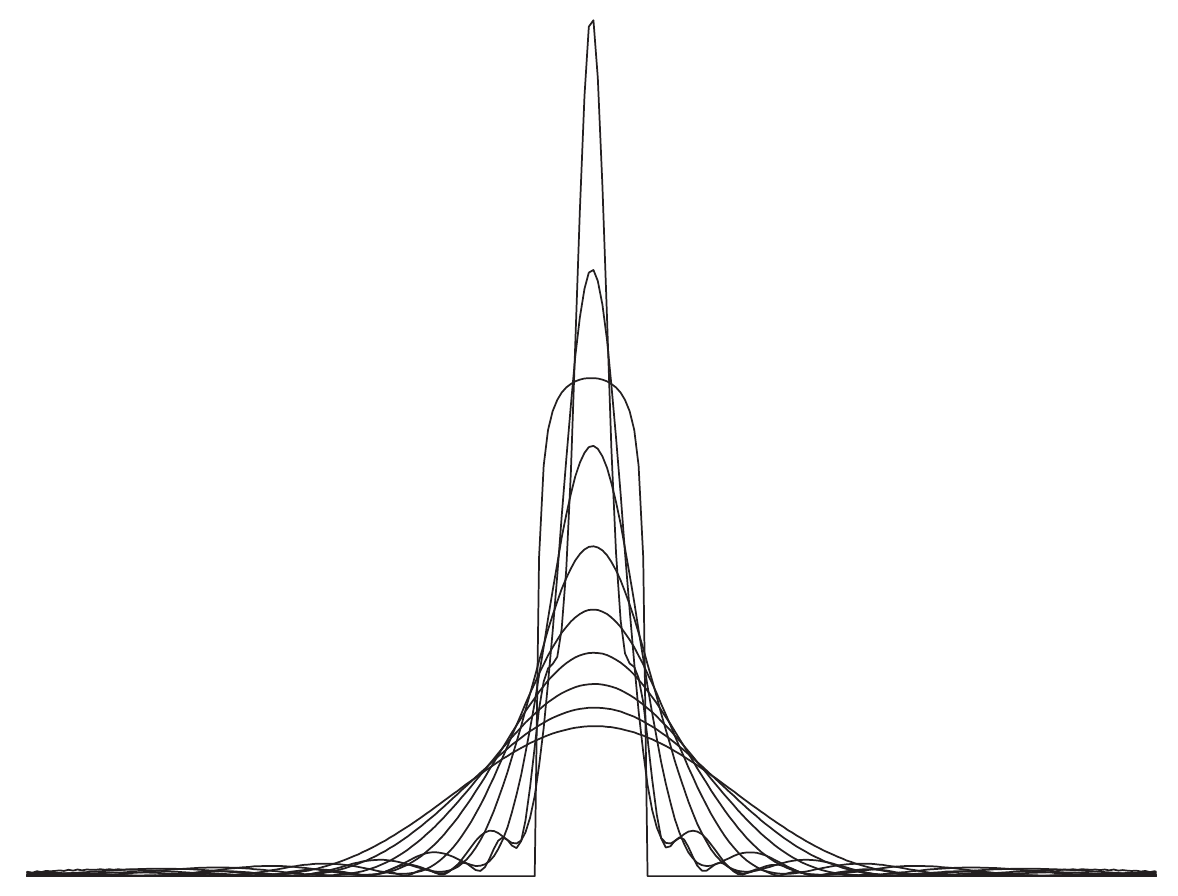}
\caption{Spreading of the density of example (\ref{Psi0}) by free propagation at different time snapshots. The initial bulge first rises in the centre, then spreads out smoothly.}
\label{fig:test-fct-spread}
\end{figure}

Rather than formally discussing the TDSE at this point let us first point out some issues one stumbles upon when trying to solve the
equation in practice. These problems then will force us later to adopt a more careful and rigorous approach. We will start to consider the simple case of one particle in one dimension
enclosed in a box of length $L$. 
The corresponding Hilbert space $\mathcal{H}$ is given by the square integrable functions in the box, or more formally $\mathcal{H} = L^2 ([0,L])$ with standard inner product
\be
\langle \Psi | \Phi \rangle = \int_{0}^{L} \diff x \, \Psi^* (x) \Phi (x)  \nonumber 
\ee
between functions $\Psi$ and $\Phi$. Using the inner product we can assign to any function $\Psi$
in the Hilbert space the norm $\| \Psi \| = \sqrt{\langle \Psi | \Psi \rangle}$. 
Let us now start by considering a simple free evolution of a single particle
wave packet. This means that we want to calculate a state $\Psi (t) \in \mathcal{H}$ at time $t$ from a given initial
state $\Psi (0) \in \mathcal{H}$. 
The TDSE (in atomic units) for this problem is given by
\be
\imagi \prt \Psi (x,t) = \hH \Psi (x,t),  \quad \quad \quad \hH= - \frac{1}{2} \frac{\diff^2}{\diff x^2 } 
\label{TDSE1}
\ee
and we specify the initial state $\Psi (0)$ at time $t=0$. We have not specified yet any properties of the initial state $\Psi (0)$ but it is reasonable to assume that it is twice differentiable such that the action of the Hamiltonian $\hH$ on it is well-defined (further conditions will follow soon). A formal solution of (\ref{TDSE1}) is given by
\bea
\Psi (x,t) &=&  \e^{-\imagi \hH t} \, \Psi (x,0) =  \sum_{n=0}^{\infty} \frac{(-\imagi t \hH)^n}{n!} \Psi (x,0) \nonumber \\
&=& 
\sum_{n=0}^{\infty} \left(  \frac{\imagi t}{2} \right)^n\frac{1}{n!} \frac{\diff^{2n}}{\diff x^{2n}} \Psi (x,0),
\label{TDSE_formal}
\eea
where the exponent of an operator is formally defined by its Taylor series. This, of course, assumes that the infinite series converges in some norm sense, which we did not check at this point
(and in fact turns out to be false in general).
One sees immediately that problems arise with (\ref{TDSE_formal}) whenever the function $\Psi (x,0)$ is only a finite times differentiable so let us assume
that $\Psi (x,0)$ is infinitely differentiable on the real line (more technically $\Psi (x,0) \in \mathcal{C}^\infty ([0,L])$). It is, of course, already suspicious that
we have to demand this infinite smoothness condition for the initial state when the TDSE only contains second spatial derivatives, but let us ignore this issue for the moment and 
simply continue. To be specific we take the initial state to be \cite{Cautionary}
\be
\Psi (x,0) = \left\{ \begin{array}{cc} \exp \left( \frac{1}{(x-a)^2-b^2} \right)  & |x-a| < b  \\ 0 & |x-a|  \geq b \end{array} \right.   
\label{Psi0}
\ee
which describes a wave packet localized in the interval $[a-b,a+b]$ where we take $a$ and $b$ to be positive real numbers such that the wave-packet is properly localized 
within $[0,L]$ (for instance we can take $a=L/2$ and $b=L/100$).
One can check that this function is infinitely differentiable and that all derivatives are zero for $|x-a| \geq b$. Let us now see what we get if we insert this initial state into the formula
of (\ref{TDSE_formal}). For $|x-a| \geq b$ the formula then tells us that at any time $\Psi (x,t)=0$ and therefore that the wave packet does not spread and
never leaves the interval $[a-b,a+b]$. This is very much in disagreement with our intuition that free wave packets do spread. 
What has gone wrong? To understand this it is useful to talk about the exponent of a linear operator $\hA$ in a more abstract sense. Let us try to derive some conditions under which the
definition
\be
\e^{\hat{A}} = \sum_{n=0}^\infty \frac{1}{n!} \hat{A}^n
\label{expA}
\ee 
makes sense. First of all when we act with with the exponential operator  on a state $\Psi$ we see that $\hA^n \Psi$ must be well-defined for all $n$.
This is certainly the case if the domain\footnote{The domain $D(\hat{A})$ of an operator $\hat{A}$ is the set of functions $\Psi$ for which $\Psi$ and $\hat{A} \Psi$ are normalizable.} of the operator $\hA$ is the whole Hilbert space, i.e.~$D(\hA)=\mathcal{H}$, since then any square integrable function is mapped to another
square integrable function and we can then apply the operator repeatedly. Let us therefore assume that $\hA$ is defined on all of $\mathcal{H}$. Then if we act with $\e^{\hA}$ on a state $\Psi$ then
every term in the sum (\ref{expA}) is well-defined. However, this does not mean that the infinite sum converges. To guarantee this we must have that
$\e^{\hA} \Psi \in \mathcal{H}$ or equivalently $\| \e^{\hA} \Psi \| < \infty$. We therefore want to make sense of the sum
\be
\| \e^{\hA} \Psi \| = \left\| \sum_{n=0}^\infty \frac{1}{n!} \hat{A}^n \Psi \right\| \leq \sum_{n=0}^\infty \frac{1}{n!} \| \hA^n \Psi \|
\label{exp_norm}
\ee
as a sufficient condition. For the right hand side to give a finite sum the terms $\| \hA^n \Psi \|$ should not grow too fast with $n$.
An estimate can be made for so-called bounded operators for which there exists a positive number $M$ such that
\be
\| \hat{A} \Psi \| \leq M \| \Psi \|
\ee
for all states $\Psi$ in the Hilbert space.
In particular, repeated use of this inequality implies that $\| \hat{A}^n \Psi \| \leq  M^n \| \Psi \|$. If we use
this in (\ref{exp_norm}) we have
\be
\| \e^{\hA} \Psi \| \leq \sum_{n=0}^\infty \frac{M^n}{n!} \|  \Psi \| = \e^M \|  \Psi \|
\label{exp_norm2}
\ee
This means that (\ref{expA}) is well-defined for bounded operators. The problem with the Hamiltonian
$\hH$ in our example is that it is not a bounded operator. Indeed a more careful analysis of the norms $\| \hH^n \Psi \|$ for our wave packet \cite{Cautionary} shows that these grow so fast with $n$ that the rightmost infinite sum in (\ref{exp_norm}) diverges if we take $\hA=-\imagi t \hH$ and $\Psi$ to be our initial wave packet (moreover there is also pointwise divergence for any $|x-a| < b$ in (\ref{TDSE_formal}) ~\cite{Cautionary}). As a consequence (\ref{TDSE_formal}) does not present the solution to our initial-value problem. Suppose now, however, that we discretize the TDSE of (\ref{TDSE1}) on
a finite spatial grid, i.e.~we replace the differential operator by a finite matrix. In that case our Hilbert space
is finite dimensional and $\hH$ becomes a bounded operator and therefore the exponential $\exp(-\imagi t\hH)$ is well-defined by the series expansion. This immediately raises the question what happens when we make our grid spacing finer and finer and take a continuum limit. It will be instructive to do this calculation.
The grid points $x_j=j \Delta x$ are labelled by an integer $j$. 
The second spatial derivative of $\Psi (x,t)$ in grid point $x_j$ can be approximated
by the finite difference formula
\be
\frac{\diff^2\Psi}{\diff x^2} (x_j,t) \approx \frac{1}{(\Delta x)^2} ( \Psi (x_{j+1},t) - 2 \Psi (x_j ,t) + \Psi (x_{j-1},t)),
\ee
where $j=1,\ldots,N$. We see that the determination of the second derivative of $\Psi$ in $N$ grid points requires the knowledge of the $\Psi$ in $N+2$ grid points. We use this feature to include the hard-wall boundary conditions. 
The first grid point is taken to be $x_0=0$ and the last will be $x_{N+1}=(N+1) \Delta x=L$ where we demand $\Psi (x_0,t)= \Psi (x_{N+1},t)=0$ for all times, leaving $N$ remaining points in between in which we have to determine $\Psi (x_j,t)$. Our Hilbert space will then be $N$-dimensional. By this discretization the Hamiltonian becomes an $N \times N$ matrix acting on the time-dependent vector $(\Psi (x_1,t), \ldots, \Psi (x_N,t))$ with the explicit form
\be
H =  -\frac{1}{2(\Delta x)^2} \left(
\begin{array}{cccccccc} -2 & 1 & 0 & \ldots & \ldots & \ldots & 0 \\  1 & -2 & 1 &  0 & \ldots & \ldots & 0  \\ 
0 & 1 & -2 & 1 & 0 & \ldots & 0\\  
\vdots & & & \ddots& & & \vdots \\
0 & \ldots &  0 & 1 & -2 & 1 & 0\\
0 & \ldots & \ldots& 0 & 1 & -2 & 1 \\
0 & \ldots & \ldots & \ldots & 0 & 1 & -2
 \end{array}\right)
\ee
while the TDSE becomes an ordinary differential equation of matrix form
\be
\imagi \prt \Psi (x_j, t) = \sum_{k=1}^N H_{jk} \Psi (x_k,t).
\ee
In this case, since the Hamiltonian is now a bounded operator, we can take the exponential of a matrix and
we find that
\bea
\Psi (x_j,t) &= \sum_{k=1}^N \left( \e^{-\imagi t H} \right)_{jk} \Psi (x_k,0) \nonumber\\
&= \sum_{k=1}^N \sum_{m=0}^\infty \frac{(-\imagi t)^m}{m!}\left( H^m \right)_{jk} \Psi (x_k,0).
\label{exp_matrix}
\eea
It remains to calculate the action of $H^m$ on the initial state. To do this we expand the initial state in the eigenvectors of $H$.
Since $H$ is a symmetric matrix it has $N$ real eigenvalues $\epsilon^{(l)}$ and the eigenvectors $\varphi^{(l)} (x_j)$ are orthogonal. These are determined from the
equation
\be
\sum_{k=1}^N H_{jk} \varphi^{(l)} (x_k) = \epsilon^{(l)} \varphi^{(l)} (x_j).
\ee
The eigenvectors turn out to be real as well for the case of symmetric matrices.
If we therefore normalize the eigenvectors such that
\be
\Delta x \sum_{j=1}^N \varphi^{(k)} (x_j) \varphi^{(l)} (x_j) = \delta_{kl},
\label{ortho}
\ee
then we find that
\bea
\varphi^{(l)} (x_j) &=& \sqrt{\frac{2}{L}} \sin \left( \frac{l \pi x_j}{L} \right),  \\
\epsilon^{(l)} &=& \frac{2}{(\Delta x)^2} \sin^2 \left(  \frac{l \pi \Delta x}{2 L} \right),
\eea
where $l=1,\ldots, N$. Let us now expand the initial state $\Psi (x_j,0)$ in terms of the eigenstates of $H$.
We have 
\be
\Psi (x_j,0) = \sum_{l=1}^N c_l \, \varphi^{(l)} (x_j),
\label{init_exp}
\ee
where due to the orthonormality condition (\ref{ortho}) we easily find that the coefficients $c_l$ are given by
\be
c_l = \Delta x \sum_{j=1}^N \varphi^{(l)} (x_j) \Psi (x_j,0) .
\label{coeff}
\ee
With these preliminaries we can continue the evaluation of the infinite sum in (\ref{exp_matrix}).
Inserting (\ref{init_exp}) into this equation we have
\bea
\Psi (x_j,t) &=& \sum_{k,l=1}^N c_l \sum_{m=0}^\infty \frac{(-\imagi t)^m}{m!}\left( H^m \right)_{jk} \varphi^{(l)} (x_k) \nonumber \\
&=&  \sum_{l=1}^N c_l \sum_{m=0}^\infty \frac{(-\imagi t)^m}{m!} (\epsilon^{(l)})^m  \varphi^{(l)} (x_j) 
=  \sum_{l=1}^N c_l \, \e^{-\imagi t \epsilon^{(l)}}  \varphi^{(l)} (x_j).  \nonumber
\eea
If we insert the explicit form of the coefficient $c_l$ of (\ref{coeff}) then we can write this as
\be
\Psi (x_j,t) = \Delta x \sum_{k,l=1}^N  \e^{-\imagi t \epsilon^{(l)}}  \varphi^{(l)} (x_j ) \varphi^{(l)} (x_k) \Psi (x_k,0).
\label{psi_t}
\ee
Or in other words
\be
(\e^{-\imagi t H})_{jk} =  \Delta x \sum_{l=1}^N  \e^{-\imagi t \epsilon^{(l)}}  \varphi^{(l)} (x_j ) \varphi^{(l)} (x_k).
\ee
Now that we have obtained an exact result for our discretized problem we can take the continuum limit.
We let $N \rightarrow \infty$ for a fixed value of $L=(N+1) \Delta x$ for the length of the box. This means
that $\Delta x \rightarrow 0$ and the sum over $k$ in (\ref{psi_t}) becomes a Riemann integral. We then get
\be
\Psi (x,t) = \sum_{l=0}^{\infty} \e^{-\imagi t \epsilon^{(l)}}  \varphi^{(l)} (x ) \int_0^L \diff x' \varphi^{(l)} (x') \Psi (x',0),
\label{psi_t2}
\ee
where
\bea
\varphi^{(l)} (x) &=& \sqrt{\frac{2}{L}} \sin \left( \frac{l \pi x}{L} \right),  \\
\epsilon^{(l)} &=& \frac{(l \pi)^2 }{2 L^2}.
\eea
We recover the well-known eigenfunctions and energies for the particle in a box.
We can rewrite (\ref{psi_t2}) as
\be
\Psi (x,t) = \int_0^L \diff x' \, U (x,x',t) \Psi (x',0),
\label{Uexp}
\ee
where formally the propagation kernel
\be
U (x,x',t) =  \sum_{l=1}^{\infty}  \e^{-i t \epsilon^{(l)}}  \varphi^{(l)} (x ) \varphi^{(l)} (x').
\label{Uexp2}
\ee
Strictly speaking the sum is not defined until after integration over $x'$ as in (\ref{psi_t2}) (as at $t=0$ it becomes the delta distribution $\delta (x-x')$)
and moreover it can depend in very complicated manner on the space and time arguments even for simple systems such as the particle in a box \cite{Fulling_box}.
We can now define by (\ref{Uexp}) an evolution operator $\hat{U} (t)$ with the property
\be
\hat{U}(t) \Psi (0) = \Psi (t).
\ee
We write this evolution operator by definition as $\hat{U}(t) = \exp(-\imagi t \hH)$. So rather, than defining the evolution operator by a Taylor series we define it by
a spectral representation involving the eigenfunctions and eigenvectors as in (\ref{Uexp}) \cite{blanchard2003,Teschl}. It also follows from (\ref{psi_t2}) that
\be
\| \hat{U} (t) \Psi (0) \|^2 = \| \Psi (t) \|^2 = \sum_{l=1}^{\infty} |\langle \varphi^{(l)} | \Psi (0) \rangle |^2 = \| \Psi (0) \|^2.
\ee
This means that $\hat{U} (t)$ is a bounded operator which preserves the norm (in other words: it is unitary). 
This is an obvious requirement from a physical point of view as the total probability of finding a particle should be conserved in time.
A more puzzling property is that that $\hat{U} (t)$ is defined on any square integrable function, including non-differentiable functions on which the action of the Hamiltonian $\hH$ is not defined. Before we go into these issues let us now go back to our example of the localized wave-packet.
If we now take $\Psi (x,0)$ as in (\ref{Psi0}) and apply (\ref{psi_t2}) we will find that the wave-packet correctly spreads in the box (see Fig.~\ref{fig:test-fct-spread}).

\begin{figure} [H]
\centering
\includegraphics[width=9cm]{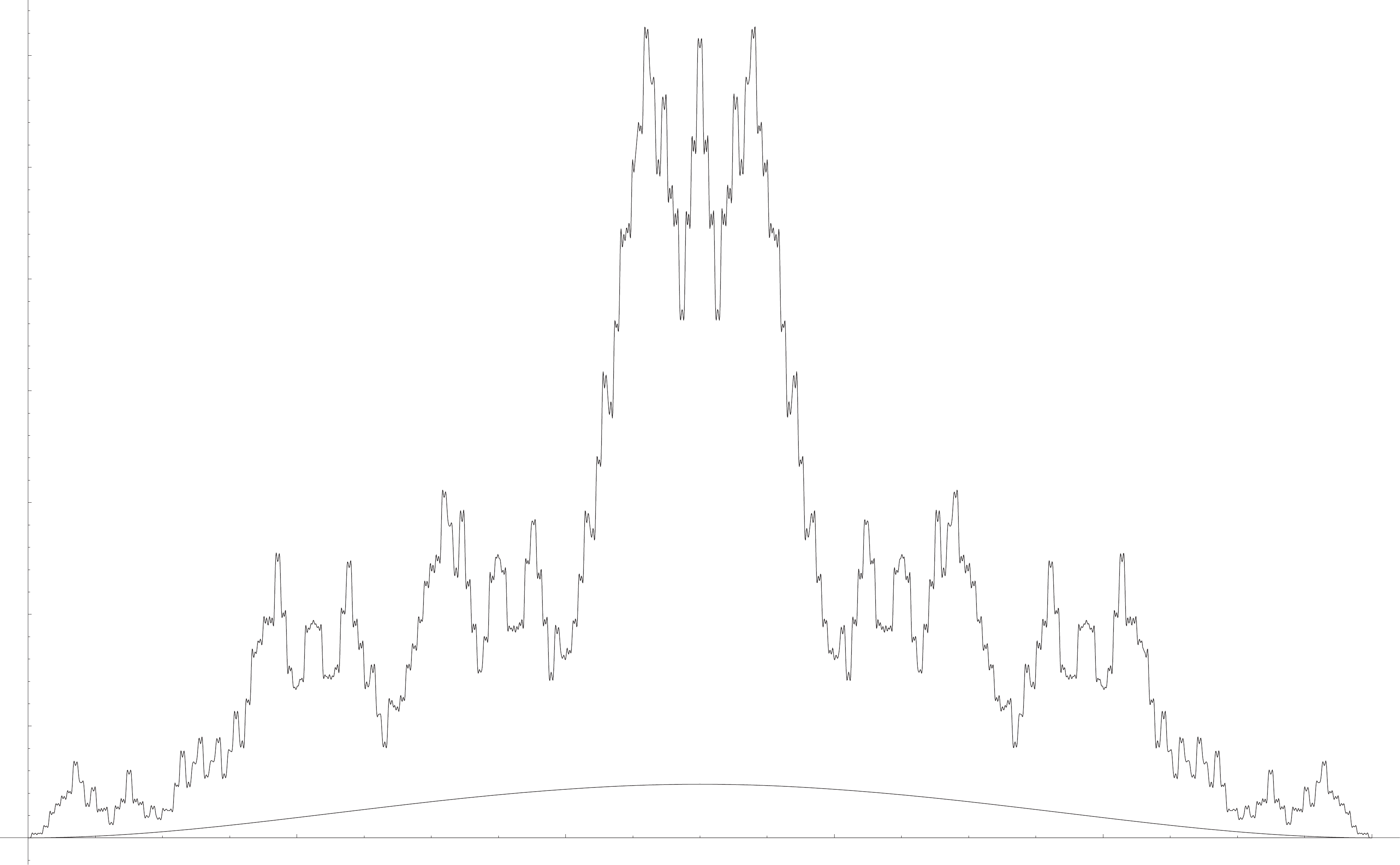}
\caption{$|\Psi(t)|^2$ in a box with $L=1$ evaluated at $t=3/4$ from (\ref{Hall_func}) with 500 terms and the smooth integrated $|\int_0^t \diff s \Psi(s) |^2$ (lower curve)}.
\label{fig:berry_example}
\end{figure}

One can in fact prove that a wave packet enclosed in
a bounded region of space at time $t=0$ will have tails reaching all over space for almost all times $t \neq 0$ \cite{hegerfeldt1980}. This is not difficult to understand from a physical point of view since the Fourier components of the initially localized wave function have momenta of arbitrarily high value allowing the particle to move arbitrarily fast. This also implies that the localized wave packet in an enclosed big box with hard walls will feel the presence of the boundary immediately. If we would have used other boundary conditions, such as the periodic one
$\Psi (0,t)=\Psi (L,t)$, $\diff\Psi (0,t)/\diff x=\diff\Psi (L,t)/\diff x$ then immediately after $t=0$ the time-evolution will be different. 
Clearly the formal series in (\ref{TDSE_formal}) which is just specified by the differentiation rule has no information on such boundary conditions as they were neither
encoded in the initial state nor in the exponential form of the time evolution operator. It is therefore no surprise that (\ref{TDSE_formal}) can not be used to predict the time-evolution correctly. \\
However, not all hope is lost in applying (\ref{TDSE_formal}). Clearly we can apply (\ref{TDSE_formal}) to a finite linear combination of eigenfunctions  of the form
\be
\Psi (x,0) = \sum_{k=1}^{N} \alpha_k \, \varphi^{(k)} (x),
\label{Psi02}
\ee
since for this initial state we obtain from (\ref{TDSE_formal}) that 
\bea
\Psi (x,t) &=& \sum_{n=0}^{\infty} \sum_{k=1}^N \alpha_k  \frac{(-\imagi t)^n}{n!} \hH^{n}  \varphi^{(k)} (x) 
\nonumber \\
&=& \sum_{n=0}^{\infty} \sum_{k=1}^N \alpha_k \frac{1}{n!}  \left( -\imagi t \epsilon^{(k)} \right)^n\varphi^{(k)} (x)
= \sum_{k=1}^N \alpha_k \e^{-\imagi t \epsilon^{(k)}} \varphi^{(k)} (x), \nonumber
\eea
which is easily checked to be a valid solution of the TDSE of (\ref{TDSE1}) with the right initial conditions.
So why did the formal approach work in this case? We first note that in this case the function $\Psi (x,t)$ is a
real-analytic function of $x$ and $t$, i.e., it has a Taylor series with non-zero convergence radius around any points $x$ and $t$ in its domain. Thus
\be
\Psi (x,t) = \sum_{k,l=0}^\infty c_{kl} \, (t-t_0)^k (x-x_0)^l
\ee
around any $(x_0, t_0)$ within its radius of convergence. In fact the function is so nice that it has infinite convergence radius and hence can be extended to the whole complex plane in $x$ and $t$.
Could it be that the formal expression of (\ref{TDSE_formal}) would work for all real-analytic functions? Since the example of our localized wave packet of (\ref{Psi0}) is non-analytic at points $x = a \pm b$ (it has a Taylor series with convergence
radius zero at these points) this would then explain in another way the failure of (\ref{TDSE_formal}).
One can prove that a function that is exactly zero on an interval of the real line can not be real
analytic unless it is the zero function and therefore any initially localized wave packet fails to be real analytic.
Let us give an example which shows that the requirement of analyticity is not sufficient to make (\ref{TDSE_formal}) work. Rather than take the interval $[0,L]$ we take the free propagation of a wave packet on the real line, i.e., our Hilbert space will be $L^2(\mathbb{R})$. The Schr{\"o}dinger equation will again be given by (\ref{TDSE1})
and as initial state we take a Lorentzian function
\be
\Psi (x,0) = \frac{1}{1+x^2} = \frac{\imagi}{2} \left( \frac{1}{x+\imagi} - \frac{1}{x-\imagi} \right) .
\label{Psi0_Lorentzian}
\ee
This is a real analytic function on the whole of the real axis with a convergence radius of at least 1 for any Taylor expansion of $\Psi (x,0)$ in powers of $x-x_0$ around $x_0$. If we insert this initial state into (\ref{TDSE_formal}) we obtain the series
\be
\Psi (x,t) = \sum_{n=0}^{\infty} \left(\frac{\imagi t}{2} \right)^n\frac{(2n)!}{n!} \frac{\imagi}{2} \left( \frac{1}{(x+\imagi)^{2n+1}} - \frac{1}{(x-\imagi)^{2n+1}} \right) .
\ee
From simple convergence criteria we see that this is a divergent series for any value of $x$ and $t$. Therefore real analyticity is not a sufficient criterion to be able to apply (\ref{TDSE_formal}). To get a sufficient condition we follow the classical derivation given by Kowalevskaya \cite{Kowalevskaya}. Define the two time-dependent functions
\be
\Psi^{(0)}(t) = \Psi (x_0,t)  \quad \mbox{and} \quad \Psi^{(1)} (t) = \frac{\diff \Psi}{\diff t} (x_0,t) 
\ee
for which we will assume that they are real analytic. Those are new initial values if we exchange the meaning of $x$ and $t$. This is necessary because \cite{Kowalevskaya} is concerned with initial-value problems where the time derivative appears in highest order.
Then the general solution of (\ref{TDSE1}) can be written as a formal power series
\bea
\Psi (x,t) &=& \sum_{\nu=0}^{\infty} \left(  \frac{\diff^\nu \Psi^{(0)}}{\diff t^\nu}(t) \left( \frac{2}{\imagi}\right)^{\nu} \frac{(x-x_0)^{2 \nu}}{(2 \nu)!}  \right. \nonumber \\
&&+ \left.  
\frac{\diff^\nu \Psi^{(1)}}{\diff t^\nu}(t) \left( \frac{2}{\imagi}\right)^{\nu} \frac{(x-x_0)^{2 \nu +1}}{(2 \nu +1)!} \right),
\label{TDSE_series}
\eea
as can be checked by insertion of this expression into (\ref{TDSE1}). 
Let now the Taylor expansions of $\Psi^{(0)} (t)$ and $\Psi^{(1)} (t)$ be given by
\bea
\Psi^{(0)} (t) = \sum_{\nu=0}^\infty c_\nu (t- t_0)^\nu, 
\label{Psi_0_series} \\
\Psi^{(1)} (t) = \sum_{\nu=0}^\infty c^\prime_\nu (t- t_0)^\nu.  
\label{Psi_1_series}
\eea
Then in terms of the coefficients $c_\nu$ and $c^\prime_\nu$ the expansion (\ref{TDSE_series}) at time $t_0$ attains the form
\bea
\Psi (x,t_0) &=&  \sum_{\nu=0}^{\infty} [ b_\nu (x-x_0)^{2 \nu} +  b^\prime_\nu (x-x_0)^{2 \nu+1}  ] 
\label{TDSE_series2}
\eea
where we defined
\be
b_\nu = \left( \frac{2}{\imagi} \right)^\nu \frac{\nu!}{(2\nu)!}  c_\nu   \quad \quad b^\prime_\nu =  \left( \frac{2}{\imagi} \right)^\nu \frac{\nu!}{(2\nu+1)!}  c^\prime_\nu  .
\ee
Since we assumed $\Psi^{(0)} (t)$ and $\Psi^{(1)} (t)$ to be real analytic functions they have finite convergence radii $R^{(0)}$ and $R^{(1)}$.
Let $R$ be a positive radius smaller than $R^{(0)}$ and $R^{(1)}$. Then since both series (\ref{Psi_0_series}) and
(\ref{Psi_1_series}) converge there exists a positive number $g$ such that
for all $\nu$
\be
|c_\nu| \leq \frac{g}{R^{\nu}}  \quad \quad |c^\prime_\nu| \leq \frac{g}{R^{\nu}} .
\ee
These conditions imply that for the coefficients $b_{\nu}$ and $b^\prime_\nu$ in (\ref{TDSE_series2}) that
\bea
|b_\nu| = \frac{\nu!}{(2 \nu)!}  2^\nu |c_\nu|  \leq \frac{\nu!}{(2 \nu)!} g \left( \frac{2}{R} \right)^\nu \quad \mbox{and}  \nonumber \\
| b^\prime_\nu| = \frac{\nu!}{(2 \nu+1)!} 2^\nu |c^\prime_\nu | \leq \frac{\nu!}{(2 \nu+1)!} g \left( \frac{2}{R} \right)^\nu  .
\label{b_constraints}
\eea
One sees from standard convergence criteria that this implies that $\Psi (x,t_0)$ as a power series in $(x-x_0)$ has an infinite convergence radius.
Now we understand what went wrong when we chose as initial state the Lorentzian function of (\ref{Psi0_Lorentzian}).
The function is real analytic but the radius of convergence is not infinite. The solution of the TDSE will then not
be time-analytic, i.e., series of the form as in (\ref{Psi_0_series}) and (\ref{Psi_1_series}) do not exist for this initial state. 
Note that the requirement of infinite convergence radius by itself is not enough, one really needs to satisfy the constraints (\ref{b_constraints}). For example the function
\be
\Psi (x,0) = \sum_{\nu=0}^\infty \frac{1}{(\nu!)^{\frac{1}{3}}} (x-x_0)^\nu
\ee
has an infinite convergence radius but does not satisfy the constraints (\ref{b_constraints}).
If we now go back to the initial state of (\ref{Psi02}) we can check that it not only has an infinite convergence radius but that it also satisfies the constraints and therefore 
the direct exponentiation worked. Another physical example in which direct exponentiation is allowed is that of an initial Gaussian wave 
packet \cite{Blinder} since also for this case the conditions (\ref{b_constraints}) are satisfied.
The issue of time non-analyticity has been raised in some papers in connection with initial states that are not differentiable at cusps \cite{YangBurke_PRA13}.
However, the analysis already carried out by Kowalewskaya shows that the situation is more severe. A Taylor expansion in time for $\Psi (x,t)$ does not even exist
for a large class of real-analytic initial states without cusps.\\
We see that by putting rather stringent conditions on the initial state we can give the expansion in (\ref{TDSE_formal}) a meaning.
However, by means of the evolution operator defined in (\ref{Uexp}) we can give meaning to time-evolution for a much larger set of initial wave functions. Let us therefore forget again about analyticity and return to this more general definition of the evolution operator.
In fact the expression is defined on any square integrable function, even on functions that are not in the
domain of the Hamiltonian operator. Let us illustrate this with an example for the particle in a box again.
We apply the evolution operator in (\ref{Uexp}) to the normalized initial state given by 
\be
\Psi (x,0) = 1/\sqrt{L}.  
\label{constPsi}
\ee
This initial state can be expanded in the
eigenstates $\varphi^{(l)}$ with expansion coefficients 
\be
\langle \varphi^{(l)} | \Psi (0) \rangle = \frac{\sqrt{2}}{l \pi} (1- (-1)^l),
\ee
which only gives a non-zero value when $l$ is odd. The time-evolution is then given according to (\ref{psi_t2}) as
\be
\Psi (x,t) = \frac{4}{\pi \sqrt{L} } \sum_{k=0}^{\infty} \frac{ \e^{-\imagi \pi^2 (2k+1)^2 t /(2L^2)} }{2k+1} \sin\left[  \frac{(2k+1) \pi x} {L}\right].
\label{Hall_func}
\ee
It turns out that this function for a given value of $x$ is continuous but nowhere differentiable with respect to time. Moreover at almost all times it is continuous but nowhere differentiable as a function of $x$ \cite{Markus_Hall,Berry}.
A snapshot of $|\Psi (x,t)|^2$ is displayed in Fig. \ref{fig:berry_example}.

This result seems puzzling at first sight, how can we get a nowhere differentiable function as a solution of a dynamics governed by a partial differential equation?
One would expect that any solution would at least be once differentiable in time and twice with respect to the spatial coordinates. The problem is that the initial state is not in the domain $D(\hH)$ of the Hamiltonian. What one can show is that if the initial state is in the domain of the Hamiltonian then a well-behaved time-evolution, i.e., remaining in the domain, is guaranteed. However, since this domain is dense in $L^2$ (meaning that any element in $L^2$ can be approached to arbitrary accuracy with an element of $D(\hH)$ as measured by the $L^2$-norm) and the evolution operator is bounded, the domain of the evolution operator can be extended by continuity to all of $L^2$, such that a time-evolution is well-defined for any square integrable initial state. Thus Schr{\"o}dinger dynamics still exists in the so-called {\em mild} sense (see also \ref{app:existence}) but to define this properly the Schr{\"o}dinger equation needs to be transformed to integral form
\bea
\Psi (x,t) &=& \Psi (x,0) - \imagi \hH \int_{0}^t \diff s \, \Psi (x,s) \nonumber \\
&=& \Psi (x, 0) + \frac{\imagi}{2} \frac{\diff^2}{\diff x^2} \int_{0}^t \diff s \, \Psi (x,s)  .
\label{mild}
\eea
A solution of this equation is called a mild solution (but later for time-dependent Hamiltonians we need to generalize the definition of mild solutions again so it will only appear here
in this form).
It turns out that when we integrate $\Psi (x,s)$ of (\ref{Hall_func}) between zero and $t$ the resulting solution becomes twice differentiable with respect to $x$ (see Fig. \ref{fig:berry_example}) and is in the
domain of the Hamiltonian such that the last term in (\ref{mild}) is well-defined (note that integration and differentiation cannot be interchanged in these cases). This is, in fact, a general feature that can be proven using semi-group theory~\cite{RenardyRogers}. So our non-differentiable   solution (\ref{Hall_func}) is a solution of the mild equation (\ref{mild}) rather than of the classical TDSE (\ref{TDSE1}).\\
What we have been vague about so far is what the domain $D(\hH)$ of the Hamiltonian actually is. It turns out that this domain is determined by
requiring $\hH$ to be self-adjoint on a domain with specific boundary conditions. 
What we want to do now is to make this more precise. We first define what we mean by a symmetric operator.
An operator $\hA$ on a Hilbert space is called {\em symmetric} when
\be
\langle \Phi | \hA \Psi \rangle = \langle \hA \Phi | \Psi \rangle
\label{def_symmetric}
\ee
for all $\Psi$ and $\Phi$ in the domain of $\hA$.
Let us give an example for the momentum operator for our particle in a box.
We define the momentum operator by
\be
\hat{p} = - \imagi \frac{\diff}{\diff x}
\ee
and let its domain be the one times continuous differentiable functions $\Psi$ on the interval $[0,L]$ with boundary conditions $\Psi (0)=\Psi (L)=0$ . We have
by partial integration
\bea
\langle \Phi | \hat{p} \Psi \rangle &=& -\imagi \int_0^L \diff x \, \Phi^* (x) \frac{\diff \Psi}{\diff x}(x) \nonumber \\
&=& \Big[ -\imagi \Phi^* (x) \Psi (x) \Big]^L_0 + \imagi  \int_0^L \diff x \, \frac{\diff \Phi^*}{\diff x} (x) \Psi (x)
= \langle \hat{p} \Phi | \Psi \rangle .
\label{momen}
\eea
So clearly $\hat{p}$ is a symmetric operator.
Note, however, that we can extend this definition to larger domains. This is easily seen from our example.
Since $\Psi(0)=\Psi (L)=0$ we do not need to put any conditions on the functions $\Phi$ at the boundary 
to make the boundary term vanish. It will for 
example suffice that they are simple square integrable and differentiable to make (\ref{momen}) valid. 
The operator $\hat{p}$ acting on $\Phi$ can therefore be regarded as an adjoint operator acting on a larger domain.
Let us make this statement more precise with a definition. 
For a given operator $\hA$ with domain $D(\hA)$ we take $D(\hat{A}^\dagger)$ to be the set of all $\Psi \in \mathcal{H}$ for which there exists a $\chi \in \mathcal{H}$ such that
\be\label{adjoint_domain}
\langle \Psi | \hA \Phi \rangle = \langle \chi | \Phi \rangle
\ee
for all $\Phi \in D(\hA)$. This defines the adjoint operator $\hat{A}^\dagger \Psi=\chi$ on $D(\hat{A}^\dagger)$.
Clearly it follows from this definition that for symmetric operators $D(\hA) \subseteq D(\hA^\dagger)$, i.e., the domain of $\hA$ is equal to or a subset of the
domain of $\hA^\dagger$. In case $D(\hA)=D(\hA^\dagger)$ the operator is called {\em self-adjoint}.
Such operators can be diagonalized (or more precisely have a spectral representation \cite{blanchard2003,Teschl}) and their eigenvalues are real which are key features used in quantum mechanics.
When $\hA$ is a self-adjoint operator a unitary time-evolution 
$\hat{U} (t) = \exp(-\imagi \hA t)$ (defined by its spectral representation) can be defined on all of the Hilbert space ($L^2([0,L])$ for our example) and 
if $\Psi (0) \in D(\hA)$ the function $\Psi (t)=\hat{U}(t) \Psi (0) \in D(\hA)$ satisfies the evolution equation
\be
\imagi \prt \Psi (t) = \hA \Psi (t).
\label{A_evolv}
\ee
This important statement is also called Stone's theorem~\cite{blanchard2003}.  
The issue of self-adjointness is thus very important for establishing the solvability of evolution equations such as the TDSE.
Let us go back to our example of the momentum operator for the particle in the box.
We already noted that the action of $\hat{p}$  on $\Phi$ in (\ref{momen}) could be defined on a bigger domain. 
This means that $\hat{p}$ is not self-adjoint. The adjoint operator $\hat{p}^\dagger$ has the same appearance as
a differentiation rule but the domain of $\hat{p}^\dagger$ is larger, $D(\hat{p}) \subset D(\hat{p}^\dagger)$
and therefore $\hat{p} \neq \hat{p}^\dagger$. A quick calculation shows that the operator $\hat{p}$ that we defined does not have eigenfunctions satisfying zero boundary conditions, as one would expect if $\hat{p}$ were self-adjoint. 
Furthermore one can see that the time-evolution equation
with $\hA=\hat{p}$ has no solution either for any initial state (since the formal solution is $\Psi(x,t) = \Psi_0(x-t)$ and will be ill-defined as soon as it hits the wall of the box). So this is a simple example how self-adjointness is important for the solvability of evolution equations.
It is clear from our example that if we put less restrictions on the functions in the domain of the momentum operator $\hat{p}$ then we reduce the domain 
of $\hat{p}^\dagger$ as we need more restrictions on $\Phi$ to make the boundary term in (\ref{momen}) vanish.
We might therefore expect that for a sufficiently large extension of the domain of $\hat{p}$ we have $D(\hat{p})=D(\hat{p}^\dagger)$
and the momentum operator becomes self-adjoint \footnote{We note that on a Hilbert space of finite dimension any symmetric 
operator is self-adjoint. If we discretize the momentum operator on a finite grid with zero boundary conditions then one finds that eigenfunctions exist but that they become ill-defined in the continuum limit while its eigenvalues diverge in the same limit \cite{Ruf}.}. The idea is to join the end points of the box and form a ring.
The most general condition under which the boundary terms in (\ref{momen}) vanishes is \cite{blanchard2003}
\be
\Psi (0,t)= \e^{\imagi \gamma} \Psi (L,t),
\ee
where $\gamma$ is a real number (note that this is indeed an extension of our earlier domain).
To make the boundary terms disappear these conditions must apply both to $\Psi$ and $\Phi$ in (\ref{momen}), unless $\Psi(0,t)=0$.
If we the  take as domain the space of functions such that both $\Psi$ and $\diff\Psi/\diff x$ are square integrable (more technically called the Sobolev space $H^1 ([0,L])$) 
then with these boundary conditions the operator $\hat{p}$ becomes self-adjoint $\hat{p}=\hat{p}^\dagger$. For each $\gamma$
we therefore find a different self-adjoint domain. The case $\gamma=0$ corresponds to the standard periodic boundary conditions (i.e.~the momentum operator of a quantum particle on a ring of circumference $L$).
If we use $\hA=\hat{p}$ in the evolution equation (\ref{A_evolv}) with this periodic domain (more precisely requiring $\Psi$ and its first derivative to be square integrable and satisfying periodic boundary conditions)  then Stone's theorem guarantees a solution for an initial state
which is in the domain of $\hat{p}$ (note that the self-adjoint ``Hamiltonian'' $\hat{p}$ is unbounded from below). \\
Let us now go back to our original TDSE problem of (\ref{TDSE1}) and discuss the self-adjoint domain of $\hH$.
A straightforward calculation shows that
\be
\langle \Phi | \hat{H} \Psi \rangle  = -\frac{1}{2} \left[ \Phi^* (x) \frac{\diff\Psi}{\diff x}  (x) -  \frac{\diff\Phi^*}{\diff x}  (x) \Psi (x) \right]^L_0  +  \langle \hat{H} \Phi | \Psi \rangle.
\label{hamil}
\ee
It is clear that the Hamiltonian is symmetric on the domain $\mathcal{C}^2 ([0,L])$ of twice continuously differentiable functions that satisfy $\Psi (0,t)=\Psi (L,t)=0$.
Since the derivatives $\diff\Psi/\diff x$ do not need to vanish at the boundary we see that also for $\Phi$ we need to require the boundary conditions $\Phi (0,t)=\Phi (L,t)=0$ in order
to make the boundary term in (\ref{hamil}) vanish. It is still not clear, however, that the specified domain would make $\hH$ self-adjoint.
Indeed the domain can be extended without violation of the boundary conditions. 
It is easily imagined that we could have a series of smooth functions $\varphi_n$ satisfying the boundary conditions which converges in the $L^2$-norm
to a function $\varphi$ such that the series $\hH \varphi_n$ would converge in $L^2$-norm to another function $\xi$. The function $\xi$ need not be continuous but merely
square integrable (in which case $\varphi$ is not in $\mathcal{C}^2([0,L])$). If this is the case then we can extend the domain of our original operator by defining $\hH \varphi=\xi$. 
Clearly for functions obtained by this limit procedure the inner products in (\ref{hamil}) are finite since due to the Cauchy-Schwarz inequality we have $|\langle \hH \Psi | \Phi \rangle | \leq \| \hH \Psi \| \| \Phi \|$
and both $\hH \Psi$ and $\Phi$ are square-integrable. Moreover, both inner products on both sides of the equality sign are identical since the series $\varphi_n$ preserves the boundary conditions.
The domain of the Hamiltonian is by this procedure the space of functions $\Psi$ for which the norm
\be
\| \Psi \|_{\hat H} = \|\Psi\| + \left\| \frac{\diff^2 \Psi}{\diff x^2} \right\|
\label{sob_norm}
\ee
is finite and where $\Psi$ satisfies the zero boundary conditions. This space
of functions is equal to the Sobolev space $H^2_0 ([0,L])$ where the superindex 2 refers to the second derivative and the subindex 0 to the zero boundary conditions~\cite{RenardyRogers,evans2010}.
One can now show that $\hH$ is self-adjoint on $H_0^2 ([0,L])$ \cite{Teschl}. Based on all these preliminaries we can finally conclude with the important statement that
a solution of the TDSE exists with the property $\Psi (t) \in H_0^2$ if the initial state is $\Psi (0) \in H_0^2$. If $\Psi (0)$ is outside $H_0^2$ but still in $L^2$ then
a trajectory $\Psi (t)$ in Hilbert space exists but $\Psi (t)$ will be outside this domain for all $t$. The solution then solves the mild version of the TDSE.
Our example of a constant function as initial state can now also be understood better. This function can be approached as close as one wants with
$\mathcal{C}^\infty$-functions $\varphi_n$ that satisfy zero boundary conditions (since they are dense in $L^2$).
 But to do so the first and second derivatives near the endpoints at $0$ and $L$ become very large
in such a way that $\| \hH \varphi_n \|$ diverges. Our initial state therefore had infinite expectation value of energy and was outside $H_0^2$ and this lead to our strange
non-differentiable solution.\\
All these things discussed here can also be extended to the case of many-particle systems with self-adjoint and time-independent Hamiltonians. The main problem then is to
know whether, for instance, the molecular Hamiltonian with Coulombic potentials actually is self-adjoint. This is answered by a famous theorem of Kato.
For the case of time-dependent Hamiltonians the situation is much more complicated. First of all, due to a lack of time-translational invariance of the Hamiltonian 
the unitary evolution operators $\hat{U}$ depend not only on the length of the time-interval of propagation, but on the initial time $t_i$ and final time $t_f$ of the time-propagation, i.e.~one writes 
$\hat{U} (t_f,t_i)$. These issues are discussed in more detail in the following sections. The main aim of this section was to serve as an introduction to these section as they 
illustrate by a simple example the subtleties of the TDSE.

\subsection{History of explicitly time-dependent initial-value problems}
 
\label{sec:history}

Before we start with the precise mathematical definition of the TDSE, let us first give a short historical overview of such explicitly time-dependent Cauchy problems. According to Kato \cite{kato1961} the first investigations of such initial-value problems with time-dependent operators date back to Phillips \cite{phillips1953} in 1953. In this work the perturbative expansion of the full evolution, i.e., the sum of all possible combinations of free propagation and the interaction with a time-dependent potential, is shown to converge under certain conditions. In contrast to this, one can also consecutively propagate along a direct path for short time intervals, assuming an evolution operator with a time-constant potential, then making those time-intervals smaller to get the desired evolution operator in the limit. This was achieved by Kato \cite{kato1953} and others, where the Hamiltonian was supposed to be \textit{maximally dissipative} at all times, a property that self-adjoint operators automatically exhibit. However, conditions on the relations of the Hamiltonian at different times can exclude typical cases of external potentials. A more up-to-date summary of these techniques can be found in Pazy \cite[ch. 5]{pazy1983} where the notion of \textit{stable families of infinitesimal generators} is used. The method of Kato \cite{kato1953} is later used in the books of Reed and Simon \cite[Th. X.70]{reed1975}, where the authors develop a theory specifically for the Schr\"odinger case for one particle, allowing also singular Coulombic potentials. This approach, which we follow in this review by and large, can be extended to the $N$-particle case with methods given in the same book.\footnote{We want to thank Prof.~Barry Simon for confirming that this generalization is actually possible.}

Other approaches \cite{lions1958, howland1974} use extended Hilbert spaces that involve time. But all these results apply to abstract operators, not beneficially taking into account any special structure of a Schr\"odinger Hamiltonian. On the other hand, W\"uller \cite{wuller1986} treats the special case of the Schr\"odinger Hamiltonian for a single quantum particle with certain moving Coulombic potentials. This specific approach unitarily transforms the equation to a static singular potential and then uses results from Tanabe \cite{tanabe1979} which in turn rely on Kato \cite{kato1953}. The study in Yajima \cite{yajima1987} resumes the perturbative approach of Phillips \cite{phillips1953} and combines it with specific properties of the Schr\"odinger Hamiltonian, e.g., Strichartz-type estimates to relate the freely evolved wave function with the initial state. The results apply to arbitrary spatial dimensions and thus allow to consider multiple particles in three-dimensional space. While the conditions on the temporal behaviour of the applied potentials are very mild (less restrictive than those employed in Reed and Simon \cite{reed1975}) the conditions on the spatial behaviour of the potentials become restrictive for more than one particle in three dimensions (Coulombic potentials are excluded for instance). Since we will use certain results of this approach, a brief introduction to \cite{yajima1987} can be found in \ref{app:existence}. A slightly more general approach than in \cite{yajima1987} was presented by D'Ancona et al.~\cite{dancona2005} just relying on a fixed-point procedure to show existence and uniqueness of Schr\"odinger dynamics rather than explicitly constructing the evolution operator as a Neumann series.

\subsection{Classical and non-classical solutions to the initial-value problem}

\label{sec:Cauchy}
%
%
%
%
%
%

In this section we will give a general discussion of the initial-value problem, also known as the Cauchy problem, of the TDSE and introduce
the notion of classical and non-classical solutions.
In (\ref{eq:TDSE_1}) we defined the TDSE for a many-electron system. The many-body wave function $\Psi (\ux,t)$ is a function of both the
space and the spin variables
and therefore the inner product with another wave function
 $\Phi (\ux,t)$
is given by an summation over spin variables as well as an integration over spatial variables, i.e.,
\be
\langle \Psi (t) | \Phi (t) \rangle=  \sum_{\us}  \int \diff \ur \, \Psi^* (\ux,t) \Phi (\ux,t) ,
\label{space_spin}
\ee
where we used the notation that $\ur=(\bbr_1,\ldots,\bbr_N)$ is the collection of spatial variables and
that $\us= (\sigma_1 ,\ldots, \sigma_N)$ is the collection of spin variables. 
For a discussion of the analytical properties of the wave function it is convenient to deal separately with the space and the spin variables.
The function $\Psi$ can be expanded as a finite expansion in terms of a product of space and spin functions. More precisely
\be
\Psi (\ux,t) = \sum_{j=1}^M  \Psi_j (\ur,t) \, \theta_j (\us)  .
\label{spin_exp}
\ee
The requirement that the wave function is anti-symmetric under simultaneous interchange of space and
spin variables implies that the functions $\Psi_j$ and $\theta_j$ transform under an $M$-dimensional representation
of the symmetric group $S_N$ of permutations of $N$ elements upon permutation of the particle labels \cite{McWeeny_book,Wilson_book}.
The spin functions $\theta_j$ are typically chosen
in such a way that the wave function is an eigenfunction of the total spin operators $S_z$ and $\mathbf{S}^2$ and the
choices of the eigenvalues determines the dimension of the representation $M$. Explicit techniques for construction of such spin functions are described in \cite{McWeeny_book,Wilson_book}.
Since the spin functions $\theta_j$ are linearly independent the initial value problem of (\ref{eq:TDSE_1}) is
equivalent to $M$ separate initial-value problems
\bea
\imagi \partial_t \Psi_j (\ur, t) &=& \hH (t) \Psi_j (\ur,t)  \label{eq:TDSE_M}\\
\Psi_j (\ur,t_0) &=& \Psi_{0,j} (\ur) \nonumber
\eea
for $j=1,\ldots,M$ where $\Psi_{0,j}$ is the $j$-th spatial component of the initial state $\Psi_0$.
Mathematically all these initial-value problems are equivalent, such that we can forget about the sub-index $j$ again.
The corresponding Hilbert space is simply the space of square integrable functions on $\mathbb{R}^{3N}$,
or in case we wish to discuss an $N$-particle system enclosed in a volume $\Omega \subset \mathbb{R}^3$, the square-integrable functions on the configuration space $\Omega^{N}$. Mathematically this Hilbert
space is denoted as $\mathcal{H}=L^2 (\Omega^N)$ with inner product
\be
\langle \Psi (t)| \Phi (t) \rangle = \int \diff \ur \, \Psi^* (\ur,t) \Phi (\ur,t),
\ee
where the integration is over $\Omega^N$ or $\mathbb{R}^{3N}$ depending on the situation of interest.
This is the setting in which we will discuss the solvability and properties of the TDSE.
When we discuss the properties of observables such as densities and currents we naturally have to remember
to properly take into account the spin-structure of the many-particle wave functions.

Since the time-variable is a parameter in quantum mechanics and not a coordinate (i.e.~there is no generic time operator) we will often suppress the spatial coordinates when writing the TDSE. The initial value problem (\ref{eq:TDSE_M}) is then written as 
\begin{eqnarray}
 \label{Cauchy}
\imagi \partial_t \Psi(t) = \hat{H}(t) \Psi(t)
\\
\Psi(t_0) = \Psi_0. \nonumber
\end{eqnarray}
The Hamiltonian $\hat{H}(t)$ is a self-adjoint operator on $\mathcal{H}$ parametrically dependent on times in the time interval $[0,T]$ (we can take without loss of generality the initial time $t_0$ to be zero). The time $T$ is positive and can be arbitrarily large but we take it to be finite since it appears in estimates in Sec.~\ref{sec:FixedPoint}.
The solution will be a trajectory $\Psi(t)$ in Hilbert space, or more mathematically a mapping from the interval $[0,T]$ to
$\mathcal{H}$. One of the most basic requirements that we can put on such a mapping is that it is continuous. Since the space $\mathcal{H}$ is a Hilbert space continuity is defined with respect to the Hilbert space norm.
This means that $\Psi (t)$ is defined to be a continuous mapping when the $L^2$-distance of two ``snapshots'' of wave functions at time $\Delta t$ apart
goes to zero as the time-difference goes to zero, or more precisely that $\| \Psi (t+ \Delta t) - \Psi (t) \| \rightarrow 0$ if $\Delta t \rightarrow 0$. The space of such continuous functions will be denoted as $\mathcal{C}^0 ([0,T],\mathcal{H})$.
In physical applications it is reasonable to further demand that the energy expectation value 
\be
E(t) = \langle \Psi (t) | \hH (t) \Psi (t) \rangle
\ee  
is finite.
To guarantee this property we need more than continuity of the mapping. As we will discuss in more detail later this is guaranteed if also 
$\partial_t \Psi \in \mathcal{C}^0 ([0,T],\mathcal{H})$
meaning that $\| \partial_t \Psi (t+ \Delta t) - \partial_t \Psi (t) \| \rightarrow 0$ if $\Delta t \rightarrow 0$. Hilbert space trajectories with this property are called
continuously differentiable mappings and to indicate that $\Psi (t)$ has this property we write $\Psi \in \mathcal{C}^1 ([0,T],\mathcal{H})$ where the super-index 1 refers to the first-order derivative with respect to time.
A solution of the TDSE which is such a $\mathcal{C}^1$-mapping is called a {\em classical solution}. 
We are, however, not always able to find a classical solution to the TDSE. 
This happens, for instance, when the initial state $\Psi_0$ is normalizable but when $\hH \Psi_0$ is not, in which case 
the initial state is not in the domain of the Hamiltonian and the expectation value of the energy might be infinite.
This was, for example, the case for the constant initial state of (\ref{constPsi}) for the example of the particle in a box discussed
in Sec.~\ref{sec:wavepacket}. 
%
%
%
%
%
%
However, if we generalize the notion of a solution, even such problems can be solved uniquely \cite{reed1975,yajima1987}.

Such generalizations of solutions to the TDSE can be found in different ways (see \ref{app:existence} for time-dependent Hamiltonians). The simplest way to do so is by first considering time-independent Hamiltonians $\hat{H}$. In this case the Hamiltonian gives rise to a unitary 
evolution operator $\hat{U}(t) = \exp(-\imagi \hat{H} t)$ defined by its spectral representation as discussed in Sec.~\ref{sec:wavepacket}.
Although the Hamiltonian is not defined on all of $\mathcal{H}$, its evolution operator is a bounded operator 
and can therefore be uniquely extended to all square-integrable wave functions. 
Consequently, we have a unique \textit{generalized solution} to the TDSE $\Psi(t) = \exp(- \imagi \hat{H} t) \Psi_0$
even if the initial state $\Psi_0$ is not in the domain of the Hamiltonian.
The mapping $\Psi (t)$ regarded as a trajectory in Hilbert space is in that case not differentiable but still continuous, i.e.~$\Psi \in \mathcal{C}^{0}([0,T], \mathcal{H})$. 
Such a trajectory will be called
a \textit{non-classical solution}. It does not solve (\ref{Cauchy}) but it might solve an equivalent but less stringent version of the TDSE, such as (\ref{mild}) which we
discussed in Sec.~\ref{sec:wavepacket}. Since unitary evolution conserves the norm it follows in the case of time-independent Hamiltonians that if $\hH \Psi_0$ is not normalizable then also $\hH \Psi (t)$ is not normalizable and we therefore can differ between classical and non-classical solutions by their initial state. 
However, in the case of explicitly time-dependent Hamiltonians $\hat{H}(t)$ we cannot straightforwardly use the same construction, since their spectral representation changes with time.

The discussion in this section has been very general as we did not discuss the actual properties of the Hamiltonian.
This will be the topic of the next section in which we will address this issue in more detail and discuss for which class of external potentials and interactions
the TDSE of the many-electron system is guaranteed to have a solution.

\subsection{Existence and uniqueness of solutions to the Schr\"odinger equation}

\label{sec:existence}

In this section we investigate under which conditions (and in which sense) we can define an evolution operator for a time-dependent Hamiltonian of a many-electron system.
Before we discuss the existence of solutions of the TDSE we want to guarantee that the Hamiltonian $\hat{H}(t)$ is a self-adjoint operator for all times $t$.
This is not only a basic requirement for any observable as dictated by the mathematical structure of quantum mechanics, but we have also seen in Sec.~\ref{sec:wavepacket}
that self-adjointness is an important property which is closely connected to the solvability of evolution equations such as the TDSE.
We first give conditions such that the time-independent Hamiltonian $\hat{H}_0$ is self-adjoint, then include also the time-dependent part 
$\hat{V}(t)$ and finally give conditions for the existence of an evolution operator.

The time-independent part $\hH_0$ of the Hamiltonian usually consists of two parts, the kinetic energy of the particles $\hat{T}$ and the interaction between the particles $\hat{W}$. The kinetic-energy operator is given by the mapping
\begin{eqnarray}
 \label{KineticRule}
(\hT \Psi) (\ur) = -\frac{1}{2} \sum_{j=1}^{N} \nabla^2_{j}  \Psi(\ur). 
\end{eqnarray}
Here the spatial derivative is meant in the weak sense\footnote{On the other hand, derivatives with respect to time are always to be understood in the classical sense.} which is defined by integration against smooth test functions. 
The defining equation simply uses the basic equation of partial integration for the product of functions.
For example, we say that a function $\Psi$ defined on an open domain $\Omega \subseteq \mathbb{R}^n$ has the weak derivative $\Phi=\nabla \Psi$
($n$-dimensional gradient) when the following equation
\be
\int \diff \ur \, \Phi (\ur) \varphi (\ur) = - \int \diff \ur  \,\Psi (\ur) \nabla \varphi (\ur)
\ee
is valid for any function $\varphi (\ur)$ which is infinitely differentiable and which is only non-zero on a bounded region (has compact support in mathematics language).
The advantage of talking about weak derivatives is that one can talk about the derivatives of functions which do not have derivatives in the classical sense.
For example, the weak derivative of the function $\Psi (x)=|x|$ in one dimension is the equivalence class of functions $\Phi (x)$ which are equal to $1$ for $x > 0$, equal to
$-1$ for $x <0$ and take an arbitrary value in $x=0$. The concept of weak derivative considerably simplifies the mathematical treatment of partial differential equations. For an extensive discussion of these issues we refer to \cite{evans2010,RenardyRogers}. 

Let us now go back to (\ref{KineticRule}).
Obviously, not every square integrable wave function is again mapped to another square integrable function. 
Therefore, the kinetic energy operator is only defined for a restricted set of functions in the Hilbert space\footnote{If an operator is defined on all of $\mathcal{H}$ and is symmetric then by the theorem of Hellinger-Toeplitz it necessarily is a bounded operator, i.e.~it has a maximal eigenvalue \cite{blanchard2003}.}, which is called its \textit{domain} and which we will denote by $D(\hat{T})$. To be a proper domain this set of functions has to be dense in the whole Hilbert space, which means that we can approximate every normalizable wave function arbitrarily close (in the $L^2$ norm) with functions of the domain. Without this condition a unique adjoint operator can, for instance, not be defined.
Now, there are two necessary conditions to make the kinetic-energy operator self-adjoint: the domain has to be equal to its adjoint domain and $\hat{T}$ has to be symmetric (see (\ref{adjoint_domain}) and (\ref{def_symmetric}) for definitions).
However, for particles restricted to a general volume $\Omega$ in three-dimensional space there are many domains that make the mapping of (\ref{KineticRule}) a self-adjoint operator.  
We can construct those different self-adjoint domains by choosing different boundary conditions, such as periodic- or zero-boundary conditions. Depending on these conditions the properties of the associated operators can change dramatically. This is clear physically as, for instance, the Hamiltonians for a free particle in a box or for the free particle on a ring have different energy eigenvalues and eigenstates.
Therefore, it is usually not enough to just prescribe the \textit{rule} of an operator, such as the differentiation rule of (\ref{KineticRule}), but one also needs to fix the domain and with it the boundary conditions. Only then we have the unique definition of a self-adjoint operator.  
The full three-dimensional space $\Omega=\mathbb{R}^3$ is an exception. In this case there is only one self-adjoint domain for the kinetic energy operator of (\ref{KineticRule}) which is the Sobolev space $D(\hat{T}) = H^2(\mathbb{R}^{3N})$ (see \ref{app:Lp} for further details on Sobolev spaces). 
By defining the self-adjoint kinetic-energy operator $\hat{T}$, we have also chosen a specific set of functions that are guaranteed to have finite kinetic energy\footnote{The most general set of finite kinetic-energy states is usually bigger than the self-adjoint domain, since it only needs to ensure that the expectation value is finite.}. However, functions that have finite kinetic energy in one self-adjoint realization of $\hat{T}$ might not have finite kinetic energy in another realization. For example the expectation value of the kinetic energy for the initial state (\ref{constPsi}) is infinite for hard wall boundary conditions but finite for periodic ones. \\
After having discussed the kinetic energy operator we turn our attention to the two-body interactions and consider the 
static part $\hH_0=\hT+\hW$ of the Hamiltonian.
The corresponding new rule for mapping wave functions is given by
\begin{eqnarray}
 \label{KineticInteractionRule}
(\hH_0 \Psi ) (\ur) \!=\! \left[ \sum_{j=1}^{N} \left(-\frac{1}{2} \nabla^2_{j}   \right) \! + \! \frac{1}{2}\sum_{i \neq j}^{N}  w(\bbr_i-\bbr_j) \right] \!\! \Psi(\ur), 
\end{eqnarray}
where two-body interaction $w(\bbr )$ is a real scalar function defined on $\mathbb{R}^3$ which is typically taken to be Coulombic, i.e.~$w(\bbr) =1/|\bbr |$.
In the following we will take it always to be a function of the inter-particle distance $|\bbr|$.
We want to ensure that the operator $\hH_0$ is self-adjoint on the same set of (physical) wave functions as the kinetic-energy operator, i.e.,  $D(\hat{H}_0)=D(\hat{T})$. Using the theory of Kato perturbations~\cite{kato1995} we can find rather simple conditions for this to hold \footnote{ In this context a Kato perturbation of a self-adjoint operator $\hat{A}$ is a symmetric operator $\hat{B}$ with $D(\hat{A}) \subset D(\hat{B})$ and real numbers $0 \leq a <1, b\geq 0$, such that $\|\hat{B} \Psi\| \leq a\|\hat{A} \Psi\| + b\|\Psi\|$ for all $\Psi \in D(\hat{A})$. From this it is obvious that every bounded operator, e.g., a multiplication with a bounded interaction potential $w$, is automatically a Kato perturbation.}.
For the case that we discuss particles in the whole three-dimensional space $\mathbb{R}^3$ the operator $\hH_0$  defines a self-adjoint operator with the same domain as the kinetic energy operator when $w$ can be written as the sum $w=w_1 + w_2$, one square integrable and the other bounded. This class of potentials is also known as the class of {\em Kato perturbations}.
In a more mathematical notation we can write  $w_1 \in L^{2}(\mathbb{R}^3)$ and $w_2 \in L^{\infty}(\mathbb{R}^3)$.
The space of functions $L^\infty (\mathbb{R}^3)$ is the set of functions $w(\bbr)$ for which there is a positive number $M$ such that $|w(\bbr)| < M$ for all $\bbr$ (technically speaking 
this has to hold almost everywhere, meaning up to a set of measure zero). The space $L^\infty$ has a norm but no inner product and is therefore not a Hilbert space, instead it is called a Banach space { (see \ref{app:Lp} for a further discussion)}.
The class of Kato potentials on $\mathbb{R}^3$ is written as $\mathcal{K}(\mathbb{R}^3) = L^2 (\mathbb{R}^3) + L^\infty (\mathbb{R}^3)$ and is again a Banach space\footnote{This Banach space has norm
$\| w \|_{\mathcal{K}}=\inf \{  \| w_1 \|_2 + \| w_2 \|_\infty | w=w_1 + w_2, w_1 \in L^2 (\mathbb{R}^3), w_2 \in L^\infty (\mathbb{R}^3)  \}$ in which $\| \cdot \|_2$ and $\| \cdot \|_{\infty}$ are the
norms on $L^2$ and $L^\infty$ respectively.}.
In the case that we discuss particles restricted to a finite volume $\Omega$ we just have $\mathcal{K} (\Omega)=L^2 (\Omega)$.
An important potential which is included in the class of Kato potentials $\mathcal{K}(\mathbb{R}^3)$ is the Coulomb potential since it can be written as
\[
\frac{1}{|\bbr| } =  \frac{\theta (1-|\bbr|)}{|\bbr|} +  \frac{\theta (|\bbr| -1)}{|\bbr|},   \nonumber
\]
where $\theta (x)=1$ for $x>0$ and is zero otherwise. The first term after the equal sign is square integrable and the second term is bounded.

In the final step we now add an explicitly time-dependent external potential
\begin{eqnarray}
 \label{ExternalEnergy}
\hat{V}(t) = \sum_{j=1}^{N} v(\bbr_j,t)
\end{eqnarray} 
to the time-independent Hamiltonian $\hat{H}_0$ to build the full Hamiltonian $\hH (t) = \hH_0 +  \hV(t)$.
Again applying the theory of Kato perturbations we find that the Hamiltonian $\hH (t)$ is a self-adjoint operator on the domain $D(\hat{T})$ for all times whenever the potential $v(\bbr,t)$ belongs to the class of Kato potentials.  We note that important physical models such as the harmonic oscillator or the dipole fields are not included in the Kato class if we consider the full three-dimensional space $\mathbb{R}^3$.
Since $\hT$ and $\hH (t)$ have the same domain it also follows that $\hH (t) \Psi$ is normalizable whenever $\hT \Psi$ is normalizable. For a Hamiltonian in which the external potentials and two-body interactions are in the Kato class the total energy expectation value is therefore finite whenever the kinetic energy expectation value is finite.

Now that we have identified the class of external potentials and two-body interactions for which the Hamiltonian is self-adjoint on the domain of the kinetic-energy operator we can start to discuss the solvability of the initial-value problem for the TDSE.

There are now several ways of investigating the existence and uniqueness of solutions to the TDSE (see Sec.~\ref{sec:history}). For the time being we restrict ourselves to an approach similar to the one presented in \cite{reed1975}. Certain details and a comparison to a different approach based on purpose-build Banach spaces of potentials and wave functions~\cite{yajima1987} are discussed in \ref{app:existence}.
From the previous considerations we have seen that if we take $w(\bbr)$ and $v(\bbr,t)$ to be Kato perturbations, the resulting Hamiltonian $\hat{H}(t)$ has the same domain $D(\hat{T})$ at every time. The task is now to prove that a well-defined time-evolution exists.
The idea of the proof is to divide the time-propagation interval $[0,T]$ into $k$ small time intervals $[t_{i}, t_{i+1}]$ where $i \in \{0,...,k-1\}$, $t_0 = 0$ and $t_k=T$ and take 
$\hat{H}(t_i) = \hat{T} + \hat{W} + \hat{V}(t_i)$ as a time-constant Hamiltonian during the time interval $[t_i,t_{i+1}]$. In each such time interval the Hamiltonian $\hH (t_i)$
defines a self-adjoint operator and we know (by Stone's theorem already employed in Sec.~\ref{sec:wavepacket}) that a well-defined evolution operator
\be
\hat{U}_{k} (t,s) = \e^{-\imagi \hat{H}(t_i)(t-s)}
\ee
exists for $s,t \in [ t_{i}, t_{i+1}]$. For arbitrary times $t$ and $s$ in $[0,T]$ we can define the evolution operator $\hat{U}_k (t,s)$ by glueing
together the evolution operators in different time intervals. If $t \in [t_n,t_{n+1}]$ and $s \in [t_m,t_{m+1}]$ we define
\be
\hat{U}_k (t,s) = \hat{U}_{k} (t,t_n) \left[  \prod_{j=m+1}^{n-1} \hat{U}_{k} (t_{j+1},t_{j}) \right] \hat{U}_k ( t_{m+1},s ),
\ee 
where in the product the operator with the latest time is always ordered to the left.
It can then be shown \cite{reed1975} that 
\begin{eqnarray}
\label{MidPointApproximation}
 \lim_{k \rightarrow \infty} \hat{U}_k(t,s) = \hat{U}(t,s)
\end{eqnarray}
(in operator norm) provided that $v \in \mathcal{C}^{1}([0,T], \mathcal{K})$.\footnote{Considering the proof of \cite[Th.~X.70]{reed1975} it seems possible that Lipschitz-continuity in time with respect to the norm of $\mathcal{K}$ is enough.} Let us elaborate on this condition. If we view the potential as a trajectory $v(t)$ in the space of Kato perturbations then (\ref{MidPointApproximation}) is valid when  $v(t)$ is a continuously-differentiable mapping with respect to the norm of $\mathcal{K}$.
The evolution operator $\hat{U}(t,0)$ is then unitary and therefore
\[
\Psi (t) =  \hat{U} (t,0) \Psi_0
\]
defines a unique continuous trajectory in Hilbert space for any normalizable initial state $\Psi_0$, or more precisely $\Psi \in \mathcal{C}^{0}([0,T], \mathcal{H})$. Therefore we have existence and uniqueness 
of a (generalized) solution for an important class of time-dependent potentials.
Such potentials include for example molecular potentials of the form of (\ref{MolecularPotential}) provided $v_{\mathrm{ext}} (t)$ is a continuous differentiable mapping to the Kato-class.
 We note that the differentiability condition with respect to time on the external potential 
excludes a sudden switch-on, but by using the technique presented in \cite{yajima1987} we can show existence and uniqueness of a generalized
solution also for such situations (see \ref{app:existence} for more details). 
Further, if the initial state $\Psi_0$ is in the domain of the kinetic-energy operator $D(\hat{T})$ 
then also $\Psi(t) \in D(\hat{T})$ for every time $t \in [0,T]$ and therefore the expectation value of the kinetic and total energy are finite.
In that case one has $\Psi \in \mathcal{C}^{1}([0,T], \mathcal{H})$ which is therefore a classical solution to the TDSE.
Let us summarize the most important results of this section. We can establish a well-defined time-evolution if the
potential is continuously differentiable in time with respect to the norm of $\mathcal{K}$.
We define this set of allowed potentials as
\be
\mathcal{V} =  \mathcal{C}^1 ([0,T], \mathcal{K} ).
\label{V_class}
\ee 
 
Now we have properly defined (including the domains) the mapping from potentials to wave functions that we discussed
in Sec.~\ref{dens_pot_map}. For a given potential $v \in \mathcal{V}$ we can solve
the TDSE for a normalizable initial state.
There are now two cases to consider.
Either the initial state is in the domain of the Hamiltonian or it is not.
In the latter case the time-evolution of the initial state defines a continuous trajectory 
$\Psi (t)$ and the trajectory regarded as functional of $v$ is given by a map
\begin{eqnarray}
\label{PsiC0mapping}
 \Psi: &\mathcal{V}  & \rightarrow  \mathcal{C}^{0}([0,T], \mathcal{H}) 
\\
& v & \; {\buildrel\rm \Psi_0 \over \mapsto }  \;  \Psi[v]  . \nonumber
\end{eqnarray}
In the case the initial state is in the domain of the Hamiltonian the time-evolution of the initial state defines a continuous differentiable trajectory $\Psi (t)$ and the trajectory regarded as functional of $v$ is given by a map
\begin{eqnarray}
 \Psi:& \mathcal{V}  & \rightarrow  \mathcal{C}^{1}([0,T], \mathcal{H} )   \label{class_map}
\\
& v & \; {\buildrel\rm \Psi_0 \over \mapsto }  \;  \Psi[v]  . \nonumber
\end{eqnarray}
We have therefore established well-defined potential to wave function mappings.
Clearly the mapping of (\ref{class_map}) is the most relevant for physical applications as it guarantees the finiteness of the expectation values of the kinetic energy and the one- and two-body interactions.

\subsection{From the wave function to observable quantities} 
 
\label{sec:observables}

After having established a well-defined mapping from potentials to wave functions we can continue with the discussion of the mapping from wave functions to physical
quantities such as densities and currents. To calculate these quantities properly we have to take into account the correct spin structure of the wave function of
(\ref{spin_exp}) and use the inner product of (\ref{space_spin}). The solvability properties of the TDSE discussed in the previous section for each of the
spatial parts of the wave function immediately imply the same solvability properties of the TDSE for full anti-symmetric space-spin function.
Let us now consider the calculation of an arbitrary physical observable. If such an observable is described by a (time-independent) self-adjoint operator $\hat{O}$
then we want to evaluate
\[
O(t) =  \braket{\Psi(t)}{\hat{O}\Psi(t)}.
\]
This expectation value is well-defined if the domain of $\hat{O}$ contains the domain of the Hamiltonian. 
We already established that for the potentials $\mathcal{V}$ given by (\ref{V_class})
the kinetic energy as well as the two-body interaction energy are finite. We therefore have well-defined functionals
\bea
T([v],t) &=& \braket{\Psi([v],t)}{\hat{T} \Psi([v],t)}, \\
W([v],t) &=& \braket{\Psi([v],t)}{\hat{W} \Psi([v],t)}, \\
V([v],t) &=&  \braket{\Psi([v],t)}{\hat{V}(t) \Psi([v],t)}
\eea
defined on the set of potentials in $\mathcal{V}$.
However, not all physical quantities are defined by self-adjoint operators on a Hilbert space. The most important ones for us are the density and the current density.
The density was already defined in (\ref{density_def}).
If the initial state is further in the domain of the kinetic-energy operator like in the case of classical solutions (\ref{class_map}) 
then the time-derivative of the wave function is well-defined and the density obeys the \textit{continuity equation}
\begin{eqnarray}
\label{ContinuityEq}
\partial_t n([v], \bbr,t) = -\nabla \cdot \mathbf{j} \, ([v],\bbr,t),
\end{eqnarray}
where 
\bea
\mathbf{j} (\bbr,t) &=& 
 \frac{N}{2 \imagi}  \sum_{\sigma} \int \diff \ux^{N-1} \, [ \Psi^* (\ux,t) \nabla  \Psi (\ux,t)  \nonumber \\
&&- (\nabla  \Psi^* (\ux,t) ) \Psi (\ux,t) ].
\label{Current}
\eea
Here we used the same notational convention as in (\ref{density_def}) and $\nabla$ is the gradient corresponding to the (non-integrated) coordinate $\bbr$.
Depending on whether we wish to consider particles in the whole space $\mathbb{R}^3$ or in a finite volume $\Omega$ the spatial
integrations in this expression are restricted to $\mathbb{R}^{3(N-1)}$ or $\Omega^{N-1}$. In the following we will just use
$\Omega$ for the volume with the understanding that possibly $\Omega=\mathbb{R}^3$. We shall mention it explicitly whenever the distinction is relevant.
The density $n(\bbr,t)$ has the obvious property that it is positive and that its integral is given by the number $N$ of electrons.
It therefore belongs to the space of $L^1 (\Omega)$ functions (see \ref{app:Lp} for a definition). The density $n(t)$ can thus be regarded as a continuous differentiable trajectory in this space. The potential to density mapping is therefore given by
\begin{eqnarray}
\label{nC1mapping}
 n:& \mathcal{V} &  \rightarrow  \mathcal{C}^{1}([0,T],L^1(\Omega)) 
\\
& v  \; & {\buildrel\rm \Psi_0 \over \mapsto }  \;  n[v]. \nonumber
\end{eqnarray}

In the following we then want to investigate under which conditions and restrictions such a mapping between $v$ and $n$ is invertible. 
Obviously $L^1(\Omega)$ is too general for the space of densities since it contains also negative functions and moreover the finiteness of the kinetic energy implies that $\nabla \sqrt{n (\bbr,t)}$ exists and is square integrable \cite{lieb1983}. The fact that $n$ and $v$ have the same degrees of freedom does at least give some hope that an inverse map may exist.
\\
To investigate this in more detail we start by considering an equation that connects $n$ and $v$ more directly. 
Formally, such an equation can be derived by combining the above continuity equation (\ref{ContinuityEq}) with the so-called \textit{local-force equation of quantum mechanics}, i.e., the time-derivative of (\ref{Current})
\begin{eqnarray}
 \partial_t \mathbf{j}([v],\bbr,t) = -n([v], \bbr,t) \nabla v(\bbr,t)  - \mathbf{Q}([v], \bbr,t),
\end{eqnarray}
where the components of the vector $\mathbf{Q}$ are given by
\begin{eqnarray}
 Q_k([v], \bbr,t) = \partial_l T_{kl}([v], \bbr,t) + W_{k}([v], \bbr,t),
\end{eqnarray}
and $\partial_k \equiv \partial/\partial r_k$ as well as summation over multiple indices is implied. The momentum-stress tensor is defined by (suppressing the dependence of the
wave function on the different variables and again using the notational convention of (\ref{density_def})) \cite{martin1959}
\bea
T_{kl} ([v], \bbr ,t) &=& \frac{N}{2} \sum_{\sigma} \int \diff^{N-1} \ux  \nonumber \\
&& \times
\left\{ (\partial_k \Psi)^* \partial_l \Psi \!+\! (\partial_l \Psi)^* \partial_k \Psi \! - \! \frac{1}{4} \partial_k \partial_l  (\Psi^* \Psi ) \right\}.
\eea
The interaction-force density is defined by
\bea
W_k ([v],\bbr ,t) &=& \frac{N}{2} \sum_{\sigma} \int \diff^{N-1} \ux \,
 \partial_k w(\bbr-\bbr_2) | \Psi (\ux,t) |^2  .
\eea 
The combination of these two  equations formally leads to the \textit{fundamental equation of TDDFT}
\begin{eqnarray}
\label{TddftFundament}
 \partial_t^2 n([v],\bbr,t) = \nabla \cdot \left[ n([v], \bbr,t) \nabla v(\bbr,t)\right] +  q([v], \bbr,t),
\end{eqnarray} 
where 
\begin{eqnarray}
\label{qoperator}
 q([v], \bbr,t) = \nabla \cdot \mathbf{Q}([v], \bbr,t).
\end{eqnarray}
It is obvious that this equation cannot hold for every possible $\Psi[v] \in \mathcal{C}^0([0,T],\mathcal{H})$, since already the continuity equation might not be well-defined for generalized solutions like in (\ref{PsiC0mapping}). We therefore need extra conditions on the potentials and initial states that guarantee that the trajectory $n(t)$ in $L^1 (\Omega)$ is twice differentiable with respect to time, i.e.~more precisely $n \in \mathcal{C}^2 ([0,T], L^1 (\Omega))$. To find those we analyse each constituent of (\ref{TddftFundament}) in detail. First, for a classical solution of the TDSE we know that $T_{kl}$ is at least in $L^1(\Omega)$ since $\Psi(t)$ is in the domain of the kinetic energy operator and thus twice differentiable. To guarantee that then the kinetic part of the operator $q[v]$ is integrable, we restrict ourselves to initial states that obey $\hat{T}^2 \Psi_0 \in \mathcal{H}$ too and to potentials that stabilize this condition, i.e., $\hat{T}^2 \Psi (t) \in \mathcal{H}$ for all times $t \in [0,T]$. This holds, for instance, if we impose periodic boundary conditions on the kinetic-energy operator and restrict ourselves to infinitely-often differentiable (in space and time) interactions and external potentials with the same boundary conditions as discussed in Ref. \cite{delort2010}. Since in this reference the question of stabilising an arbitrary number of derivatives is considered, we expect that for our case weaker conditions are sufficient. Based on the conditions for the stability of the domain $D(\hat{T})$ under time-evolution in the proofs of \cite{reed1975, yajima1987} we conjecture that it is enough that $v, \nabla v$ and $\nabla^2 v$ are in $\mathcal{C}^1 ([0,T], \mathcal{K}(\Omega))$. 
We point out, that physically such conditions are quite reasonable. If we, for instance, assume that the external potential is due to a $L^2(\Omega)$ charge distribution and hence determined from the Poisson equation (see for instance (\ref{MaxwellEquation}) where Coulomb gauge on $\mathbb{R}^3$ is employed), then $v(t), \nabla v(t)$ and $\nabla^2 v(t)$ are in $L^2( \Omega ) \subset \mathcal{K}( \Omega )$. 
For the case $\Omega=\mathbb{R}^3$ such a potential is of the form
\begin{equation}
v(\bbr , t) = -\int \diff \bbr' \, \frac{\rho (\bbr',t)}{|\bbr - \bbr'|}
\end{equation}
where $\rho (\bbr,t)$ is a square-integrable time-differentiable charge distribution (which excludes the case of external point charges which are described by delta distributions). For example, such potentials arise when in molecules the atomic nuclei are represented by finite charge distributions (for a more extensive discussion see \cite{andrae_2000}). 
For the interaction $w$ it suffices that it is in the Kato class of potentials, since the derivatives with respect to $\bbr$ in the definition of $W_{k}[v]$ can be expressed as derivatives with respect to $\bbr_2$ and thus by 
partial integration be absorbed by the wave function. For the interactions the important case of pure Coulomb potentials is therefore allowed.
These conditions make $q[v]$ integrable but note that since we rely on Hilbert space techniques in the later discussion on the inversion problem in Sec.~\ref{sec:SturmLiouville} a $L^2(\Omega)$ condition arises there.

Finally we consider under which conditions the external-force term $\nabla \cdot ( n[v] \nabla v)$ is at least integrable as well. By the product rule we can express the external-force expression in two terms $\nabla n[v] \cdot \nabla v$ and $n[v] \nabla^2 v$. Under the above assumptions of $\nabla v(t)$ and $\nabla^2 v(t)$ being in $\mathcal{K}(\Omega)$ it holds that they are individually integrable \footnote{ For instance, for Kato perturbations $\nabla^2 v$ of $\hat{T}$ it holds that for $\Psi$ in the self-adjoint domain $\|\Psi \nabla^2 v\| \leq (a \|\hat{T} \Psi\| + b \|\Psi\|) < \infty$, and thus $\|n \nabla^2 v\|_1 \leq N \|\Psi^{*} \Psi \nabla^2 v\|_1 \leq N \|\Psi \| \|\Psi \nabla^2 v\| < \infty$.}.

We will in the following denote such a set of potentials $v$ for which an initial state with the property $\hat{T}^2 \Psi_0 \in \mathcal{H}$ implies $\hat{T}^2 \Psi ([v],t) \in \mathcal{H}$ by $\mathfrak{V}$.
Then we have a mapping
\begin{eqnarray}
\label{nC2mapping}
 n: &\mathfrak{V}  & \rightarrow  \mathfrak{N} \subset \mathcal{C}^{2}([0,T],L^1(\Omega)) 
\\
& v \;& {\buildrel\rm \Psi_0 \over \mapsto }  \;  n[v], \nonumber
\end{eqnarray}
where $\mathfrak{N}$ is the set of densities $n[v]$ generated by all possible potentials in $\mathfrak{V}$. These conditions guarantee that the internal and external forces (as well as their divergences) are finite. On the basis of these domains we can discuss the bijectivity of the mapping $v \mapsto n$.

\section{The density-potential mapping}
 
\subsection{Exemplification}
 
\label{sec:Example}

So far we have discussed the potential-density mapping $v \mapsto n$, where $v \in \mathcal{V}$ is the set of potentials for which we have a unique (possibly generalized) solution of the TDSE. 
However, in order to invert this mapping we need to also show injectivity, i.e., that every density $n$ has at most one potential associated. Then and only then the potential-density mapping is bijective which allows to define its inverse, the density-potential mapping $n \mapsto v$. 

As pointed out in the previous section, we will employ the fundamental equation of TDDFT (\ref{TddftFundament}) to establish injectivity and thus bijectivity. However, in order to do so we need to restrict the set of allowed potentials to the smaller set $\mathfrak{V} \subset \mathcal{V}$ that guarantees that the individual terms of the fundamental equation are all well-defined. But this does not imply that we could not have a bijective mapping (and can thus invert the mapping) for a more general set of potentials. For instance, if we restrict to all those potentials that guarantee a classical solution of the TDSE we can for a one-dimensional noninteracting spin-singlet problem with periodic boundary conditions construct the density-potential mapping explicitly \cite{ruggenthaler2013}.
\\

The example we want to invert is
\begin{eqnarray}
\label{2periodicTDSE}
 \hspace{-0.5cm} \imagi \partial_t \Phi(x_1 \sigma_1, x_2 \sigma_2,t) =  \sum_{n=1}^2  \left[ -\frac{1}{2} \partial^2_{x_{n}} +  v_s(x_n,t) \right] \Phi(x_1 \sigma_1, x_2 \sigma_2,t),
\end{eqnarray}
with 
\be
\Phi_0(x_1 \sigma_1, x_2 \sigma_2) = \varphi_0(x_1)\varphi_0(x_2)   \theta (\sigma_1,\sigma_2) , 
\ee
where the $\theta$ is the anti-symmetric singlet spin-function with the explicit form
\be
 \theta (\sigma_1,\sigma_2)  = \frac{1}{\sqrt{2}} [\delta_{\sigma_1,\uparrow} \delta_{ \sigma_2,\downarrow} - \delta_{\sigma_2,\uparrow} \delta_{\sigma_1,\downarrow}].
\ee
So for this particular two-electron case the expansion in spin functions of (\ref{spin_exp}) has only one term.
For the spatial part of the wave function we impose
periodic boundary conditions on the interval $\Omega = [0,L]$. This system is routinely investigated in TDDFT due to its simplicity. The TDSE for two (noninteracting) particles can then be rewritten into the single-orbital TDSE
\begin{eqnarray}
\label{periodicTDSE}
 \imagi \partial_t \varphi(x,t) = \left[ -\frac{1}{2}\partial_x^2 + v_s(x,t) \right] \varphi(x,t),
\end{eqnarray}
with an initial state in the domain of the self-adjoint kinetic-energy operator with periodic boundary conditions that thus obeys $\varphi_0(0) = \varphi_0(L)$ and $\partial_x \varphi_0(x)|_{x=0} = \partial_x \varphi_0(x)|_{x=L}$. We can rewrite the initial state in its unique polar representation
\begin{eqnarray}
 \varphi_0(x) = \sqrt{\frac{n_0(x)}{2}} \e^{-\imagi S_0(x)},
\end{eqnarray}
provided $n_0(x)>0$ everywhere for a unique phase $S_0$. Therefore the density $n_0(x)$ obeys the above periodicity conditions and
\begin{eqnarray}
\label{quasiperiodic}
 S_0(0)= S_0(L)+2 \pi m \qquad \partial_x S_0(x)|_{x=0} = \partial_x S_0(x)|_{x=L},
\end{eqnarray}
where $m \in \mathbb{Z}$. Now, every (classical) solution $\varphi([v_s],x,t)$ gives rise to a temporally continuously-differentiable density $n([v_s],x,t) = 2 |\varphi([v_s],x,t)|^2$ that obeys the same boundary conditions as the initial density. Further, the resulting phases $S([v_s],x,t)$ obey the same boundary conditions as the initial phase in (\ref{quasiperiodic}). A sudden jump from $m$ to $m'$ is not allowed, since it needs an external potential that is proportional to a delta-distribution in time, which is not part of $\mathcal{V}$. Under these conditions we can explicitly invert the potential-density mapping $v_s \mapsto n$ (at least for some finite time-interval $[0,T]$).

The construction of the density-potential mapping in this case is now based on the continuity equation, which in the above polar representation reads as
\begin{eqnarray}
 \hspace{-0.5cm} - \partial_x j([v_s],x,t) = - \partial_x\left[n([v_s],x,t)\partial_x S([v_s],x,t)\right]= \partial_t n([v_s],x,t).
\end{eqnarray}
We can interpret the continuity equation as a Sturm-Liouville equation for the phase $S$ for a given time-dependent density $n$
\begin{eqnarray}
\label{periodicSLphase}
- \partial_x\left[n(x,t)\partial_x S([m,n],x,t)\right]= \partial_t n(x,t),
\end{eqnarray}
supplemented with the boundary conditions of (\ref{quasiperiodic}) for a fixed value of $m$. The unique solution for a periodic density $n(x,t)>0$ therefore becomes \cite{ruggenthaler2013, ruggenthaler2012}
\begin{eqnarray}
\label{Thetam}
S ([m,n], x,t) &=& \int_0^L \diff y \, K_t (x,y) \partial_t n(y, t)  \\
&+& \frac{2 \pi m}{\int_{0}^{L} \frac{\diff z}{ n(z, t)}} \int_{0}^{x} \frac{\diff z}{n(z,t)}, \nonumber
\end{eqnarray}
where
\begin{eqnarray}
K_t ([n], x,y) &=&  \frac{1}{2} [ \theta (y-x) - \theta (x-y) ] \int_y^x \frac{\diff z}{n (z, t)} \nonumber \\
 &-& \frac{\eta (x,t) \eta (y,t)}{\int_0^L \frac{\diff z}{n(z,t)}} , \nonumber
\end{eqnarray}
with $\theta$ the Heaviside function and 
\begin{eqnarray}
\eta (xt) = \frac{1}{2} \left( \int_0^x \frac{\diff y}{ n (y, t)} + \int_L^x\frac{ \diff y}{ n (y, t)}\right) . \nonumber
\end{eqnarray}
Consequently, we have infinitely many orbitals $\varphi[m,n]=\sqrt{n/2}\exp(-\imagi S[m,n])$ which are labelled by $m$ and correspond to realizations of the same time-dependent density $n$ starting from different initial states. If we interpret the periodic system as a quantum ring of length $L$, the different realizations correspond to different rotations of the ring \cite{ruggenthaler2013}. Finally we can invert (\ref{periodicTDSE}) for $v$ and employ the continuity equation, which allows us to express the potential as a functional of the initial state and the density
\begin{eqnarray}
\label{vfunctional}
v_s([m,n], x, t) =& \frac{1}{2} \frac{\partial_x^2 \sqrt{n(x,t)}}{\sqrt{n(x,t)}}  - \partial_t S([m,n],x,t)  \nonumber
\\
&- \frac{1}{2}\left( \partial_x S([m,n],x,t) \right)^2.
\end{eqnarray}
Thus, we have inverted the usual map from potentials to densities and explicitly constructed a density-potential map $n \mapsto v_s$. We note, that a similar construction for radially symmetric problems can be found in \cite{damico1999}.
\\

Let us consider a few consequences of the existence of a density-potential mapping with the help of this explicit example. First of all, such a mapping implies that the wave functions and thus also all observables are functionals of the initial state and the density by the composite mapping $n \mapsto v \mapsto \Psi$. This is one of the main implications of the famous Runge-Gross result \cite{runge1984} and the very foundation of TDDFT. It allows us to determine any physical quantity by only knowing the density and the initial state (at least in principle). In our example we can give an explicit realization of the Runge-Gross result since by the above construction we can express the wave function $\Phi[m,n]$ as a functional of the initial state (labelled by $m$) and the time-dependent density $n$. Any observable $\hat{O}$ inherits the functional dependence by
\begin{eqnarray}
 O([m,n],t) = \braket{\Phi([m,n],t)}{\hat{O} \Phi([m,n],t)}.
\end{eqnarray}
For instance the kinetic energy as a functional of the initial state and the density reads as
\begin{eqnarray}
\label{kineticweizsacker}
\hspace{-1.5cm}T([m,n],t) = \frac{1}{8} \int_0^{L} \diff x \frac{\left(\partial_x n(x,t)\right)^2}{n(x,t)} + \frac{1}{2} \int_0^{L} \diff x\; n(x,t) \left(\partial_x S([m,n],x,t)  \right)^2,
\end{eqnarray}
where the first term is the Weizs\"acker energy functional and the second term is an initial-state dependent velocity contribution.

A further detail of the density-potential mapping is the \textit{initial-state dependence}. For every different possible initial state we have a different density-potential mapping and consequently different wave functions that generate the same density in time. This makes the construction of universal time-dependent density-functionals a lot more complicated, since in principle we would need to incorporate the initial-state dependence as well. Therefore, one usually employs ground-state DFT to get rid of the initial-state dependence \cite{ullrich2012, marques2012, maitra2002, maitra2010}. By the Hohenberg-Kohn theorem, for any ground-state density there is (usually) a unique ground-state wave function that minimizes the Hohenberg-Kohn functional, i.e., the kinetic and interaction energy. Thus by restricting to ground-states as the only allowed initial states one can ignore the initial state dependence and have a ``pure'' density-functional. This restriction excludes initial densities that have nodes or densities with $\partial_t n(x,t)|_{t=0} \neq 0$. As we can then also see in our explicit example, the Hohenberg-Kohn functional corresponding to (\ref{kineticweizsacker}), i.e.,
\begin{eqnarray}
T[m,n] = \frac{1}{8} \int_0^{L} \diff x \frac{\left(\partial_x n_0(x)\right)^2}{n_0(x)} + \frac{2 \pi^2 m^2}{\int_0^L \frac{\diff x}{n_0(x)}},
\end{eqnarray}
has a unique minimum for $m=0$ which singles out the ground state corresponding to the chosen density $n_0(x)$. While from a purely formal point of view it seems desirable that we ``only'' need to approximate the dependence of functionals on the density, the initial-state dependence can also be an advantage in practice, e.g., in the case of charge-transfer problems, which are briefly discussed later in this section. 

The main approach to perform practical TDDFT calculations is the \textit{time-dependent KS scheme} discussed in Sec.~\ref{sec:KohnSham}. By employing two different density-potential mappings we can determine the time-dependent density of a quantum system by solving an auxiliary non-linear problem. While usually this is done to determine the density of an interacting many-particle problem by a non-interacting auxiliary system, the KS construction allows to connect any two different systems. 
We can, for instance, connect two different non-interacting systems with two different initial states. In that case we will still have an xc potential but this then, in the absence of two-particle interactions, is purely generated by initial state dependence.
In our example at hand we can determine the density found by solving (\ref{2periodicTDSE}) for a fixed $v_s(x,t)$ starting from an initial state characterized by $\varphi_0 = \sqrt{n_0/2} \exp(-\imagi S_0[m,n])$, by solving a non-linear auxiliary problem of the form
\begin{eqnarray}
\imagi \partial_t \varphi (x,t) &=&  \Big( -\frac{1}{2} \partial_{x}^2  +  v_s(x,t) + v_{\mathrm{xc}} ([m,m',n],x,t) \Big) \varphi (x,t), \label{KS1} \\
  n(x,t)   &=&  2 |\varphi (x,t)|^2 ,
\end{eqnarray}
with a different initial state $\varphi_0' = \sqrt{n_0/2} \exp(-\imagi S_0[m',n])$. The xc potential in this case\footnote{We note that since we look at two non-interacting problems the Hartree term is zero by construction.} is determined by
\begin{eqnarray}
\label{periodicKSpotential}
 &v_{\mathrm{xc}}([m,m',n],x,t)  = v_s ([m',n],x,t)-v_s ([m,n],x,t) \nonumber \\
 & = 2 \pi (m-m') \; \partial_t \left(\frac{ \int_{0}^{x} \frac{\diff z}{n(z,t)}}{\int_{0}^{L} \frac{\diff z}{ n(z, t)}}\right) \nonumber
  + \frac{2 \pi^2 (m^2- m'^2)}{\left( n(x,t) \,\int_{0}^{L} \frac{\diff z}{ n(z, t)}\right)^2} \nonumber \\
 & + \frac{2 \pi (m- m')}{\int_{0}^{L} \frac{\diff z}{ n(z, t)}} \frac{ \partial_x S([ 0,n], x,t)}{n(x, t)} .
\end{eqnarray}
Note, that with help of the current and the continuity equation we can express the terms $\partial_t n$ time-locally by the orbital $\varphi$. Thus we do not need any further information to uniquely solve the KS equation than the chosen potential of the original problem $v_s(x,t)$ and the initial states. If the xc potential would depend on higher-order derivatives, we might need further information to determine the unique solution.

Even though the above density-potential mapping $v_s[m,n]$ and the example of an xc potential $v_{\mathrm{xc}}[m,m',n]$ do only depend on the instantaneous density (and its time-derivatives), this time-local behaviour is not a generic feature. Actually, the density potential-mappings usually depend not only on the initial state but also on the density at previous times. This property is termed \textit{memory} \cite{maitra2002, maitra2010, ullrich2012, marques2012} and for fixed initial state $\Psi_0$ it is formally expressed by
\begin{eqnarray}
\hspace{-1.5cm} \frac{\delta v([n], \bbr,t)}{\delta n(\bbr',t')} \equiv \chi^{-1}([n], \bbr,t , \bbr',t') \neq \delta(t-t') f([n], \bbr,t , \bbr',t'),
\end{eqnarray}
where the inverse linear-response kernel $\chi^{-1}[n]$ is assumed to obey
\begin{eqnarray}
\hspace{-1.5cm} \delta v([n,\Delta n],\bbr,t) = \int_0^{t} \diff t' \int_{\Omega} \diff\bbr \, \chi^{-1}([n], \bbr,t , \bbr',t')   \Delta n(\bbr',t').
\end{eqnarray}
Here $\delta v[n,\Delta n]$ is understood as the Fr\'echet derivative (see \ref{Frechet} for the case of $\delta\Psi[v,\Delta v]$). The inverse linear-response kernel takes a specifically simple form if we assume that we start from a ground-state. As in the case of the linear-response kernel $\chi[v] = \delta n[v]/ \delta v$ the inverse response kernel then only depends on the time-difference, i.e., $\chi^{-1}([n], \bbr, \bbr',t-t')$ \cite{ullrich2012, marques2012}. This form also shows most clearly why memory is a necessity of most density-potential mappings, especially in the context of the KS construction. If we Laplace-transform (related to the Fourier-transform with a step-function) the (inverse) linear-response kernel from $t-t'$ to the frequency $\omega$ we find that the linear-response kernel has poles at the eigenfrequencies of the (time-independent) system that has the initial density $n_0$ as its ground-state density. The inverse linear-response function has zeros at these frequencies. If we then want to simulate the linear-response of an interacting system (starting from its ground state) by the linear-response of a KS system \cite{ullrich2012, marques2012}, we see that the (linear-response of the) xc potential needs to cancel the poles of the auxiliary system and generate the poles of the original interacting system (see \cite{ruggenthaler2013} for an analytic example of these properties and \cite{thiele2014} for a numerical reconstruction). Thus, for any KS construction where we want to simulate an interacting by a non-interacting system, the xc potential necessarily will have memory (since the spectrum of their respective Hamiltonians are very different). This argument also illustrates, why we get away with such a simple (time-local) xc potential in our example above: we simulate one system by another system with a similar Hamiltonian (and thus with a similar spectrum).

This result together with the knowledge that \textit{memory and initial-state dependence are closely related} \cite{maitra2002}, i.e., it is possible to replace $v([\Psi_0,n],x,t)$ by $v([\Psi(t'),n], x,t)$ for all $0 < t' < t$ using the functional relation $\Psi(t') = \Psi([\Psi_0,v],t')$, can be used to simplify the KS scheme and make it more reliable. Firstly, by choosing a KS system that already has some of the properties of the interacting many-body system one wants to simulate, the complexity of the xc potential can obviously be reduced. The more similar the original system and the KS system are, the less the density-potential mappings will differ and the xc potential will become less signficant. For instance, one could simulate an interacting problem by a system with a different interaction that can be treated in a numerically efficient way (similar to hybrid functionals \cite{engel2011, marques2012, ullrich2012}). A further simplification is possible, if one employs the initial-state dependence and chooses an initial KS state that incorporates some of the physical properties of the interacting system \cite{elliot2012,ruggenthaler2013}. For instance, one of the major challenges the current (usually time-local) approximations to the xc potential face, is the proper description of charge-transfer reactions \cite{casida2000,dreuw2003,tozer2003,gritsenko2004,stein2009,hesselmann2009,fuks2011}. If the simulations are started from the ground state it is well-known \cite{raghunathan2011,fuks2013} that one needs time-nonlocal (frequency-dependent) xc functionals to get the transfer process right. However, recent results \cite{rozzi2013,falke2014,fuks2014,fuks2014b} show that if one starts from an excited state, the usual simple (time-local) approximations can reproduce a charge-transfer reaction reasonably well.

\begin{figure} [H]
\centering
\includegraphics[width=9cm]{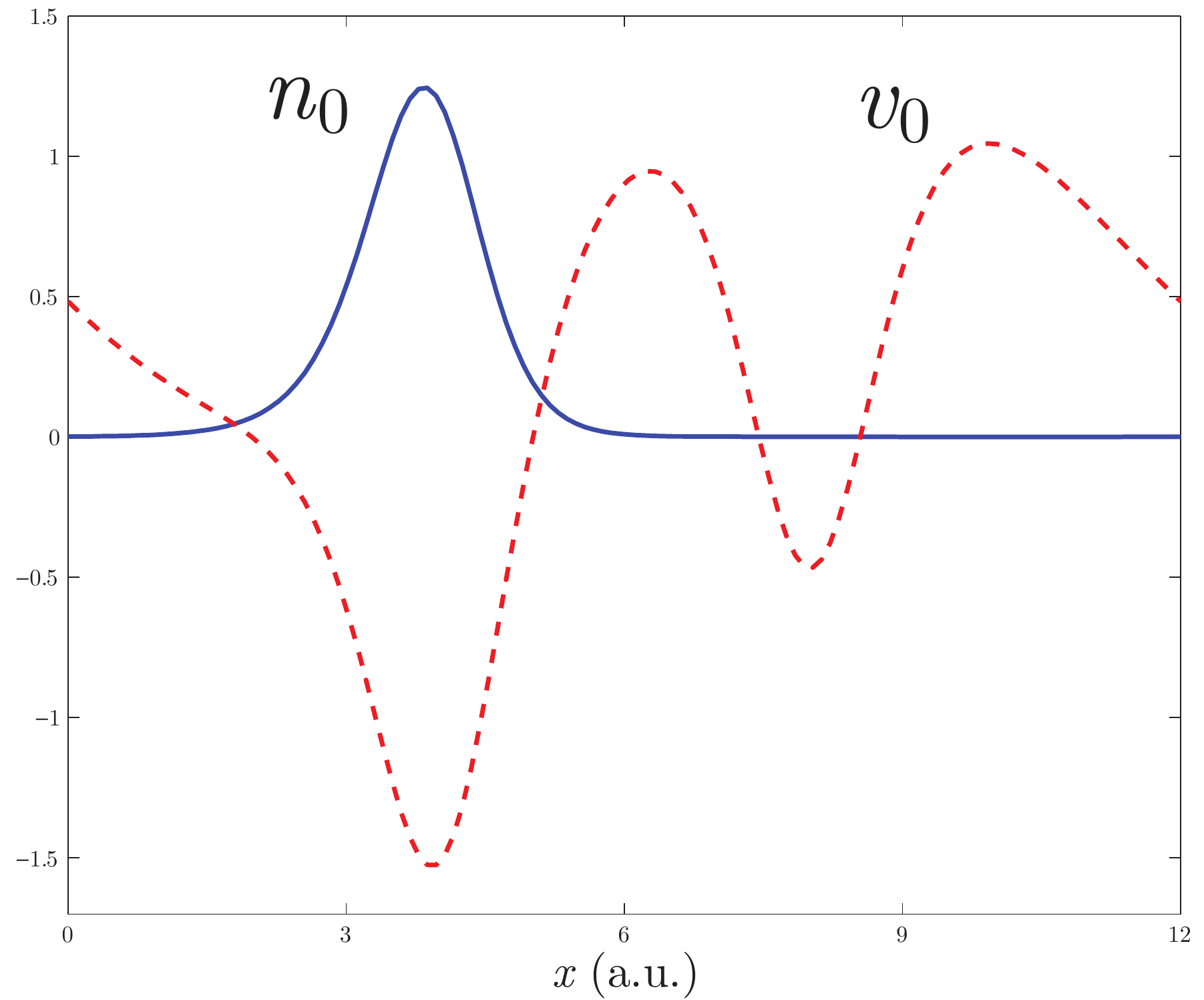}
\caption{The static potential $v_0$ (dashed red line) and the ground-state density $n_0$ (solid blue line) of the interacting system.}
\label{fig:DensityPotential}
\end{figure}

Finally, let us make use of our analytical expression $v_s[m,n]$ from (\ref{vfunctional}) and give an example of the xc potential for two different values of $m$. Since we have been briefly discussing the challenge of charge-transfer problems within TDDFT, we will consider a very simple toy model of such a process. The interacting reference system is a two-electron system on a ring $\Omega = [0,12]$ which has the periodic interaction $w(x) =  \cos(2 \pi x/12)/2$ and starts in the ground state of the time-independent potential 
\begin{eqnarray}
\label{StaticExample}
 v_0(x) = -\frac{2}{\cosh(x-4)^2} - \frac{2}{\cosh(x-8)^2} + 0.7 \cos\left(\frac{2 \pi (x-8)}{12}\right),
\end{eqnarray}
which is displayed in red in Fig.~\ref{fig:DensityPotential}. 
The first two terms in this expression describe wells around $x=4$ and $x=8$ while the last term reduces the depth of the well around $x=8$.
In the same figure we also have given the ground-state density $n_0$ which is localized in the left well.

\begin{figure} [H]
\centering
\includegraphics[width=9cm]{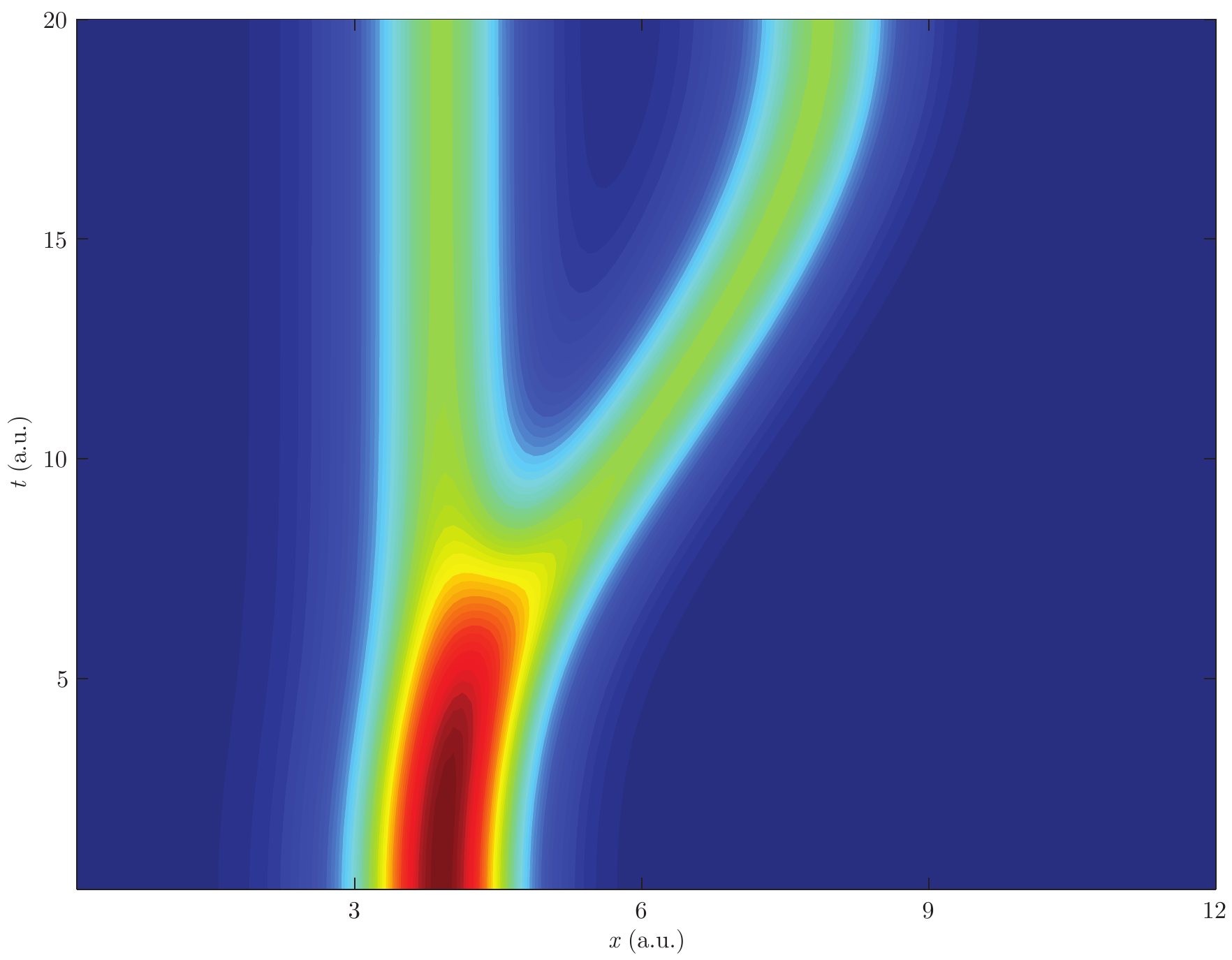}
\caption{The density profile $n$ used in this example. Half of the density of the initial state is moved to the right potential well.}
\label{fig:Density}
\end{figure}

Now, we prescribe a time-dependent density profile evolving from the ground state density $n_0$ in which we split the two-particle density into two parts and move half of the density to the right well in $T=20$ as is displayed in Fig.~\ref{fig:Density}.

\begin{figure} [H]
\centering
\includegraphics[width=\textwidth]{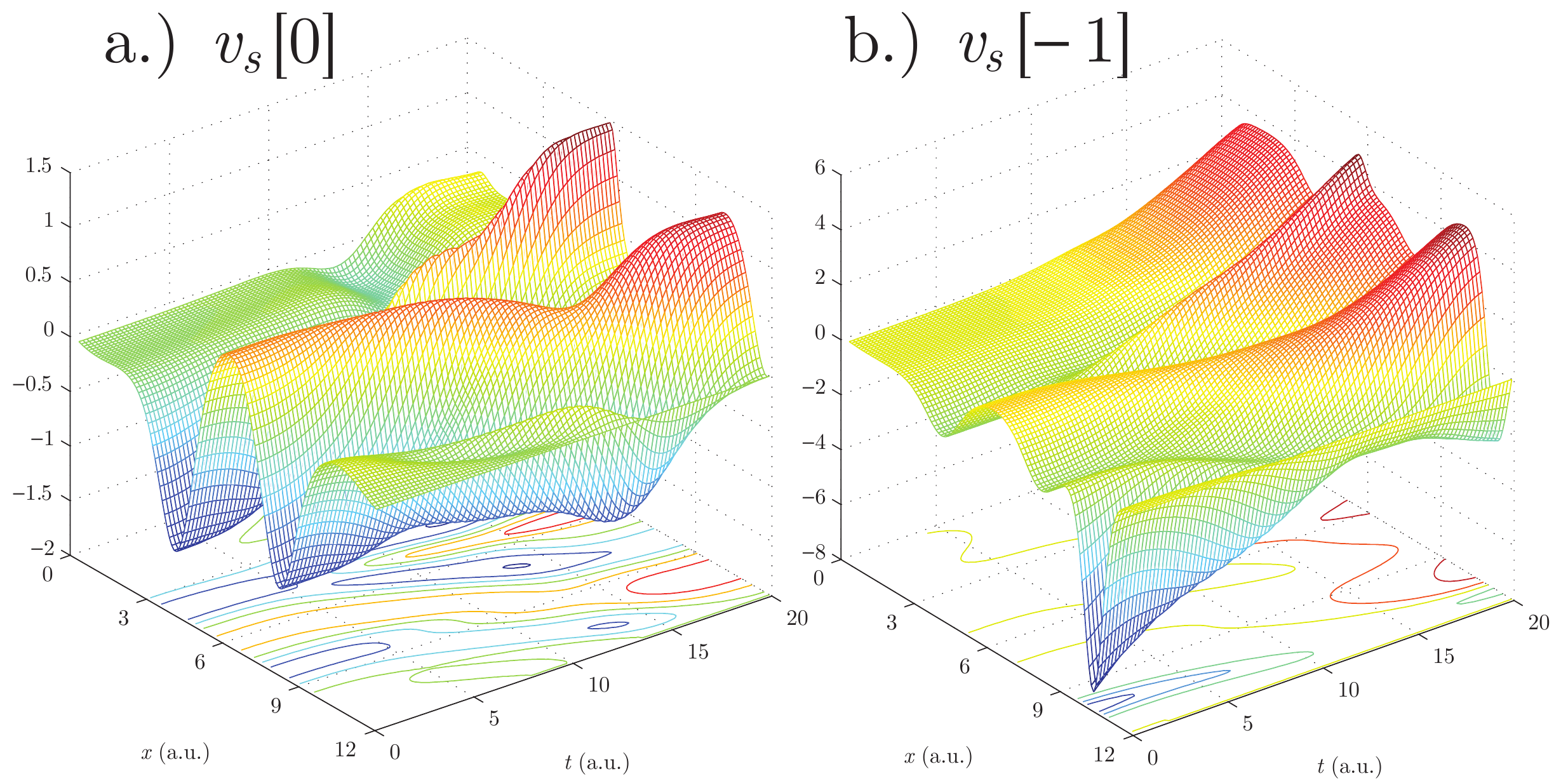}
\caption{The potential $v_s[0,n]$ and $v_s[-1,n]$ that splits the charge on the ring.}
\label{fig:vm0vm1}
\end{figure}

From our analytic formula we can then determine the external potential $v_s[m,n]$ that does this for different initial states of the KS system, i.e., different values of $m$. The results for $m=0$ and $m=-1$ are shown in Fig~\ref{fig:vm0vm1}. 

\begin{figure} [H]
\centering
\includegraphics[width=\textwidth]{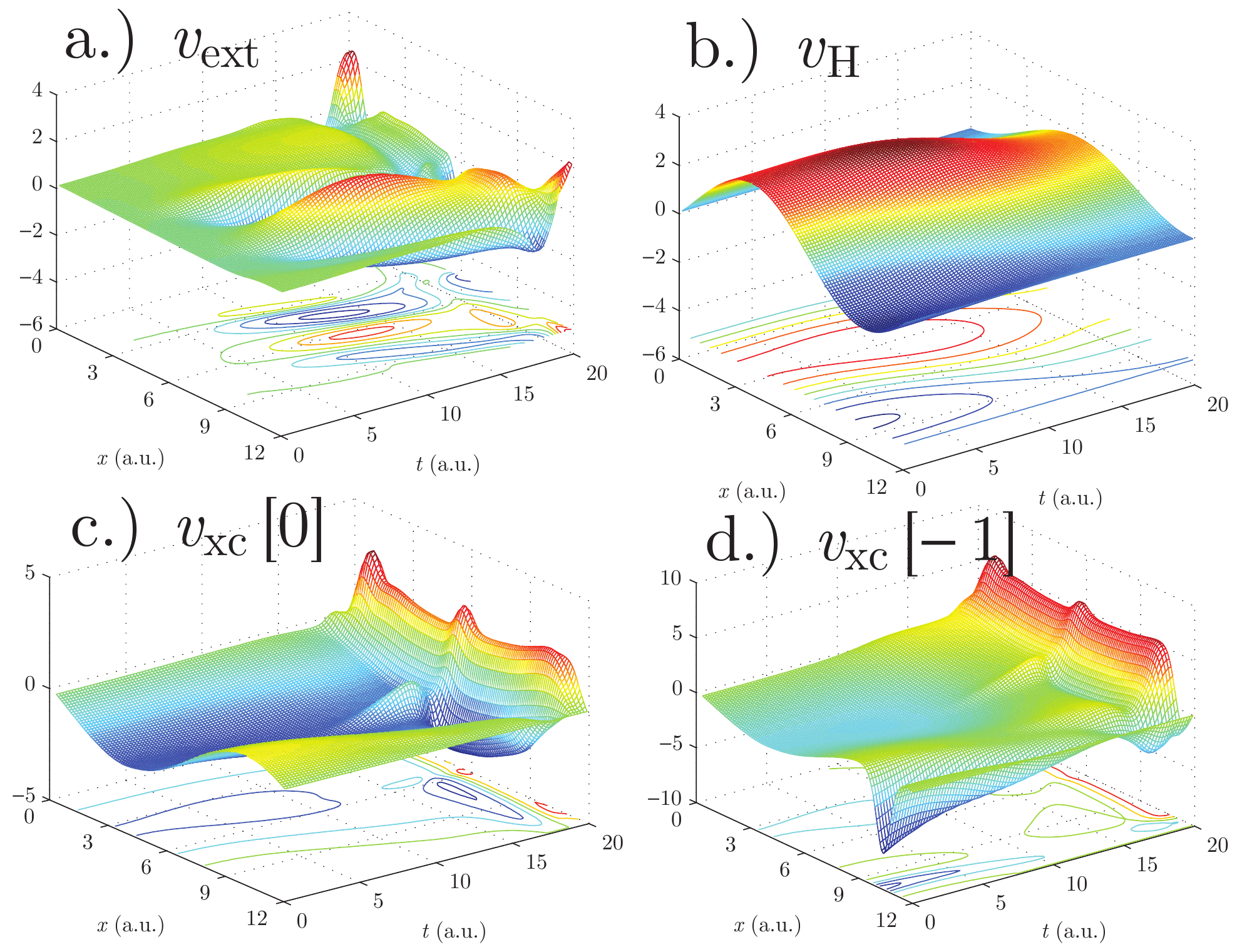}
\caption{a.) The external potential $v_{\mathrm{ext}} = v[\Psi_0,n] - v_0$ that splits the density on the ring in the interacting system. b.) The Hartree potential and the xc potentials for c.) the $m=0$ and for d.) the $m=-1$  initial state.}
\label{fig:PotHxc0xc1}
\end{figure}

To determine the xc potential for these two different initial states we finally also need to know the external potential $v [\Psi_0,n]$ of the interacting
reference system with interaction $w$ that generates the same density via time propagation (see (\ref{xcPotential})) starting from the interacting
ground state $\Psi_0$.
While we do not have an analytical formula in this more complex situation (except for the initial time when it is equal to $v_0$) we can determine this potential from the numerical procedure that will be discussed in 
Sec.~\ref{sec:numerics}. The results are then displayed in Fig.~\ref{fig:PotHxc0xc1}. 

Here we see the external time-dependent potential $v_{\mathrm{ext}}$ (where we have subtracted the time-independent part, i.e., $v_{\mathrm{ext}} = v[\Psi_0,n] - v_0$) that forces the interacting two-particle system to obey the above prescribed rigid charge transfer. The Hartree potential $v_{H}$ is the same for both initial states, since it only depends on the instantaneous density. All the memory and initial-state dependence is found in the xc potentials given by 
\begin{equation}
v_\xc [m,\Psi_0,n] = v_s [m,n]  - v_\mathrm{H} [n] - v [\Psi_0,n] .
\end{equation}
Obviously simple time-local approximations to the xc potential cannot capture the rich structure of $v_{\mathrm{xc}}$ in this case.

\subsection{The local-force equation approach to the density-potential mapping}

\label{sec:IterativeScheme}

Let us now consider the general case of a density-potential mapping.
In Sec.~\ref{sec:observables} we have seen that the local-force equation makes a connection between the density of a many-electron system
and the potential $v$ that generates the density $n$ by time-propagation of the TDSE. Here we will outline how this equation can be used
to show invertibility of the mapping $v \mapsto n$ for a given initial state, i.e., the existence of the density-potential mapping, and 
how it can be used in an iterative way to calculate the potential $v$ that generates a {\em given} density $n$. 
\\

We first rewrite (\ref{TddftFundament}) as
\begin{equation}
- \nabla \cdot \left[ n ([v],\bbr,t) \nabla v(\bbr,t)\right] = q ([v], \bbr,t) - \partial_t^2 n([v],\bbr,t).
\end{equation}
Here both $n([v],\bbr, t)$ and $q([v],\bbr, t)$ are functionals of the potentials. Suppose now, however,
that we fix the density $n(\bbr,t)$. Then we have a non-linear equation for the potential $v$, i.e.,
\begin{equation}
\label{nonlinear}
- \nabla \cdot \left[ n (\bbr,t) \nabla v(\bbr,t) \right] = q([v],\bbr,t) - \partial_t^2 n (\bbr,t).
\end{equation}
If we have prescribed a density that we have generated by time propagation of the initial state
$\Psi_0$ with an external potential $v$, we can use this equation to ask whether a density-potential mapping
exists. Namely, if we can show that the only potential that solves this equation is the potential $v$ that generated the
density $n(\bbr,t)$, we
can show that the mapping $v \mapsto n$ is injective and hence invertible. To answer whether $v$ is the
only solution, we linearize the above non-linear equation by an iterative procedure 
\begin{equation}
- \nabla \cdot \left[ n (\bbr,t) \nabla v_{k+1}(\bbr,t)\right] = q([v_k],\bbr,t) - \partial_t^2 n (\bbr,t).
\label{iterate}
\end{equation}
Here we determine the inhomogeneity $q[v_k]$ from $\Psi[v_k]$ for a given initial state $\Psi_0$.
Now, a solution to the non-linear equation is a fixed point of the linearized one, i.e.,
if we use $v$ to propagate $\Psi_0$ and determine the unknown $q[v]$ from $\Psi[v]$ the inversion 
of $-\nabla \cdot (n \nabla)$ gives back the same $v$ in (\ref{iterate}). Thus for the existence
of a density-potential mapping we need to show that the only fixed point of the iterative equation is $v$.
This will be done in Sec.~\ref{sec:Uniqueness} and Sec.~\ref{sec:FixedPoint}.

On the other hand, if we prescribe a density $n$ for which we do not know \textit{a priori} that it is generated by solving the TDSE we can try  to employ (\ref{iterate}) to find an appropriate $v$.
So we start by making a guess for an initial potential $v_0 (\bbr, t)$ on a time interval $[0,T]$.
Then we propagate the TDSE with this potential and a given initial state (compatible with the initial density) to calculate $q[v_0]$. Then we can calculate a new potential $v_1 (\bbr,t)$ by solving (\ref{iterate}). With $v_1$ we can repeat the procedure to find a new potential $v_2$, etc. 
In this way we have constructed the mapping 
\begin{eqnarray}
\label{Fmapping}
 \mathcal{F}: v_{k} \mapsto q[v_k] \mapsto v_{k+1} = \left[- \nabla \cdot (n \nabla) \right]^{-1} \left( q[v_k] - \partial_t^2 n  \right)
\end{eqnarray} 
which defines a series $\left\{ v_k \right\}$ of potentials.
The goal is now to show that 
under certain assumptions $v_k \rightarrow v$ (in some norm sense) for $k \rightarrow \infty$, i.e., that $v$ is a fixed point
of (\ref{iterate}). However, if we can invert $-\nabla \cdot (n \nabla)$ then at $t=0$ we are 
already converged after the first iteration, i.e., 
$v_1(\mathbf{r},0) = v(\mathbf{r},0)$ since $q([v_0], \mathbf{r},0)$ is given in terms of the initial state only and therefore independent of $v_0$.
Under which conditions we can invert this operator will be discussed in detail in Sec.~\ref{sec:SturmLiouville}. 
Assuming this for the moment we can conclude that for small enough times already $v_1$ will be close to the exact $v$. Therefore it seems likely that the next step in the iteration will be even closer to $v$ unless the $q[v_1]$ is very different from the exact $q[v]$. This, however, can only 
happen if the internal-force densities generated by two potentials that are arbitrarily close differ strongly. Such a situation would be quite unphysical, since arbitrarily small changes in a potential would lead to totally different dynamics within very short times (making any prediction for real system impossible). Although one could imagine such situations (for instance in the case of non-classical solutions to the TDSE discussed in Sec.~\ref{sec:Cauchy}), we exclude them from our considerations and will discuss this in more detail in Sec.~\ref{sec:FixedPoint}. 
Summarising we therefore conclude that for short enough time intervals the iteration scheme based on (\ref{iterate}) is expected to converge fast, which is an important feature that is also used in our numerical implementation of Sec.~\ref{sec:numerics}.

However, even if we assume that $v_k \rightarrow v$, does this guarantee that $v$ really generates the prescribed density by propagation of the initial state $\Psi_0$? The potential that we found is by construction a fixed point of (\ref{iterate}) but does also obey its own local-force equation (\ref{TddftFundament}). If we subtract both equations and denote $\rho = n[v]-n$ the difference between the density generated via propagation of $v$ and the prescribed density $n$, we find \cite{ruggenthaler2011, ruggenthaler2012}
\begin{eqnarray}
\label{RhoEquation}
 \partial_t^2 \rho(\bbr,t) = \nabla\cdot\left[\rho(\bbr,t) \nabla v(\bbr, t)\right].
\end{eqnarray}
Now, if we assume that the prescribed density obeys the minimal restrictions
\begin{eqnarray}
n(\bbr,0) &= \braket{\Psi_0}{\hat{n}(\bbr) \Psi_0} = n([v], \bbr,0),
\\
\partial_t n(\bbr, t)|_{t=0} &= -\nabla \cdot \braket{\Psi_0}{\hat{\mathbf{j}}(\bbr) \Psi_0} = \partial_t n([v],\bbr, t)|_{t=0},
\end{eqnarray}
(here the current operator is defined by $\hat{\mathbf{j}}(\bbr) = 1/(2 \imagi )\sum_{k=1}^{N}(\delta(\bbr-\bbr_k) \overrightarrow{\nabla}_{k} - \overleftarrow{\nabla}_{k} \delta(\bbr-\bbr_k))$) the above equation is a linear evolution equation with initial conditions $\rho(\bbr,0) = \partial_t \rho(\bbr, t)|_{t=0} = 0$.
Thus the question whether $v$ generates the prescribed density via propagation reduces to the question whether (\ref{RhoEquation}) has only $\rho = 0$ as unique solution for the above initial conditions. While for analytic potentials $v$ one can rigorously show by the classical considerations of Kowalevskaya \cite{Kowalevskaya, evans2010} that this is true, we are not aware of a general proof for this statement. However, due to the fact that we can impose further conditions on $\rho$ \cite{ruggenthaler2011, ruggenthaler2012},
e.g., $\int \diff \bbr\, \rho(\bbr,t) = 0$ for all times, we presume in the following that it holds true as it indeed seems highly probable. Under this assumption our iteration scheme, if it converges, does indeed reproduce the prescribed $n$.

Finally we note that the above iteration procedure in terms of $v_k$ also gives rise to iterations in $\Psi_k=\Psi[v_k]$. Thus one could make the TDSE part in the above iteration explicit by introducing
\begin{eqnarray}
 \imagi \partial_t \Psi_{k+1}(t) = \left(\hat{H}_0 + \hat{V}([\Psi_k],t) \right) \Psi_{k+1}(t)
\end{eqnarray}
for the initial state $\Psi_0$, where the individual potentials in $\hat{V}([\Psi_k],t)$ are determined by (\ref{iterate}).
If we converge then $\Psi_{k} \rightarrow \Psi$ and then the resulting wave function solves the non-linear TDSE
\begin{eqnarray}
 \imagi \partial_t \Psi(t) = \left(\hat{H}_0 + \hat{V}([\Psi],t) \right) \Psi(t),
\end{eqnarray}
where accordingly the $v[\Psi]$ are determined by (\ref{nonlinear}). We therefore see that the fixed-point procedure is equivalent to the non-linear-TDSE approach to TDDFT \cite{tokatly2007,tokatly2009,maitra2010,tokatly2011b}.

\subsection{The Sturm-Liouville operator and its invertibility}

\label{sec:SturmLiouville}

%
%
We see from (\ref{nonlinear}) that, when we denote the right-hand side by $\zeta (\bbr,t)$, that we need to solve an equation for $v$ of the form
\be
\label{GeneralSturmLiouville}
- \nabla \cdot \left[  n (\bbr,t) \nabla v (\bbr,t) \right] = \zeta (\bbr,t) 
\ee
for a given inhomogeneity $\zeta$ with the property
\begin{eqnarray}
\label{IntegralZeta}
 \int_{\Omega} \diff \bbr\, \zeta (\bbr,t) =0
\end{eqnarray}
The fact that the inhomogeneity integrates to zero is due to the fact that $q$ is a divergence and that the total number of particles is conserved.
From the previous considerations we have seen that the invertibility of this Sturm-Liouville equation (\ref{GeneralSturmLiouville}) (usually this terminology is only used in the one-dimensional case but we will employ it also for higher dimensionality)
is fundamental to the construction of a density-potential mapping. This equation appears in the iterative sequence of potentials for a given density $n$ of (\ref{iterate}) as well as in the definition of the (equivalent) non-linear TDSE approach (which itself gives rise to a iterative sequence of wave functions). We therefore provide in this subsection a detailed discussion about the properties of the Sturm-Liouville operator $-\nabla \cdot  [ n (\bbr,t) \nabla ]$ and conditions for the invertibility of the Sturm-Liouville equation (\ref{GeneralSturmLiouville}). The discussion will be general in the sense that we will not use the explicit form of the inhomogeneity $\zeta$ in terms of the density and the divergence of the local forces. Possibly stronger results may be obtained
by taking into account this specific structure but we will leave this for future work.
\\
For simplicity we start with the one-dimensional version of (\ref{GeneralSturmLiouville})
\begin{eqnarray}
\label{OneDimensionalSturmLiouville}
- \partial_x\left[n(x,t)\partial_x v(x,t)\right]= \zeta(x,t).
\end{eqnarray}
To discuss this equation in a general setting we will in the following regard the Sturm-Liouville operator as a linear operator in the Hilbert space of square integrable functions on either some finite interval $[a,b]$ or one whole real line. In this setting it is then important that both
$v$ and $\zeta$ are in the Hilbert space and therefore square-integrable. The issue of a unique inversion up to a pure gauge is then
equivalent to the question of the (possible) self-adjoint domains which have the purely time-dependent function as the unique square integrable eigenfunction with zero eigenvalue. 
In the following we therefore want to say something about the eigenvalues and eigenfunctions of the Sturm-Liouville operator (and whether they exist) for certain types of boundary conditions.
Let us start by a few simple manipulations to discover some general features. By integration of (\ref{OneDimensionalSturmLiouville}) we find
\be
n(x,t) \partial_x v(x,t) = n (a,t) \partial_x v(a,t) - \int_a^x \diff y \, \zeta (y,t).
\ee
Clearly we can find an equation for $\partial_x v(x,t)$ provided we are allowed to divide by the density. This is allowed everywhere except at the boundaries of our interval where the density may go to zero. Let us, however, assume that we
are allowed to divide by the density. Then we can do another integration to write
\begin{eqnarray}\label{v_green}
v(x,t) &=& v(a,t) + n (a,t) \partial_x v(a,t) \int_{a}^x \diff y \, \frac{1}{n(y,t)} \nonumber \\
&&- \int_{a}^x \diff y \, \frac{1}{n(y,t)} \int_a^y \diff z \, \zeta (z,t).
\end{eqnarray}
In case that $\zeta=0$ we see that the general form of the solution is
\be
\label{v_kernel}
v(x,t) = c(t) + d(t)  \int_{a}^x \diff y \, \frac{1}{n(y,t)}.
\ee
This corresponds to the eigenfunction $\phi_0 (x,t)$ of the Sturm-Liouville operator with zero eigenvalue. 
We see that the zero eigenfunction is more general than just a pure gauge $v(x,t)=c(t)$. We can always add
$\phi_0 (x,t)$ to a particular solution of (\ref{OneDimensionalSturmLiouville}) and it will be another
solution.  To make the inversion unique up a gauge
we therefore have to impose boundary conditions such that the second term in (\ref{v_kernel}) vanishes.
It turns out that which boundary conditions we can choose and which eigenspectrum we can obtain 
depends very much on the behavior of the  function $\int_a^x \diff y/n(y,t)$ where $a$ is a boundary point. The simplest
case is when $n(x,t) \geq \varepsilon >0$ for some positive number $\varepsilon$. In this case
division by the density is no problem and the mathematics is the simplest. 
The most relevant physical case in which this happens is the
case of a system with periodic boundary conditions, such as is the case for particles on a ring.
In this case it is then natural to also impose periodic boundary conditions on the solutions of
the Sturm-Liouville equation, and a quick calculation then shows that the only possibility for the zero eigenfunction
in (\ref{v_kernel}) to be periodic is to demand that $d(t)=0$.
In this case we can choose periodic boundary conditions also on the potentials, making the Sturm-Liouville operator self-adjoint with a purely discrete spectrum \cite{bailey2001, zettl2010, ruggenthaler2012}, i.e., 
\begin{eqnarray}
\label{PeriodicOneDimensionalSturmLiouville}
 - \partial_x ( n\partial_x ) \equiv \sum_{k=0}^{\infty} \lambda_k \ket{\phi_k}\bra{\phi_k},\\
0= \lambda_0 < \lambda_1 \leq \lambda_2 \leq \lambda_3 \leq \ldots \nonumber
\end{eqnarray}
Since $\zeta$ by assumption~(\ref{IntegralZeta}) is perpendicular to $\phi_0 (x,t)= c(t)$ the (pseudo-) inverse operator is well-defined and bounded, i.e.,
\begin{eqnarray}
 \| v \|^2 =& \left \| \sum_{k=1}^{\infty}\frac{1}{\lambda_k} \ket{\phi_k}\braket{\phi_k}{\zeta} \right \|^2 \leq \sum_{k=1}^{\infty} \left|\frac{1}{\lambda_k} \right|^2 \left| \braket{\phi_k}{\zeta} \right|^2
\\
   & \leq  D  \sum_{k=1}^{\infty} \left| \braket{\phi_k}{\zeta} \right|^2 = D \| \zeta \|^2,  
\end{eqnarray}
where $D=\lambda_1^{-2}$. Alternatively, one can show boundedness also from (\ref{Thetam}) for $m=0$ (see also \cite{ruggenthaler2012}), since 
\be
\left[ - \partial_x (n \partial_x)\right] ^{-1} \equiv \sum_{k=1}^{\infty}\frac{1}{\lambda_k} \ket{\phi_k}\bra{\phi_k}
\label{inverse_SL}
\ee
is the spectral form of the Green's function of (\ref{Thetam})\footnote{To be precise, since for $\Omega$ bounded the spectral form of the inverse is defined on $L^2(\Omega) \subset L^1(\Omega)$ we can define it as the restriction of the Green's function of (\ref{Thetam}) onto $L^2(\Omega)$. $\left[ - \partial_x ( n \partial_x)\right] ^{-1}$ is a bounded operator on $L^2(\Omega)$ as well as on $L^1(\Omega)$. We further note, that the specific number of the bound $D$ might depend on whether we consider the operator on $L^2(\Omega)$ or on $L^1(\Omega)$.}. This shows that for every $\zeta$ which is perpendicular to $\phi_0 = d$ we have a well-defined $v$, which is
obtained by the action of the inverse operator in (\ref{inverse_SL}) on $\zeta$. \\
This brings us to the more difficult case in which the density can become zero at one of the boundary points. 
To be more definite we take zero boundary conditions on the interval $[a,b]$ where $n(a,t)=n(b,t)=0$.
We consider a class of densities such that $n(x,t) \sim (x-a)^2$ close to the boundary (and similarly in point $b$) which
is the most generic case for particles in a box.
For this case the integral $\int_a^x \diff y/n(y,t)$ diverges and we have to use so-called singular Sturm-Liouville theory \cite{zettl2010}.
%
%
%
%
%
%
Within this theory the boundary points of our problem
are so-called \textit{limit-point endpoints} of the Sturm-Liouville equation \cite{bailey2001, zettl2010, ruggenthaler2012}. In this case we only have one possible self-adjoint domain (for details we refer to \cite{zettl2010}).
Any twice-differentiable potential $v$ with a behaviour at the boundaries which is less singular than $v(x) \sim (x-a)^{-1/2}$ will be in this domain\footnote{We point out, that although from the condition on the differentiability of $n\partial_x v$ potentials as singular as $v(x) \sim (x-a)^{-1}$ are possible, they are no longer in $L^2(\Omega)$ and thus outside of the self-adjoint domain.}. The self-adjoint Sturm-Liouville operator then (possibly) has also a continuum in the spectrum and its spectral representation is
\begin{eqnarray}
 - \partial_x (n \partial_x ) \equiv \int_{\mathbb{R}}  \, \diff k\; \lambda (k) \, \ket{\phi_{k}}\bra{\phi_{k}}.
\end{eqnarray}
In our case of limit-point endpoints the unique zero eigenfunction is given by $\phi_0 (x,t)=c(t)$ and since $n(x) \sim (x-a)^2$ at the boundaries, the continuum is gapped away from zero by some $\delta > 0$ \cite{ruggenthaler2012} and we can define a (pseudo-) inverse $\left[- \partial_x (n \partial_x )\right]^{-1} \equiv \int  \, \diff k \; \lambda (k)^{-1} \, \ket{\phi_{k}}\bra{\phi_{k}}$, which determines $v$ in terms of the inhomogeneity $\zeta$ up to the physical gauge freedom. Again, the inverse is a bounded operator on the space perpendicular to the constant function with bound $D = \delta^{-2} < \infty$. This is no longer the case, however, if $n(x) \sim (x-a)^{p}$ where $p>2$ at the boundaries \cite{ruggenthaler2012}. Nevertheless, since we only consider potentials that are less singular than $ v(x) \sim (x-a)^{-1/2}$, we expect that the generated density goes at most as $n(x) \sim (x-a)^{2}$ and thus such densities allow for a unique inversion (up to a gauge).\\
In the case that we consider the Sturm-Liouville problem on the whole real axis  we are again in the case of limit-point endpoints, 
and there is again only one self-adjoint realization of the Sturm-Liouville operator.
%
%
%
Unfortunately it is not known under which conditions on the density the self-adjoint operator has a spectral gap around zero and thus allows for a (pseudo-) inverse of $-\partial_x (n \partial_x)$. However, from considerations similar to \cite{runge1984} one can presume that for most non-vanishing densities a unique inversion should be possible. 
\\

So far we have seen that the one-dimensional case is already quite involved. Turning back to (\ref{GeneralSturmLiouville}) in more dimensions we pose again the question of invertibility to solve for $v$ in a certain class of potentials, respecting the appropriate boundary conditions. By turning to a weak formulation of the problem, these immediately arise. To this end we adjoin a scalar field $u$, thought of being from the same class as the potential $v$, by means of the standard $L^2(\Omega)$ inner product.
\begin{equation}
-\langle u | \nabla \cdot (n \nabla v) \rangle = \langle u| \zeta \rangle
\end{equation}

The possible time dependence of all quantities is now suppressed, the equation is to hold at every instant. Now if the class of potentials is assumed to have zero or periodic boundary conditions partial integration defines a symmetric bilinear form $Q$ by
\begin{equation}\label{sl-weak}
Q(u,v) = \langle \nabla u| n \nabla v \rangle = \langle u| \zeta \rangle.
\end{equation}

We employed the specific boundary conditions to have a vanishing boundary term after partial integration. Note that in the case of periodic potentials also the density $n$ and $\nabla v$ have to obey this periodicity. Another option would have been to demand $n=0$ at the border like in the original Runge-Gross proof \cite{runge1984}.\\

The theorem of Lax-Milgram \cite{blanchard1992} now gives a direct and positive answer to the question of existence and uniqueness of a solution $v$ of (\ref{sl-weak}). Moreover the solution depends continuously on the given data $\zeta$ thus the inverse operator is bounded. For this theorem to hold, the bilinear form has to fulfil for all $u \in \mathcal{H}_Q$
\begin{equation}
\begin{array}{ll}
Q(u,u) \geq c_1 \|u\|^2_{\mathcal{H}_Q} &\quad \mbox{(coercivity)} \\
Q(u,v) \leq c_2 \|u\|_{\mathcal{H}_Q} \|v\|_{\mathcal{H}_Q} &\quad \mbox{(continuity)}
\end{array}
\end{equation}
for fixed constants $c_1,c_2>0$. The only open problem is then to choose an appropriate Hilbert space $\mathcal{H}_Q$ of potentials such that these condition on the bilinear form defined by the density $n$ are fulfilled. The simplest case is if for almost all $x \in \Omega$ it holds $c_1 \leq n(x) \leq c_2$. Then
\begin{equation}
\begin{array}{l}
Q(u,u) \geq c_1 \langle \nabla u|\nabla u \rangle = c_1 \|\nabla u\|^2 \quad \mbox{and} \\
Q(u,v) \leq c_2 \|\nabla u\| \|\nabla v\|
\end{array}
\end{equation}
and thus $Q$ is automatically continuous on the Sobolev space $H^1(\Omega)$ with the additional boundary conditions and norm $\|u\|_{1,2} = \|u\|_2 + \|\nabla u\|_2$. To show coercivity we need to rely on Poincar\'e's inequality \cite[6.30]{adams2003} $\|u\|_2 \leq c(\Omega) \|\nabla u\|_2$ that is true on bounded domains $\Omega$ (or domains that can be fully enclosed between two parallel hyperplanes) and again zero or periodic boundary conditions. Thus a unique solution of the Sturm-Liouville problem can be guaranteed to lie in the given Sobolev space.\\

The strategy for more general densities is similar but more involved, a detailed account is given in \cite{penz2011}. We construct a density-adapted weighted Sobolev space with norm $\|u\|_{1,2,n} = \|u\|_2 + \|\sqrt{n}\nabla u\|_2$ and one shows that for this space the bilinear form $Q$ is naturally continuous and coercive if the density fulfils $n^{-s} \in L^1(\Omega)$ with $s > \frac{d}{2}$, $d$ being the dimensionality of $\Omega$.\footnote{Note that the restriction $s > \frac{d}{2}$ was missing in the main theorem of \cite{penz2011}. Still everything is correct in the typical case $d=3$ where the smallest applicable integer value is $s=2$ as stated there.} Note that the restrictions to a bounded domain $\Omega$ (real boundedness because one does not only rely on Poincar\'e's inequality) and zero or periodic boundary conditions are still active. A further practical consequence is that the weighted Sobolev space is compactly embedded in $L^2(\Omega)$. Nevertheless the condition $n^{-s} \in L^1(\Omega)$ seems harsh, especially if the problem is not periodic, and will not be fulfilled by natural densities with zero boundary like a particle in a box.
\\

While the above considerations give us conditions on $n$ and $\zeta$ such that we can uniquely solve the Sturm-Liouville equation, we did not discuss whether the resulting potential is regular enough to generate a well-defined $q[v]$. However, in order to set up an iterative scheme as proposed in Sec.~\ref{sec:IterativeScheme}, we need to guarantee that $q[v]$ exists. Details about conditions on the initial state, interaction and potentials will be given in Sec.~\ref{sec:FixedPoint}. For the moment assume that for an appropriate $v_0$ we can determine via propagation a well-defined $q[v_0]$. If we are given a density $n$ for which the Sturm-Liouville operator allows for an inversion we can construct $v_1$ uniquely by
\begin{eqnarray}
 v_1(\bbr,t) = \left[-\nabla \cdot (n(\bbr,t) \nabla ) \right]^{-1} \left( q([v_0], \bbr,t) - \partial_t^2 n(\bbr,t) \right) .
\end{eqnarray}
In a next step we then can construct $v_2$ accordingly and we find from the boundedness of $[-\nabla \cdot (n(t) \nabla )]^{-1} $ by a time-dependent constant $D_t$ that
\begin{eqnarray}
  \| v_2(t) - v_1(t) \| \leq D_t \| q([v_1],t) - q([v_0],t) \|.
\end{eqnarray} 
We point out that this inequality does not necessarily need to use the same norms for $v$ and $q[v]$. Thus if we denote the norm on $v$ by $\|\cdot\|$ and the norm on $q[v]$ by $\|\cdot\|'$, we have alternatively
\begin{eqnarray}
\label{FirstInequality}
  \| v_2(t) - v_1(t) \| \leq D_t \| q([v_1],t) - q([v_0],t) \|',
\end{eqnarray}  
where the constant $D_t$ depends on the norms used.
This inequality will play an important role later in the fixed-point procedure. The possible norm $\| \cdot \|'$ that we can employ on the space of $q[v]$'s depends on
the situation. Under some conditions the $L^2$-norm was used \cite{ruggenthaler2011} but for one-dimensional periodic systems it was possible to use the $L^1$-norm
instead \cite{ruggenthaler2012}.

\subsection{Proof based on Taylor expansion} 

\label{sec:Uniqueness}

In the previous section we have discussed the Sturm-Liouville problem. With this we can approach the issue of proving the existence of a density-potential mapping, which is basically equivalent to showing that the mapping $v \mapsto n$ is injective for an initial state $\Psi_0$ and some well-defined set $\mathfrak{V}$. While to show the uniqueness of a fixed point of the approach presented in Sec.~\ref{sec:IterativeScheme} is quite involved, there is a very clever trick to establish the injectivity of $v \mapsto n$ without too much ado. The trick was first presented in the seminal work \cite{runge1984} and forms the basis of the \textit{Runge-Gross theorem} and its many variants for different physical situations \cite{xu1985, li1985, tong1986, ghosh1988, liu1989, wacker1994, rajagopal1994, vignale2004, burke2005, diventra2007, appel2009, yuen2009, yuen2010, ruggenthaler2011b, vanleeuwen2012, tempel2012, tokatly2013, mosquera2013, mosquera2014, huang2014, ruggenthaler2014}.
\\

We first assume an initial state that is spatially twice differentiable, i.e., $\Psi_0 \in D(\hat{T})$, and has an initial density that is zero at most at the boundaries. This is fulfilled, for instance, if we start with the ground-state of a quantum system. Then, for any two different potentials $v,v' \in \mathfrak{V}$ (the set defined for the mapping (\ref{nC2mapping})) the fundamental equation (\ref{TddftFundament}) gives rise to a well-defined Sturm-Liouville equation. Since at $t=0$ both systems have the same density, one can subtract both equations and finds
\begin{eqnarray}
\hspace{-1.5cm} - \nabla \cdot\left[ n(\bbr,0) \nabla \left( v(\bbr,0) - v'(\bbr,0) \right) \right] = \zeta([v], \bbr,0) - \zeta([v'], \bbr,0).
\end{eqnarray}
Now, if both potentials differ by more than a gauge constant, the left hand side of the equation is non-zero. If $\Psi_0$ is at least four-times differentiable, i.e., $\Psi_0 \in D(\hat{T}^2)$, and the interaction square-integrable\footnote{We point out that by partial integration all the derivatives of the interaction can be shifted to derivatives on the wave function in the definition of $q[v]$.} such that $\braket{\Psi_0}{\hat{q}(\bbr) \Psi_0}$ is well-defined and due to $q([v], \bbr,0) = q([v'], \bbr,0) = \braket{\Psi_0}{\hat{q}(\bbr)\Psi_0}$ we have
\begin{eqnarray}
\hspace{-1.5cm} - \nabla \cdot\left[ n(\bbr,0) \nabla \left( v(\bbr,0) - v'(\bbr,0) \right) \right] = \partial_t^2 n ([v'], \bbr,0) - \partial_t^2 n([v], \bbr,0) \neq 0.
\end{eqnarray}
As a consequence, the densities will be different infinitesimally later in time. Therefore, any two potentials that differ by more than a gauge at $t=0$ will generate different densities. Note that if one of the two potentials would not obey the conditions imposed by the invertibility of $-\nabla \cdot (n \nabla)$, then we could not make this conclusion. In this case the application of the Sturm-Liouville operator to the difference $v-v'$ would not be defined and the equation would not exist \footnote{Remember that if the potential is outside of the domain of the operator, then by construction $\| \nabla \cdot ( n \nabla v) \|' \rightarrow \infty$.}. Such a problem appears, for instance, in the example of \cite{xu1985}. There a difference function of the form $v(\bbr,0)-v'(\bbr,0) \sim \exp(|\bbr|) \notin L^2(\mathbb{R}^3)$ is used, for which the Sturm-Liouville operator is clearly not defined. The initial resolution of this alleged counter example \cite{gross1990}, i.e., to only allow for potentials that are generated by a finite charge distribution (and are thus square-integrable), is in clear accordance with the necessary conditions to ensure invertibility of the Sturm-Liouville equation. We therefore see how an exact mathematical formulation of the density-potential mapping helps to avoid erroneous conclusions.

The trick of Runge and Gross is now based on the consecutive application of the above result to the Taylor expansion in time of the different potentials. Obviously this is a first restriction, since not all possible potentials for the solution of the TDSE need to be infinitely-often differentiable with respect to time (with appropriate boundary conditions). If we assume that the wave functions are infinitely-often differentiable in time as well, then we formally find \cite{vanleeuwen1999, vanleeuwen2001, ullrich2012, marques2012}
\begin{eqnarray}
\label{ConsecutivePotential}
 \nabla \cdot &\left[ n^{(0)}(\bbr) \nabla v^{(k)}(\bbr)   \right] = n^{(k+2)}(\bbr) - q^{(k)}(\bbr)
\nonumber\\
&- \sum_{l=0}^{k-1} \left( \begin{array}{c} k \\ l \end{array}
\right) \nabla \cdot \left[ n^{(k-l)}(\bbr) \nabla v'^{(l)}(\bbr) \right],
\end{eqnarray}
where we write the $k$-th derivative in time at $t=0$ of the different functions by as 
\begin{equation*}
\left. \partial_t^k v(\bbr,t)\right|_{t=0} = v^{(k)}(\bbr) \quad \mbox{etc.}
\end{equation*}
Since the different $q^{(k)}$ and $n^{(k)}$ can be calculated from only knowing $v^{(l)}$ with $l< k$ (due to the Heisenberg equations for these operators), we find that if two potentials differ in order $k$ (while they are the same for $l< k$) we have
\begin{eqnarray}
\label{RungeGross}
\hspace{-1.5cm} - \nabla \cdot\left[ n^{(0)}(\bbr) \nabla \left( v^{(k)}(\bbr) - v'^{(k)}(\bbr) \right) \right] = n^{(k+2)} ([v'], \bbr) - n^{(k+2)}([v], \bbr) \neq 0.
\end{eqnarray}
Thus the two potentials will necessarily lead to different densities. We point out that in order for this conclusion to be made we need to ensure that all the functions in (\ref{ConsecutivePotential}) exist. If this would not be the case, we could not subtract (\ref{ConsecutivePotential}) for two different potentials and rearrange them in the form of (\ref{RungeGross}). A necessary condition for this to hold is that the initial state obeys $\Psi_0 \in D(\hat T^k)$ for all $k \in \mathbb{N}$, such that the kinetic-energy operator can be repeatedly applied from which infinite differentiability follows. Further, due to the repeated application of the Heisenberg equation in the definition of $q^{(k)}$, we demand that the potentials and interactions are infinitely-often differentiable with respect to space as well (although for the interaction $w$ it might be enough that it is in the Kato class of potentials, which includes the Coulomb potential). The constraint on the initial state excludes initial states with cusps in the density. Such initial states occur, for instance, if we solve for the ground state of the static Schr{\"o}dinger equation with external Coulomb potentials generated by point charge nuclei. 
However, if we soften the external potential by using finite nuclei \cite{andrae_2000} in an infinitely differentiable way the cusps in the ground state density will vanish
and by using the corresponding ground state as an initial state the Runge-Gross proof is valid  without changing essential physics (in fact finite nuclei are more realistic than point nuclei).
 
A final loophole we have to close is that there are still infinitely-often differentiable potentials which are different but all their derivatives are the same at one point, e.g., $v(\bbr,t)-v'(\bbr,t) = f(\bbr) \exp(-t^{-2})$ at $t=0$. So we cannot conclude for these type of potentials that they will necessarily lead to different densities. To overcome this problem we restrict to only those potentials that have a converging Taylor expansion for some finite time $t>0$, i.e., $v(\bbr,t) = \sum_k v^{(k)}(\bbr)t^k/k!$. 
Therefore, if we assume an appropriate initial state, the mapping from the set of Taylor-expandable potentials (with the appropriate boundary conditions) to densities is invertible\footnote{All these conditions are fulfilled, for instance, in the case of an infinitely-often differentiable initial state (with $n_0 \geq \epsilon > 0 $) on a torus, i.e., a periodic system. In this case the mapping from all spatially smooth and temporally Taylor-expandable potentials to their respective densities is invertible.}. This is the statement of the famous Runge-Gross theorem and provides the foundation of TDDFT.
\\

The Runge-Gross result enables us to perform a density-functionalization of time-dependent quantum mechanics. Instead of solving the full TDSE for a given initial state $\Psi_0$ and an external potential $v$ we can self-consistently solve the equivalent non-linear evolution equation
\begin{eqnarray}
\label{OrbitalFreeTDDFT1}
 \partial_t^2 n(\bbr,t) = \nabla\cdot\left[n(\bbr,t)\nabla v(\bbr,t) \right] + q([\Psi_0,n], \bbr,t),
\end{eqnarray}
with the initial conditions $n(\bbr,0) = \braket{\Psi_0}{\hat{n}(\bbr) \Psi_0}$ and $\partial_t n(\bbr,t)|_{t=0} = - \braket{\Psi_0}{\nabla\cdot\mathbf{\hat j}(\bbr) \Psi_0}$. However, there are two related problems we encounter at this point. Firstly, we do not know the set of $v$-representable densities we are allowed to vary over in search for the (existing) self-consistent solution. We would need a precise specification of the set of densities associated with the given set of potentials. The second problem is that while this equation would be in principle enough to do (orbital-free) TDDFT, similar to the minimization of the energy functional in ground-state DFT, it is extremely challenging to find good approximations to the operator $q[\Psi_0,n]$. Especially the kinetic part of the operator (see (\ref{qoperator})) is notoriously hard to approximate in terms of the density and initial state only. Therefore one usually wants to use an approximation to the $q$ operator based on an auxiliary quantum system.

Both problems are related to the question of $v$-representability. If we know the set of $v$-representable densities, then we can solve (\ref{OrbitalFreeTDDFT1}) by varying over this set, and if we can show that two different Hamiltonians have the same set of $v$-representable densities, then we can connect both systems by a KS construction. This allows us to approximate the (divergence of the) internal forces of an interacting problem by the $q_{\mathrm{s}}$ of a non-interacting problem. In this case, the so-called Hartree-exchange-correlation potential $v_{\mathrm{Hxc}}[\Psi_0,\Phi_0,n] = v_{\mathrm{s}}[\Phi_0,n] - v[\Psi_0, n]$ would be defined (assuming that both systems have initial states $\Psi_0$ and $\Phi_0$ with the same initial density and first time-derivative of the density) by
\begin{eqnarray}
\label{OrbitalFreeTDDFT}
 \nabla\cdot\left[n(\bbr,t)\nabla v_{\mathrm{Hxc}}(\bbr,t) \right] = q_{\mathrm{s}}([\Phi_0,n], \bbr,t) - q([\Psi_0,n], \bbr,t).
\end{eqnarray}
Now, can we learn something about the set of $v$-representable densities in the Runge-Gross approach? First of all, the densities generated under the above assumptions are infinitely often differentiable in time and space at $t=0$. From (\ref{ConsecutivePotential}) we even have the Taylor coefficients of the density in terms of $v$ and $\Psi_0$. However, it is not clear whether the series converges and thus that the density is analytic. It does definitely not hold in general, since one can find counter examples (see Sec.~\ref{sec:wavepacket} and \cite{holstein1972, yang2012}). On the other hand, using (\ref{ConsecutivePotential}) we can also construct the Taylor coefficients of the associated potential from a time-analytic density $n$ and the initial state $\Psi_0$. Again, we cannot guarantee that the resulting series converges, i.e., that the density is $v$-representable by a time-analytic potential. However, if we assume that (under very restrictive conditions) a Taylor-expandable density gives rise to a Taylor-expandable potential (and vice versa) we have a specific characterization of a set of $v$-representable densities. This forms the basis of the \textit{extended Runge-Gross approach} presented in \cite{vanleeuwen1999}, which also has been applied to different physical situations \cite{vignale2004, burke2005, diventra2007, appel2009, yuen2009, yuen2010, ruggenthaler2011b, vanleeuwen2012, tempel2012, tokatly2013, mosquera2013, mosquera2014, huang2014, ruggenthaler2014}. In this case we can vary over all time-analytic densities to solve (\ref{OrbitalFreeTDDFT}) and we can explicitly construct the Hartree-exchange-correlation potential by 
\begin{eqnarray}
\label{ConsecutiveHxc}
 \nabla \cdot &\left[ n^{(0)}(r) \nabla v^{(k)}_{\mathrm{Hxc}}(r) \right] = q_{\mathrm{s}}^{(k)}(r) - q^{(k)}(r)
\nonumber\\
&- \sum_{l=0}^{k-1} \left( \begin{array}{c} k \\ l \end{array}
\right) \nabla \cdot \left[ n^{(k-l)}(r) \nabla v^{(l)}_{\mathrm{Hxc}}(r) \right],
\end{eqnarray}
where $n^{(k)}$ is determined by $v^{(l)}$ and $v^{(l)}_{\mathrm{Hxc}}$ for $l < k$ \cite{ruggenthaler2009}.
In this way we can construct the Hartree-exchange-correlation potential from its Taylor expansion in time
\be
v_\mathrm{Hxc} (\bbr,t) = \sum_{k=0}^\infty \frac{1}{k!} v^{(k)}_{\mathrm{Hxc}} (\bbr) (t-t_0)^k.
\ee
This proof therefore gives a construction of the xc potential and forms the theoretical basis for the existence
of a KS system corresponding to a given interacting system. Note that the construction can also be carried out for connecting systems
with two different interactions \cite{vanleeuwen1999}. Again it will be important that at least these interactions are in the Kato class of potentials.

To conclude, the above Taylor-expansion approach of Runge and Gross \cite{runge1984} provides the existence of a density-potential mapping for infinitely-often differentiable initial states and potentials which are Taylor-expandable in time and infinitely-often differentiable in space. However, while it tells us that a density-potential mapping exists under these rather strong conditions, it does not provide us with a route to construct this mapping. For this we need to use the extended Runge-Gross approach \cite{vanleeuwen1999} which assumes Taylor-expandability in time of the density as well. The considerations in Sec.~\ref{sec:wavepacket}, however, show that this can only be true under further very restrictive conditions. Hence, to extend the density-potential mapping to more general situations and also have a constructive procedure at the same time we employ in the following the iterative approach introduced in Sec.~\ref{sec:IterativeScheme}.

\subsection{Proof based on a fixed-point scheme}
 
\label{sec:FixedPoint}

We have started our general discussion of the density-potential mapping in Sec.~\ref{sec:IterativeScheme} with a quite intuitive approach. Under the  assumption that the Sturm-Liouville operator $-\nabla \cdot (n \nabla)$ is invertible and that if two external potentials are close then their respective internal forces are close, we employed the fundamental equation (\ref{TddftFundament}) for a fixed time-dependent density to generate an iterative sequence of external potentials. These potentials were supposed to reproduce the prescribed density better and better via time propagation as we proceed in the iterative procedure. In Sec.~\ref{sec:SturmLiouville} we presented conditions such that the first part of the assumption, i.e., that the operator $-\nabla \cdot (n \nabla)$ is invertible, holds. These considerations were enough to show the existence of a density-potential mapping for Taylor-expandable potentials, i.e., to prove the classical Runge-Gross theorem. While the original work \cite{runge1984} did not have this iterative approach in mind, the result is directly applicable to this sequence of potentials. It tells us that if the sequence converges to a potential, then the fixed-point is unique within the set of Taylor-expandable potentials. Of course there could still be a second fixed-point outside of this set.

While these results provide the existence of a density-potential mapping, they do not give us precise conditions for a density to be $v$-representable. However, for a solution of (\ref{OrbitalFreeTDDFT1}) or the KS approach to TDDFT we need a characterization of these densities. Put differently, we would like to know under which conditions a fixed point of the iterative procedure introduced in Sec.~\ref{sec:IterativeScheme} exists. In this section we will discuss the existence (as well as an extension of the uniqueness results) of a fixed point with the help of a combination of (\ref{FirstInequality}) and an inequality, which makes the statement that two potentials which are close generate similar internal forces, precise.
\\

Let us start by first discussing under which conditions two external potentials generate similar internal forces. From our considerations in Sec.~\ref{sec:observables} we know that in order to have a well-defined divergence of the internal-force density $q[v]$ the corresponding wave function $\Psi[v]$ needs to be at least four-times spatially differentiable. Consequently we impose this condition on the initial state $\Psi_0$ and only allow for external potentials $v$ that stabilise this property when propagating with the associated evolution operator $\hat{U}([v],t,0)$. We denoted this set of potentials by $\mathfrak{V}$. While from the discussion preceding (\ref{nC2mapping}) we presume that potentials which are twice differentiable in space and which have some regularity with respect to the time variable will fulfil this condition, we only know it explicitly for infinitely-often differentiable potentials in space and time \cite{delort2010}. For simplicity we impose the same conditions on the interaction, although since we can shift the derivatives with respect to $w$ in the definition of $q[v]$ to derivatives of the wave function, we expect that any square-integrable interaction (for instance the Coulomb interaction) should be possible as well. Under these conditions the different potentials $v$ are bounded functions\footnote{By different Sobolev-embedding theorems this can also be shown for weakly twice-differentiable potentials. The important case of $\Omega = \mathbb{R}^3$ is discussed, for instance, on p.~316 in \cite{blanchard2003}.} on $\Omega$ and give rise to well-defined (divergence of) internal-force densities $q[v]$.

Now, for bounded potentials the wave functions $\Psi[v]$ are Fr\'echet differentiable with respect to the external potential $v$ (see \ref{app:existence} or \cite{penz2014}). This implies that changing the potential slightly by $\Delta v$ will affect the evolution of the initial state even less. Further, in this case the fundamental theorem of calculus for Banach spaces holds. Provided the wave functions are regular enough we can apply the fundamental theorem also to $q[v]$ by
\begin{eqnarray}
\hspace{-1.5cm} q([v_2], \bbr,t) - q([v_1], \bbr,t) = \int_0^{1} \diff \lambda \int_{0}^{t} \diff t' \int_{\Omega} \diff \bbr' \,\frac{\delta q([v_{\lambda}], \bbr,t)}{\delta v_{\lambda}(\bbr',t')} \Delta v_{12}(\bbr',t'),
\end{eqnarray}
where $\Delta v_{12} = v_2 -v_1$ and $v_{\lambda} = v_1 + \lambda \Delta v_{12}$. The linear-response kernel can be expressed explicitly in the Heisenberg picture of quantum mechanics by a commutator of the form \cite{ruggenthaler2011,ruggenthaler2012}
\begin{equation}
\frac{\delta q([v],\bbr,t)}{\delta v(\bbr',t')} = -\imagi \braket{\Psi_0}{[\hat{q}(\bbr,t), \hat{n}(\bbr',t')] \Psi_0} . 
\end{equation}
With an estimate for the linear-response kernel on the connecting line from $v_1$ to $v_2$, we have
\begin{eqnarray}
\label{SecondInequality}
 \| q([v_2],t) - q([v_1], t)\|' \leq C_t[v_2,v_1] \int_{0}^{t} \diff t' \|v_{2}(t')-v_{1}(t')\|,
\end{eqnarray}
with the constant $C_t[v_2,v_1]<\infty$ depending on $v_2$, $v_1$ and time, where possible norms $\| \cdot \|'$ where discussed below equation (\ref{FirstInequality}). The constant  $C_t[v_2,v_1]$ stands for the bound of the linear-response kernel along the straight line from $v_1$ to $v_2$. Inequality (\ref{SecondInequality}) shows that if two potentials are close in $\|\cdot\|$ norm  over time, then their respective internal forces are close in $\|\cdot\|'$ norm as well. 
\\

In the following we combine (\ref{FirstInequality}) and (\ref{SecondInequality}) to provide (under certain conditions) uniqueness and existence of a fixed point of the iterative sequence introduced in Sec.~\ref{sec:IterativeScheme}. We first employ a well-known trick from the theory of differential and integral equations \cite{walter1990,bielecki1956,light1990}. By using the so-called \textit{Bielecki norm} for an $0 \leq \alpha < \infty$ that changes the balance of the norm towards earlier times \footnote{Note that a different but equivalent norm was used in \cite{ruggenthaler2011}}
\begin{eqnarray*}
 \|v\|_{\alpha} = \sup_{t \in [0,T]} \left( \e^{-\alpha t} \| v(t) \| \right),
\end{eqnarray*}
we can deduce from (\ref{SecondInequality}) with the help of
\begin{eqnarray*}
\hspace{-0cm}\int_{0}^T \diff t \, \|  v(t) \| &= \int_{0}^T \diff t \, \e^{-\alpha t}  \e^{\alpha t}\|  v(t) \| 
\\
&\leq \| v \|_{\alpha} \int_{0}^T \diff t \, \e^{\alpha t} \leq \| v \|_{\alpha} \frac{\e^{\alpha T}}{\alpha} 
\end{eqnarray*}
that 
\begin{eqnarray}
\label{qvInequality}
 \| q[v_1] - q[v_2]\|'_{\alpha} \leq \frac{C[v_2,v_1]}{\alpha}   \|v_{2}-v_{1}\|_{\alpha},
\end{eqnarray}
where we denote $C[v_2,v_1]  := \sup_{t \in [0,T]}C_t[v_2,v_1] < \infty$. Also (\ref{FirstInequality}) can be rewritten in terms of the Bielecki norm as
\begin{eqnarray}
\label{vqInequality}
 \|v_{2}-v_{1}\|_{\alpha} \leq D \| q[v_1] - q[v_0]\|'_{\alpha} ,
\end{eqnarray}
with $D  := \sup_{t \in [0,T]}D_t < \infty$. These inequalities imply for the iteration $\mathcal{F}[v_k] = v_{k+1}$ that
\begin{eqnarray}
\label{Inequality}
 \hspace{-1.5cm} \|v_{k+1}-v_{k}\|_{\alpha} \leq D \| q[v_k] - q[v_{k-1}]\|'_{\alpha} \leq \frac{ D C[v_{k},v_{k-1}] }{\alpha} \|v_{k}-v_{k-1}\|_{\alpha}  
\end{eqnarray}
We note that all $\alpha$ norms are equivalent since
\begin{eqnarray*}
 \e^{-\alpha T } \|v \|_{0} \leq \|v\|_{\alpha} \leq \|v\|_{0},
\end{eqnarray*}
and thus they all define the same Banach space of potentials.

Assume now that we would have two fixed points of our iterative procedure $\mathcal{F}[v]=v$ and $\mathcal{F}[u]=u$ which differ by more than just a gauge. Then by taking $\alpha = 2 D C[v,u]$ and (\ref{Inequality}) we have
\begin{eqnarray}
 \|v-u\|_{\alpha}  \leq \frac{1}{2} \|v - u\|_{\alpha}. 
\end{eqnarray}
However, this can only be true if $\|u-v\|_{\alpha} = 0$ and therefore $u=v$. As a consequence the iterative procedure $\mathcal{F}[v_k] = v_{k+1}$ has at most one fixed-point in $\mathfrak{V}$, which is equivalent to state that the mapping $v \mapsto n$ is invertible on this domain.

Finally we consider the \textit{existence} of a fixed point. While before we already knew that the density $n[v]$ is associated with a potential, since we wanted to ensure the uniqueness of a $v$-representable density, now we do not \textit{a priori} know that the respective density is generated from a TDSE. In the case of a KS scheme, for instance, we want to ensure that a potential that was generated by a different TDSE, i.e., with a different interaction and initial state, can also be reproduced by a non-interacting system. The most basic conditions we need to impose on the density is that in accordance to Sec.~\ref{sec:IterativeScheme} the density obeys the initial conditions
\begin{eqnarray}
\label{InitialCondition1}
 n(\bbr,0) &= \braket{\Psi_0}{\hat{n}(\bbr) \, \Psi_0},
\\
\label{InitialCondition2}
 \left. \partial_t n(\bbr,t)\right|_{t=0} &= - \braket{\Psi_0}{\nabla \cdot \mathbf{\hat{j}}(\bbr) \, \Psi_0}.
\end{eqnarray}
Further, the initial state and the density have to be chosen such that the Sturm-Liouville operator $-\nabla\cdot (n \nabla)$ is self-adjoint and invertible (up to a gauge) for the whole time interval $[0,T]$ (see Sec.~\ref{sec:SturmLiouville} for details). A further minimal condition is that $n \in L^1(\Omega)$ is at least four-times differentiable in space (with the appropriate boundary conditions) and two-times differentiable in time, since it should correspond to a density that is generated from a $\Psi[v]$ which obeys the fundamental (\ref{TddftFundament}).

Now, to set up a well-defined iterative sequence, we need to ensure that every $v_k \in \mathfrak{V}$ gives rise to an appropriate $q[v_k]$ from which we can construct $\mathcal{F}: v_k \mapsto v_{k+1}$ as defined in (\ref{Fmapping})
and which is again in the same set. For instance, if the initial state, $v_k$ and $n$ are infinitely-often differentiable (in space and time) then $q[v_k]$ is infinitely-often differentiable \cite{delort2010}, and consequently $v_{k+1}$ has the same properties. For a more general set $\mathfrak{V}$ we do not know whether $q[v_k]$ is regular enough to guarantee $v_{k+1} \in \mathfrak{V}$. However, from (\ref{vfunctional}) we see that a $v_{k}$ twice differentiable in space and continuous in time generates a temporally continuous $\partial_t^2 n[v_k]$ and $q[v_k]$, which would in turn generate a $v_{k+1}$ having the same properties. Also the proof strategy of \cite[Th. X.70]{reed1975} makes it plausible that potentials which are twice differentiable in space and continuous in time generate a continuous $q[v_k]$ which would be enough to guarantee that $v_{k+1}$ obeys the same conditions. These considerations make it plausible that the fixed-point approach applies also under the above less restrictive conditions.

Under the above assumptions we can apply (\ref{Inequality}) to the iterative sequence. The constants $C[v_{k},v_{k-1}]$ are all estimates on the lines connecting two successive iteration points $v_{k-1}$ and $v_{k}$. If we assume $\alpha$ big enough such that $\alpha \geq 2 D C[v_{j},v_{j-1}]$ for all iterations $j \leq k$ we discover that its distance from the starting point $v_0$ is indeed limited by the $\alpha$-norm of the very first step.
\begin{eqnarray}\label{iteration-geom-series}
\|v_{k+1}-v_0\|_\alpha &\leq \sum_{j=0}^k \|v_{j+1}-v_j\|_\alpha \leq \sum_{j=0}^k \left(\frac{1}{2}\right)^{j} \|v_1-v_0\|_\alpha \nonumber\\
&= 2 \left(1-\left(\frac{1}{2}\right)^{k+1}\right) \|\mathcal{F}[v_0]-v_0\|_\alpha
\end{eqnarray}
The next constant $C[v_{k+1},v_{k}]$ can be chosen as a maximal estimate of the linear-response kernel over a set of potentials big enough such as to include the connecting line from $v_k$ to $v_{k+1}$. From (\ref{iteration-geom-series}) we see that both $v_k$ and $v_{k+1}$ will surely be included in the convex set $B(v_0) =\{ v \in \mathfrak{V} \,|\, \|v-v_0\|_\alpha \leq 2 \|\mathcal{F}[v_0]-v_0\|_\alpha \}$ and this is true for all further steps if we assume that there exists a constant
\begin{eqnarray}
 C = \sup\{C[v,u] \,|\, v,u \in B(v_0)\} < \infty
\end{eqnarray}
and take $\alpha = 2DC$. Note that this assumption is still needed here because such a supremum does not necessarily exist on a ball in a infinitely dimensional Banach space such as the space of potentials. Then $v_k$ is a Cauchy sequence in the Banach space equipped with the Bielecki norm and therefore the resulting fixed point $v=\lim_{k \rightarrow \infty} v_k$ solves the local-force equation
\begin{eqnarray}
- \nabla \cdot \left[n(\bbr,t) \nabla v(\bbr,t)\right] = q([v], \bbr,t) -\partial_t^2 n(\bbr,t)
\end{eqnarray}
and by the argument of (\ref{RhoEquation}) generates the prescribed $n$ by propagation of the initial state $\Psi_0$.
\\

In summary, for the possible $v$-representability of a density via the fixed-point procedure we need that the initial state $\Psi_0$ is at least four-times differentiable and that the prescribed density obeys the initial conditions of (\ref{InitialCondition1}) and (\ref{InitialCondition2}). Further, the associated Sturm-Liouville operator has to be invertible, i.e., $n$ is strictly positive except possibly at the boundary, $\partial_t^2 n$ is at least continuous in time and $\nabla^4 n$ is integrable, since it should be associated with a $\Psi[v]$ which is four-times differentiable as well. Although finally we adopted stronger assumptions on the initial states and densities (infinite differentiability) to make the iteration well-defined it seems likely that the iterative procedure can be extended to these less restrictive assumptions.

\section{The density-potential mapping in lattice systems} 
 
\label{sec:Lattice}

In the course of this review we have seen that a lot of subtleties arise due to the standard formulation of quantum mechanics in terms of unbounded operators on an infinite-dimensional Hilbert space. The same subtleties are of course also found in the density-potential mappings. In order to avoid the resulting problems of the standard formulation, we can consider an approximate treatment of quantum mechanics from the start. There are two different (but closely related) approaches to do so: We can either stay within an infinite-dimensional Hilbert space of quantum states but restrict ourselves to bounded operators only (which allows to consider the $C^*$-algebra of bounded and self-adjoint operators \cite{blanchard2003,mackey1963}), or we can consider quantum mechanics on a finite dimensional Hilbert space (which makes all operators automatically bounded). Both approximation strategies are quite general, and their explicit realizations depend strongly on the actual physical situation we want to model \cite{stefanucci2013}. Most clearly we see the relation between both approximation strategies if we consider discretized models, so-called \textit{lattice systems}. For instance, we can divide $\mathbb{R}^3$ into (infinitely many) small boxes and approximate the square-integrable functions by their mean-value in these boxes. By then approximating the kinetic-energy operator by a finite-difference expression in terms of these mean values we have a bounded operator on the appropriate (infinite-dimensional) sequence space of square-summable functions (see for instance \cite{chayes1985}). Also the interaction energy can be expressed as a bounded operator in this sequence space and the external-energy operator is given in terms of a (bounded) multiplication operator (also called the on-site potential). By then restricting further to a finite part of the lattice we have a finite-dimensional approximation. The resulting TDSE (without loss of generality we again disregard the spin-degrees of freedom) for $N$ particles on the lattice with $M$ sites for every particle reads \cite{li2008, baer2008, tokatly2011}
\begin{eqnarray}
\label{discreteHamiltonian}
\hspace{-2.5cm} \imagi \partial_t \Psi(\vec{z}_1,...,\vec{z}_N,t) = & - \sum_{n=1}^{N} \sum_{\vec{y}_{n}} T(\vec{z}_n, \vec{y}_{n}) \Psi(...,\vec{y}_n,...,t)  + \sum_{n=1}^{N} v(\vec{z}_n,t) \Psi(\vec{z}_1,...,\vec{z}_N,t) \nonumber
\\
& + \sum_{i>j}^{N} w(|\vec{z}_i -\vec{z_j}|) \Psi(\vec{z}_1,...,\vec{z}_N,t), 
\end{eqnarray}
where the hopping rate obeys $T(\vec{z}_n, \vec{y}_{n}) = T(\vec{y}_{n}, \vec{z}_n)$ and $T(\vec{z}_n, \vec{z}_{n}) = 0$ can be assumed (since it amounts to a constant shift in the on-site potential, i.e., a gauge transformation). The existence of a unique solution $\Psi \in \mathcal{C}^{1}([0,T], \mathcal{H}_d)$ on the discretized Hilbert space $\mathcal{H}_d$ for every $\Psi_0 \in \mathcal{H}_{d}$ can then be based on the Picard-Lindel\"of theorem of ordinary-differential equations (even for the case of infinitely many sites). The Picard-Lindel\"of theorem (which itself is an application of the Banach fixed-point theorem) implies that an iterative mapping
\begin{eqnarray}
\label{DisceretePsiIteration}
\Psi_{k+1}(t) = \Psi_0 + \int_{0}^{t} \diff s \, f(s, \Psi_k(s))  
\end{eqnarray}
converges to a unique solution provided that $f$ is continuous in its first argument and there is a constant $L < \infty$ such that
\begin{eqnarray}
\label{Lipshitz}
 \| f(s,\Psi_1) - f(s, \Psi_2) \| \leq L \| \Psi_1 - \Psi_2\|
\end{eqnarray}
for all $s \in [0,T]$ and $\Psi_1, \Psi_2 \in \mathcal{H}_d$. Since in our case $f(s, \Psi(s)) = \hat{H}_d(s) \Psi(s)$ and $\hat{H}_d$ is the bounded discretized Hamiltonian of (\ref{discreteHamiltonian}), the Lipschitz constant $L$ is the highest eigenvalue of the Hamiltonian. Equivalently one could use similar ideas as presented in the Sec.~(\ref{sec:existence}) to show the existence of a unique evolution operator.

Thus, for the set of continuous and bounded on-site potentials $\mathcal{V}_{d}$ we have a mapping
\begin{eqnarray}
 \Psi:& \mathcal{V}_d &\rightarrow \mathcal{C}^{1}([0,T], \mathcal{H}_d)
\\
& v &\; {\buildrel\rm \Psi_0 \over \mapsto }  \; \Psi[v]  \nonumber
\end{eqnarray}
for every initial state $\Psi_0$. Accordingly we can then define mappings for observable quantities like the \textit{(on-site) density} 
\begin{eqnarray}
 n(\vec{z},t) = N \sum_{\vec{z}_2,...,\vec{z}_N} |\Psi(\vec{z}, \vec{z}_2,...,\vec{z}_N,t)|^2,
\end{eqnarray}
the \textit{density matrix}
\begin{eqnarray}
 \rho(\vec{z}, \vec{y},t) = N \sum_{\vec{z}_2,...,\vec{z}_N} \Psi^{*}(\vec{z}, \vec{z}_2,...,\vec{z}_N,t) \Psi(\vec{y}, \vec{z}_2,...,\vec{z}_N,t)
\end{eqnarray}
or the \textit{link current}
\begin{eqnarray}
J(\vec{z}, \vec{y},t) =  2 \,\Im\left\{ T(\vec{z}, \vec{y}) \rho(\vec{z}, \vec{y},t) \right\}
\end{eqnarray}
provided we have anti-symmetrized the wave function appropriately. The question whether we can establish a density-potential mapping for this discretized problem will again be based on an equation that connects $n$ and $v$ explicitly. In analogy to the continuum we first determine the discretized version of the \textit{continuity equation} \cite{kurth2011,tokatly2011,farzanehpour2012}
\begin{eqnarray}
\label{LatticeContinuity}
 \partial_t n(\vec{z},t) = - \sum_{\vec{y}} J(\vec{z}, \vec{y},t).
\end{eqnarray}
By construction this equation is well-defined for every $\Psi[v]$ (which is not automatically true for the continuum case since there we are dealing with unbounded operators). By then calculating the equation of motion for the link-current $J$ and then combining it with the continuity equation we find the \textit{fundamental equation of lattice TDDFT}
\begin{eqnarray}
\label{LatticeFundament}
 \partial_t^2 n(\vec{z},t) = \sum_{\vec{y}} k(\vec{z}, \vec{y},t) v(\vec{y},t) + q(\vec{z},t),
\end{eqnarray}
where
\begin{eqnarray}
 k(\vec{z}, \vec{y}, t) = 2 \Re \left\{ T(\vec{z}, \vec{y}) \rho(\vec{z}, \vec{y},t) - \delta_{\vec{z}, \vec{y}} \sum_{\vec{x}} T(\vec{z}, \vec{x}) \rho(\vec{z}, \vec{x},t) \right\}
\end{eqnarray}
and 
\begin{eqnarray}
\hspace{-1.5cm}q(\vec{z},t) = -2 \Re &\left\{ \sum_{\vec{x}, \vec{y}}  T(\vec{x}, \vec{y}) \left\{ \left[   \rho_2(\vec{z}, \vec{y}, \vec{x},t) \left(w(|\vec{z}-\vec{x}|) - w(|\vec{y}-\vec{x}|)  \right) \right]  \right. \right.
\\
& \left. \left. \hspace{0.5cm} + \left[T(\vec{x}, \vec{y}) \rho(\vec{z}, \vec{x},t) - T(\vec{x}, \vec{z}) \rho(\vec{y}, \vec{x},t) \right] \right\} \right\}. \nonumber
\end{eqnarray}
Here we have defined the two-particle density matrix by
\begin{eqnarray}
\hspace{-1.5cm}\rho_2(\vec{x}, \vec{y}, \vec{z},t) = N (N-1) \sum_{\vec{z}_3,...\vec{z}_N}  \Psi^{*}(\vec{x}, \vec{y}, \vec{z}_3,...,\vec{z}_N,t) \Psi(\vec{z}, \vec{y},\vec{z}_3,...,\vec{z}_N,t).
\end{eqnarray}
Equation (\ref{LatticeFundament}) is the lattice equivalent to (\ref{TddftFundament}) of the continuum version. We point out, that a sufficient condition for this equation to hold is to guarantee that $\Psi \in \mathcal{C}^2([0,T], \mathcal{H}_d)$. This can be shown to be true (at least) if $v$ is one-times continuously differentiable in time, which restricts the set of allowed on-site potentials. Can we now define (maybe without all the mathematical trouble we face in the continuum case) a lattice density-potential mapping based on this equation?

Before we answer this question, we point out that it was realized by different authors \cite{baer2008, li2008} that there is an evident lack of $v$-representability for certain time-dependent densities on a lattice. These issues can be demonstrated most easily for a simple two-site model. The Hamiltonian for this problem is
\begin{eqnarray}
 \hat{H}(t) = - T_\kin \hat{\sigma}_{x} + v(t) \hat{\sigma}_{z} 
 = \left( \begin{array}{cc} v(t) & -T_\kin \\ -T_\kin & - v(t) \end{array} \right)
\end{eqnarray}
where $\{\hat{\sigma}_{x}, \hat{\sigma}_{y}, \hat{\sigma}_{x}\}$ are the Pauli matrices, $T_\kin>0$ a hopping parameter, and the Hilbert space is simply $\mathcal{H}_{d} = \mathbb{C}^2$. The function $2 v(t)$ is actually the potential difference between site 1 and 2, thus we have effectively fixed a gauge. The conjugate observable to the potential difference is the density (difference) operator $\hat\sigma_z(t)$ and therefore we would like to investigate whether we can define a $\sigma_z \mapsto v$ mapping with $\sigma_z(t) = 
\braket{\Psi(t)}{\hat\sigma_z \Psi(t)}$. We do so by considering the equivalent equations to (\ref{LatticeContinuity}) and (\ref{LatticeFundament}).
\begin{eqnarray}
 \partial_t \sigma_z(t) &= -2 T_\kin \sigma_y(t)
\\
 \partial_t^2 \sigma_z(t) &= -4 T_\kin^2 \sigma_z(t) - 4 T v(t) \sigma_x(t)
\end{eqnarray}
Provided $v \in \mathcal{C}^{1}([0,T], \mathbb{R})$ any solution of the above two-site TDSE obeys these two equations. Obviously, if we want to realize a density-potential mapping for a fixed initial state $\Psi_0$, we need to obey the initial conditions that are implied by these two equations. However, these equations also immediately give other conditions on $v$-representable densities. Firstly, from the lattice continuity equation we see that only those densities are possible to achieve, which obey a maximality condition of the form
\begin{eqnarray}
\label{Maximality}
 |\partial_t \sigma_z(t)| \leq 2 T_\kin.
\end{eqnarray}
The hopping parameter $T_\kin$ restricts the maximal change of density at each site. Further, if we want to follow the Runge-Gross idea by restricting to the set of Taylor-expandable $v$, we need to ensure that $\sigma_x(0) \neq 0$ for the initial state (which excludes for instance states of the form $(1,0)$ or $(0,1)$). In the general case of (\ref{LatticeFundament}) this condition is equivalent to guarantee the invertibility of the $M \times M$ symmetric matrix $\hat{K}(0)$ with entries $k(\vec{x}, \vec{y}, 0)$ (in a space perpendicular to the constant function, i.e., the gauge freedom of the on-site potential). To circumvent some of these issues, in \cite{kurth2011,tokatly2011} time-dependent link-current density-functional theory was developed. However, by restricting to initial states that obey the invertibility condition on $\hat{K}(0)$ a TDDFT on the lattice can be formulated \cite{farzanehpour2012}.

We start by choosing an initial state $\Psi_0$ for which $\hat{K}(0)$ is invertible (up to a constant function), e.g., the ground state of a connected lattice \cite{farzanehpour2012}. In the above two-site example this amounts to demand $\sigma_x(0) \neq 0$. The initial state is then part of an open set $\mathcal{B} \subset \mathcal{H}_d$ of states that allow an inversion of $\hat{K}$. Any solution of the TDSE $\Psi(t)$ starting from $\Psi_0 \in \mathcal{B}$ stays within this set for a finite amount of time, say up to $0<t^{*}$ (the exact time depends on the external field applied). Now, for every continuously-differentiable on-site potential $v$ the solution to the lattice TDSE $\Psi[v]$ solves (\ref{LatticeFundament}), which can be rewritten in matrix form as
\begin{eqnarray}
\label{LatticeFundamentMatrix}
 \hat{K}([\Psi],t) V(t) = S([\Psi],t).
\end{eqnarray}
Here $S([\Psi],t) = \partial_t^2 N([\Psi],t)+Q([\Psi],t)$ are the according $M$-vectors. The two-site version of this equation reads simply as
\begin{eqnarray}
\label{2sFundamental}
 4 T_\kin \sigma_{x}([\Psi],t)v(t) = - \partial_t^{2}\sigma_z([\Psi],t) - 4 T_\kin^2 \sigma_z([\Psi],t).
\end{eqnarray}
This equation can be used as a functional equation for $v(\vec{z},t)$ for a fixed density $n(\vec{z},t)$, i.e.,
\begin{eqnarray}
\label{PsiSelfconsistent}
 \hat{K}([\Psi],t) V(t) = S([n,\Psi],t).
\end{eqnarray}
In a next step we could now express the dependence on $\Psi$ in terms of $v$ and the initial state $\Psi_0$ (which is merely a functional variable change), which would lead to the lattice equivalent of (\ref{iterate}). A self-consistent solution of the resulting functional equation merely in terms of $v$ would amount to the lattice version of the fixed-point approach of Sec.~\ref{sec:FixedPoint}. We will investigate this scheme at the end of this section.

Before we do so, we follow \cite{farzanehpour2012} and use (\ref{PsiSelfconsistent}) to set up a non-linear TDSE (at least until $t^*$) by expressing 
\begin{eqnarray}
 V([n, \Psi],t) = \hat{K}^{-1}([\Psi],t) S([n,\Psi],t),
\end{eqnarray}
and employing the resulting $v[n, \Psi]$ as the (non-linear) on-site potential.
\begin{eqnarray}
 \label{discreteNonlinear}
\hspace{-2.5cm} \imagi \partial_t \Psi(\vec{z}_1,...,\vec{z}_N,t) = & - \sum_{n=1}^{N} \sum_{\vec{y}_{n}} T_{\vec{z}_n, \vec{y}_{n}} \Psi(...,\vec{y}_n,...,t)  + \sum_{n=1}^{N} v([n, \Psi],\vec{z}_n,t) \Psi(\vec{z}_1,...,\vec{z}_N,t) \nonumber
\\
& + \sum_{i>j}^{N} w(|\vec{z}_i -\vec{z_j}|) \Psi(\vec{z}_1,...,\vec{z}_N,t)
\end{eqnarray}
This non-linear TDSE is the lattice equivalent to the non-linear TDSE introduced in \cite{tokatly2007,tokatly2009,maitra2010,tokatly2011b} and also discussed in Sec.~\ref{sec:IterativeScheme}. Now, remember the Picard-Lindel\"of theorem that we used to guarantee existence and uniqueness of solutions to the original TDSE (and thus the potential-density mapping). We only needed to show that the right-hand side of the TDSE obeys (\ref{Lipshitz}). Provided that we are within $\mathcal{B} \subset \mathcal{H}_d$, the inversion of $\hat{K}(t)$ perpendicular to the constant on-site potential is possible and thus the right hand-side of (\ref{discreteNonlinear}) is bounded. This local boundedness is enough for a local version of the Picard-Lindel\"of theorem, which guarantees the existence and uniqueness of a solution to (\ref{discreteNonlinear}). The only restriction is that $[0,T]$ is supposed to be such that the iterative sequence defined by (\ref{DisceretePsiIteration}) does not leave $\mathcal{B} \subset \mathcal{H}_d$, thus $T$ depends on the chosen $n$.

What happens if we hit this $v$-representability boundary $\partial \mathcal{B}$ at some time $0<t^*$ is most easily seen in the two-site model. The resulting non-linear TDSE is due to (\ref{2sFundamental}) given by
\begin{eqnarray}
\label{2sNLTDSE}
\imagi \partial_t \Psi(t) = -T_\kin \hat{\sigma}_x \Psi(t) - \frac{\partial_t^2 \sigma_z(t) + 4 T_\kin^2 \sigma_z([\Psi],t) }{4T_\kin \sigma_x([\Psi],t)} \hat{\sigma}_{z} \Psi(t). 
\end{eqnarray}
Obviously, if we leave $\mathcal{B}$ the non-linearity becomes infinite since $\sigma_x([\Psi],t^*) = 0 $. This excludes the application of the Picard-Lindel\"of theorem across the boundary to guarantee a unique solution. However, in this model-system we can analyse the behaviour at the $v$-representability boundary in more detail. By a construction similar to the construction of the explicit density-potential mapping in the continuum of Sec.~\ref{sec:Example}, where we expressed the complex wave function in terms of its polar representation, we can deduce an explicit form of the two-site density-potential mapping in terms of $\sigma_z(t)$ by \cite{farzanehpour2012}
\begin{eqnarray}
\label{explicit2s}
 v^{\pm}([n],t) = \pm \frac{\partial_t^2 n_1(t) + 2 T_\kin^2 \sigma_z(t)}{2 \sqrt{4T_\kin^2 n_1(t) n_2(t) - \left(\partial_t n_1(t)\right)^2}},
\end{eqnarray}
where $n_1(t)= (\sigma_z(t)+1)/2$ and $n_2(t) = (1- \sigma_z(t))/2$ are the densities at each site. The $\pm$ stands for two different choices of the initial-states phase function. Whenever $4T_\kin^2 n_1(t) n_2(t) - (\partial_t n_1(t))^2 \rightarrow 0$ (which is equivalent to the maximality condition of (\ref{Maximality})) we are at the boundary between the $v^{+}$ and $v^{-}$ realization of the density-potential mapping \cite{farzanehpour2012}. It would therefore be highly desirable to somehow extend the mapping across this boundary uniquely.

At this point we can profit from the fixed-point approach introduced in the previous section. Firstly we point out, that a \textit{lattice fixed-point approach} amounts to consider an iterative mapping based on
\begin{eqnarray}
\label{LatticeFixedPoint}
  \hat{K}([\Psi_0,v],t) V(t) = \partial_t^2 N(t) + Q([\Psi_0,v],t),
\end{eqnarray}
which is the variable-transformed version of (\ref{PsiSelfconsistent}). Obviously we need to impose the same assumption about invertibility of $\hat{K}(t)$ as before. We therefore can set up an iterative sequence
\begin{eqnarray}
 V_{k+1}(t) = \hat{K}^{-1}([\Psi_0,v_k],t) \left( \partial_t^2 N(t) + Q([\Psi_0,v_k],t) \right),
\end{eqnarray}
which is similar to the iterative sequence defined by (\ref{DisceretePsiIteration}), where every new $\Psi_k(t)$ defines by construction a new $v_k$. Thus, while the above non-linear TDSE approach investigates the existence and uniqueness of a fixed point in terms of $\Psi_k(t)$ the lattice version of the fixed point approach of Sec.~\ref{sec:FixedPoint} considers convergence in terms of $v_k$. If we choose the starting points of these two iterations such that $\Psi_0(t)$ leads to $v[n, \Psi_0] = v_0$ then both iterations are exactly the same. Therefore, both iterations lead to the same unique solution $v[\Psi_0,n]$ for some (appropriately short) time interval $[0,T]$. But how do we guarantee that the limiting potential really reproduces the prescribed $n$? In particular, the Picard-Lindel\"of theorem implies convergence also for densities which are unphysical, e.g., for an $n$ with wrong initial conditions. To guarantee that the solution $\Psi[\Psi_0,n]$ and therefore also $v[\Psi_0,n]$ actually reproduces the prescribed density $n$ we need an equation similar to (\ref{RhoEquation}) in the continuum case. This equation can be found in the lattice situation provided that $v(\vec{z},t)$ is continuously-differentiable such that we can subtract (\ref{LatticeFundamentMatrix}) from (\ref{LatticeFixedPoint}) and end up with
\begin{eqnarray}
 \partial_t^2 \rho(\vec{z},t) = 0,
\end{eqnarray}
where $\rho(\vec{z},t) = n([\Psi_0, v], \vec{z},t)- n(\vec{z},t)$. If we then assume that the given $n$ and the propagated $n[\Psi_0,v]$ have the same initial conditions, i.e., $\rho(\vec{z},0) = \partial_t \rho(\vec{z},t)|_{t=0} = 0$, then necessarily $n = n[\Psi_0,v]$. 

As a consequence we can conclude, that if the unique solution $\Psi[\Psi_0,n]$ produces a continuously-differentiable $v[\Psi_0,n]$ by (\ref{DisceretePsiIteration}), it reproduces the prescribed $n$. This implies, that if a potential $v[\Psi_0,n]$ crosses a $v$-representability boundary $\partial \mathcal{B}$, then only a continuously-differentiable extension across this boundary is a sensible choice. Thus, in the situation of the two-site problem one needs to choose the extension of $v^{\pm}([n],t)$ according to this condition whenever $\sigma_x \rightarrow 0$. Whether this can always be done is not clear. Nevertheless, for any initial state that has an invertible $\hat{K}(0)$ (for instance the ground-states of connected lattices \cite{farzanehpour2012}) and an appropriately short time interval we have a well-defined \textit{lattice density-potential mapping} $n \mapsto v$.

\section{Numerical realization of the density-potential mapping}

\label{sec:numerics}
 
%
%

The main ideas and concepts of density-potential mappings were developed alongside those of DFT and TDDFT. However, while DFT and TDDFT without the basic density-potential mappings would not be possible, these mappings, on the other hand, do not rely on DFT methods. Actually, they can be put to use also in other areas of physics and chemistry. For instance, in the context of quantum control theory they augment already existing techniques \cite{gross1993, zhu1998, zhu2003, serban2005} to steer the dynamics of quantum systems \cite{nielsen2013, nielsen2014}. In this section we discuss the density-potential mappings from the point of view of quantum control theory and consider their numerical construction.
\\

We usually employ quantum mechanics to \textit{predict} the behaviour of a microscopic system. Such a system is modelled by the initial state $\Psi_0$ and the external potential $v$ which acts on it (see for instance (\ref{MolecularPotential})). Then we can, in principle, determine $\Psi[v]$ from the resulting TDSE and calculate all physical observables $O(t) = \braket{\Psi(t)}{\hat{O} \Psi(t)}$. However, we can also use quantum mechanics to \textit{control} the behaviour of a microscopic system \cite{brif2010}, i.e., we can try to determine a $v$ that forces the wave function to show a previously specified behaviour. This can be done in two different ways: 

The first approach optimizes a chosen observable $\hat{O}$ in time by varying over all possible wave functions starting from a given $\Psi_0$ in the functional \cite{serban2005}
\begin{eqnarray}
\label{OptimizeFunctional}
 J[\Psi] = \frac{1}{T} \int_0^{T} \diff t \; \braket{\Psi(t)}{\hat{O} \Psi(t)}.
\end{eqnarray}
For instance, we could start with the ground state of a quantum system and then try to maximize the occupation of the first excited state, i.e., $\hat{O} = \ket{\Psi_1} \bra{\Psi_1}$. This approach is called \textit{quantum optimal-control theory} \cite{serban2005, brif2010}. In our search for an optimal wave function we cannot allow all $\Psi \in \mathcal{C}^0([0,T], \mathcal{H})$, since not all of them are connected via the solution of the TDSE to the initial state. To enforce that we only vary over wave functions that are solutions to the TDSE we can either work with a Langrange multiplier on the wave functions, which in this case is a wave function itself\footnote{To be precise, the Lagragian multiplier will be part of the dual space of the Banach space of wave functions. Therefore in the case of optimal control theory with a Lagrangian multiplier it might be more convenient that one considers the wave functions as part of the self-dual Hilbert space of time and space $L^2([0,T],\mathcal{H})$, such that both, the TDSE wave function and its Lagragian multiplier, are within the same function space.} and obeys a time-reversed TDSE \cite{serban2005}, or one can employ that all wave functions are labelled uniquely by their respective external potentials and vary with respect to the potentials \cite{castro2011}. This formulation of optimal control theory \cite{castro2011} employs the mapping $v \mapsto \Psi$ introduced in Sec.~\ref{sec:existence} directly. Either way, the resulting control equations to determine an optimal external potential $v$ are numerically extremely demanding, since they usually imply hundreds if not thousands of global iterative solutions of the full TDSE \cite{serban2005}. 

The second approach avoids these numerically expensive global iterations by (instead of optimizing) prescribing the physical observable $O(t)$ at every time and then try to find a $\Psi$ that reproduces this as a solution of the TDSE with some potential $v$. However, not every path $O(t)$ can be reproduced by a TDSE with a local potential only. For instance, the control of the non-local observable $\hat{O} = \ket{\Psi_1} \bra{\Psi_1}$ would need a non-local potential that can project $\Psi(t)$ directly onto the first initial state. On the other hand, if we would like to find an external potential $v$ that generates a prescribed dipole moment, we will find multiple solutions. In abstract terms, the mapping $v \mapsto O$ is usually not invertible and we cannot guarantee that a prescribed path $O(t)$ is $v$-representable\footnote{For $v$-representability both, the control objective and the control field need to have the same degrees of freedom, i.e, the size of their respective sets need to be the same.}. If we, however, restrict ourselves to controlling the density, these issues can be avoided. This has to do with the fact, that the Runge-Gross theorem discussed in Sec.~\ref{sec:Uniqueness} guarantees the invertibility of the mapping $v \mapsto n$, and thus almost all densities which are consistent with the initial state are $v$-representable. In principle we can then apply the basic procedure of this so-called \textit{local control theory}\footnote{We point out that controllability of the prescribed observable is a major challenge in standard local-control schemes. However, local-control schemes that only enforce a monotonic increase/decrease of an observable can overcome most of these problems.} \cite{gross1993, zhu1998, zhu2003}, where we discretize time and determine an appropriate $\hat{H}(0)$ from solving
\begin{eqnarray}
 -\nabla \cdot (n \nabla v) =   q[\Psi_0] - \partial_t^2 n,
\end{eqnarray}
where $n$ and $\partial_t^2n$ are prescribed and $q[\Psi_0]$ is determined from the initial state. This Hamiltonian is then used to make an Euler time step $\Psi(t_{1}) = \Psi_0 - \imagi \Delta t \hat{H}(0) \Psi_0$\footnote{Remembering all the intricacies of the TDSE discussed in Sec.~\ref{sec:wavepacket} one should be a little suspicious about approximating the evolution of the wave function by repeatedly applying the Hamiltonian. We will discuss this issue a little later.} which gives a state that has the prescribed density (without multiple global iterations as was the case in optimal control theory). While the Euler method works well in practice for simple control objectives and small enough time steps $\Delta t$, it is numerically very inefficient and can fail in practice due to round-off errors, which it does in the density case. This is due to the fact that the density (at a given point) may change by orders of magnitude, so that we have to be very precise to stay correct. If we do not, an extremely strong artificial potential is needed to compensate for the error in the next time step, which makes the resulting algorithm unstable. How we can stabilize this local-control algorithm by employing the iterative scheme introduced in Sec.~\ref{sec:IterativeScheme}, is discussed below.

Finally we can combine local and optimal control theory, provided that we have a way to efficiently calculate $\Psi$ for a given $n$ (besides the local-control approach, in certain physical situations one can also use other schemes \cite{lein2005, verdozzi2008,tempel2012,ramsden2012}). Due to the Runge-Gross theorem we can label the wave functions in terms of their respective densities and thus we can vary with respect to the density in the optimal-control functional of (\ref{OptimizeFunctional}), i.e.,
\begin{eqnarray}
 J[n] = \frac{1}{T} \int_0^{T} \diff t \; \braket{\Psi([n],t)}{\hat{O} \Psi([n],t)}.
\end{eqnarray}
While the numerical cost to optimize the functional is still the same as in the standard optimal-control approach, the restriction of the search space becomes simpler than in the previous cases. Based on one's physical intuition one can set up a basis of possible densities and then optimize with respect to these (finitely many) degrees of freedom \cite{nielsen2014}. Further, in the rare cases that observables can be expressed (approximately) in terms of the density (see also Sec.~\ref{sec:Example}), one can determine first an optimal density from (\ref{OptimizeFunctional}) and afterwards use the density-potential mapping to calculate the respective $v$. 
\\

Now, to stabilize the above local-control algorithm, we start by defining again the iterative procedure
\begin{eqnarray}
\label{FullIteration}
-\nabla \cdot (n \nabla v_{k+1}) =   q[v_k] - \partial_t^2 n  .
\end{eqnarray}
We choose a density $n$ (strictly positive on $\Omega$) that satisfies the initial conditions of (\ref{InitialCondition1}) and (\ref{InitialCondition2}) on some time interval $[0,T]$. Then we propagate the initial state with $v_k$ and then calculating $q[v_k]$ as defined in Sec.~\ref{sec:observables}. A first important detail is the fact that the iterative procedure can be performed not only on the whole time interval $[0,T]$, on which we prescribe the density, but also successively in every subinterval of length $\Delta t$ (where we use the converged potential of the previous subinterval to determine the new initial state via propagation of the previous initial state).

This partitioning of the time interval is also needed to numerically perform the time propagation \cite{leforestier1991}. If we take the time intervals small enough we can approximate the exact evolution by a time-stepping procedure with time-constant Hamiltonians (see Sec.~\ref{sec:existence}). In principle it is then possible to determine the eigenfunctions of the Hamiltonian and calculate the propagator $\exp(-\imagi \hat{H} \Delta t)$ for the time step $\Delta t$. Since the Hamiltonian (and hence the eigenfunctions) change in time, this procedure has to be repeated for all the successive time steps. In practice such a procedure is impossible and hence one usually adopts the approximation
\begin{eqnarray}
\label{TaylorApproximation}
\e^{-\imagi \hat{H} \Delta t} \simeq \sum_{k=0}^{K} \frac{\left(-\imagi \hat{H}  \Delta t \right)^{k}}{k!},
\end{eqnarray}
for some arbitrary $K \in \mathbb{N}$. While this approximation is well-defined if we have discretized our Hamiltonian (to represent the problem on our computer), analytically the Taylor approximation (\ref{TaylorApproximation}) is usually not well-defined as has been discussed in detail in Sec.~\ref{sec:wavepacket}. Therefore we have to be specifically careful that this approximation does not violate any analytical constraints. Otherwise, as can be seen from the examples of Sec.~\ref{sec:wavepacket}, the discretised TDSE is not a proper representation of the continuum TDSE. For instance, we need to make sure that the wave function obeys the boundary conditions at all times, i.e., it stays within the domain of the Hamiltonian. Hence for periodic boundary conditions on $[0,L]$ (the multi-dimensional case is straightforward) the wave function always has to stay periodic as dictated by the eigenfunctions of the self-adjoint domain $\exp(\imagi 2 \pi k x/L )$, and for the zero-boundary case the wave function has to stay odd across the boundaries and periodic on the double domain $[-L,L]$ due to $\sin(\pi k x/L)$ \cite{nielsen2014}. Thus any external potential (as well as interaction) that is applied via the Taylor approximation (\ref{TaylorApproximation}) to the wave function needs to keep this symmetry. Consequently we restrict in the following to strictly periodic potentials in the case of a periodic quantum system and in the case of zero boundary conditions we restrict to potentials that are periodic on the double domain and even across the boundaries $0$ and $L$ \cite{nielsen2014}. Hence we see, how the rather abstract mathematical concepts discussed in Sec.~\ref{sec:wavepacket} and \ref{sec:existence} become important in practice when numerically solving the TDSE.

While these conditions come from the propagation of the wave functions and \textit{not} from the iterative procedure, they are also necessary to make the iterations well-defined. This is the case, since they make sure that we can uniquely (up to a gauge) invert the Sturm-Liouville operator $-\nabla \cdot (n \nabla)$ and find a new $v_{k+1}$ that again allows the above time-stepping strategy (see also Sec.~\ref{sec:FixedPoint}). To make this more precise, we first consider the case of a periodic system. For strictly positive densities we can impose periodic boundary conditions on the Sturm-Liouville operator and invert it uniquely (see Sec.~\ref{sec:SturmLiouville} for details). Therefore we can propagate with the new $v_{k+1}$ without violating the boundary conditions and perform the next iteration step. In the case of zero boundary conditions on the wave functions, we can invert the Sturm-Liouville operator provided the density does not go faster than $x^{2}$ to zero (see Sec.~\ref{sec:SturmLiouville} for details). If we have made sure that the wave function stays odd across $x=0$ then $\Psi(x) \sim x$ near the boundary and hence we can invert the problem. Since $q$ and $n$ are even across $x=0$ the iterated potential $v_{k+1}$ is even about the boundaries too. The resulting potential is therefore periodic on the double domain and even about the inner and the outer boundaries as required.

In a next step we get rid of the somewhat complicated term $q[v_k]$ by employing (\ref{TddftFundament}), which leads to
\begin{eqnarray}
\label{qSubstitution}
 q[v_k] = \partial_t^2 n[v_k] - \nabla \cdot (n[v_k] \nabla v_k ).
\end{eqnarray}
Since we only make one time step, the target density $n$ and the iterated density $n[v_k]$ are usually (even in the first iteration step, where we just take the converged potential of the previous subinterval) very close, such that we can approximately write (use (\ref{qSubstitution}) in (\ref{FullIteration}))
\begin{eqnarray} 
- \nabla \cdot \left[ n \nabla \left(v_{k+1} - v_{k} \right) \right] \simeq
\partial_t^2 (n[v_k] - n).
\end{eqnarray}
Here a further important detail to stabilize the numerical procedure has to be taken into account. The above update formula changes the potential according to how strongly the iterated density differs from the target density. In terms of the iterated wave function this means that we directly control the modulus of the wave function, however its phase we only control indirectly. This indirect control only holds in the exact case, where the phase of a wave function corresponds to its current $\mathbf{j}$ (see (\ref{Current}) for the definition), and is determined by the modulus via the continuity equation (\ref{ContinuityEq}). However, since we will employ a discretization of $\Omega$ and will have numerical errors in our algorithm, the continuity equation will not be fulfilled exactly. To avoid that the density is almost exact in the iteration (and numerically we would interpret the potential as converged) while the current is still far off we make the control of the phase explicit by 
\begin{eqnarray} 
\label{UpdateExact}
\hspace{-1.5cm} -\nabla \cdot \left[ n \nabla \left( v_{k+1} - v_k \right) \right] \simeq
(1-\mu) \partial_t^2 \left( n[v_k] - n \right) - \mu \partial_t \left( \nabla \cdot \mathbf{j}[v_k] + \partial_t n \right),
\end{eqnarray}
where $\mu$ is a non-zero parameter at our disposal. The last term on the right-hand side of (\ref{UpdateExact}) therefore measures how well we obey the continuity equation.

Now we make the time-stepping explicit. We use instead of the simple on-point Hamiltonian in the original local-control algorithm a mid-point Hamiltonian and thus a mid-point potential $\bar{v}_k(\bbr,t_{i+1}) = v_k(\bbr,(t_{i+1} + t_i)/2)$ to make a time-step $\Delta t= t_{i+1}-t_{i}$ \cite{nielsen2013, nielsen2014}. For the finite-difference approximation to the time-derivatives in (\ref{UpdateExact}) we only employ times prior to the current time. Since for all prior times by assumptions we have converged to the exact density and current we end up with
\begin{eqnarray} 
\label{UpdateNumerical}
\hspace{-1.5cm}- \nabla \cdot \left[ \bar{n} \nabla \left( \bar{v}_{k+1} - \bar{v}_k \right) \right] \Delta t^2  = A \left( n[v_k] - n \right) - B \Delta t \left( \nabla \cdot \mathbf{j}[v_k] + \partial_t n \right) ,
\end{eqnarray}
where $\bar{n}$ is the mid-point density and $A$ and $B$ are constants depending on the discretization scheme of the time-derivatives and the $\mu$ of (\ref{UpdateExact}), which effectively leaves the choice of their values at our disposal. Usual choices are $A$ and $B$ between 0.5 and 1 \cite{nielsen2013, nielsen2014}.

Finally we (equidistantly) discretize $\Omega$ and use a (usually seven-point) finite-difference approximation for the spatial derivatives in the Hamiltonian as well as in the above update formula. Due to the fact that we can treat the zero-boundary case in the same manner as a periodic system with double the period, we have the same accuracy in the derivatives at every point of our grid. This allows us to determine the wave functions and the respective iterated potentials to a high accuracy everywhere (also at the boundaries). Especially when the density changes by orders of magnitude (at a point) we need to be very precise, since errors are compensated by the iterative algorithm in the next time step with a large and unphysical potential, which can lead to instabilities \cite{nielsen2013, nielsen2014}. By smoothing the iterated potentials, unphysical and extreme differences can be suppressed. For the actual numerical propagation of the wave function the Lanzcos method is employed, since it is the most versatile and numerically cheap approach. The inversion of the discretized Sturm-Liouville operator can be performed with relaxation methods or more efficiently with multi-grid methods \cite{press2007}.
\\

The above local-control algorithm is stable and can treat rapidly changing (by orders of magnitude) densities. It is independent of the dimension of the problem, the number of particles and the initial state as well as the interaction. In practice, the main obstacle to perform this local-control scheme is to store (and then propagate) the interacting many-body wave function, since one quickly runs out of computer memory. The individual update-cycles (\ref{UpdateNumerical}) are usually converged (with respect to the change of the potential difference $\bar{v}_{k+1} - \bar{v}_k$) within a few iterations \cite{nielsen2013, nielsen2014}. However, for non-interacting problems, where the propagation can be performed efficiently, a combination with TDDFT approximations to the xc potential allows to find also approximate potentials for large interacting systems \cite{nielsen2014}. 
\\

\begin{figure} [H]
\centering
\includegraphics[width=9.0cm]{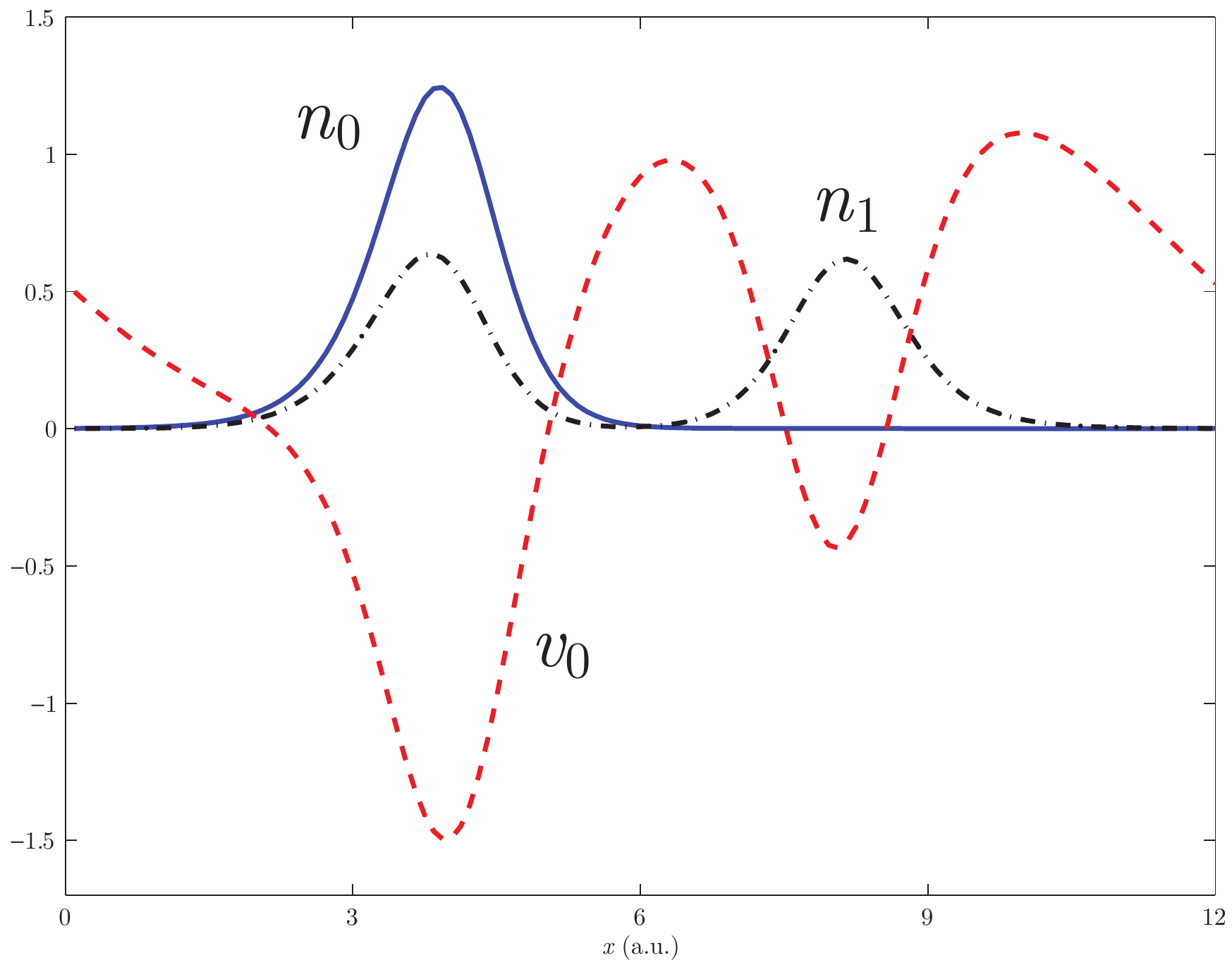}
\caption{The static potential $v_0$ (dashed red line) and the ground-state density $n_0$ (solid blue line) and the first excited-state density $n_1$ (dashed dotted black line) of the interacting system.}
\label{fig:DensitiesPotential}
\end{figure}

In the following we present a few illustrative examples of the density-potential mapping constructed via the above algorithm. We first consider the system of the two interacting particles already introduced in Sec.~\ref{sec:Example}. While before we were interested in the rigid charge transfer, we now want to do a little more and force that the density of the interacting two-particle system changes from the ground-state density $n_0$ (of the potential $v_0$ given in (\ref{StaticExample})) over time to the density $n_1$ of the first excited state (displayed in Fig.~\ref{fig:DensitiesPotential}). The first excited state is a charge-transfer state, where roughly half of the density is at the left site and the other half is on the right site.

\begin{figure} [H]
\centering
\includegraphics[width=9cm]{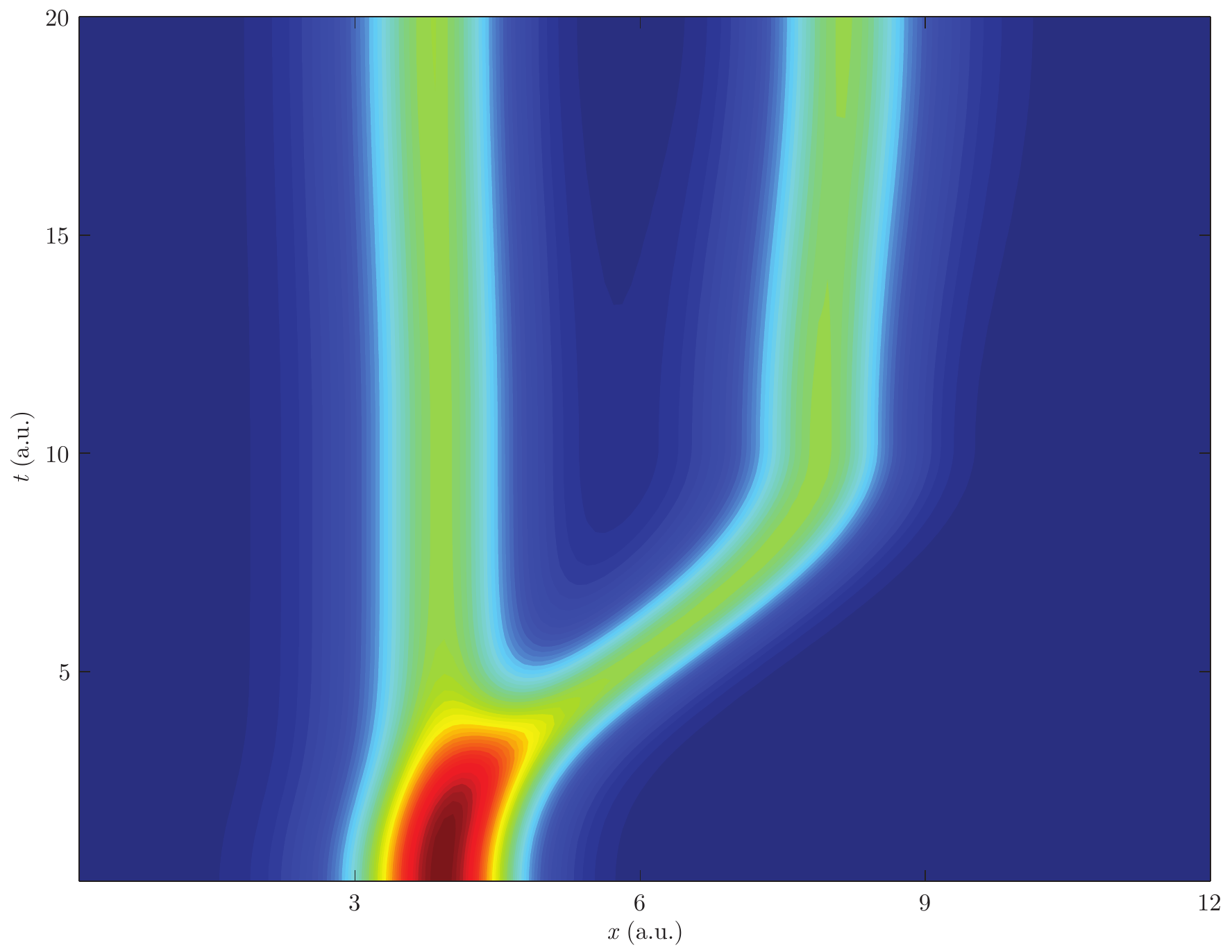}
\caption{The density profile $n$ used in this example. First the density of the initial state is split into two halves and then it changes to the form of the excited-state density $n_1$.}
\label{fig:Density2}
\end{figure}

We first split the charge in a similar manner as has been done in the example of Sec.~\ref{sec:Example} and then slowly change the density to the one of the charge-transfer state (see Fig.~\ref{fig:Density2}).

\begin{figure} [H]
\centering
\includegraphics[width=9.0cm]{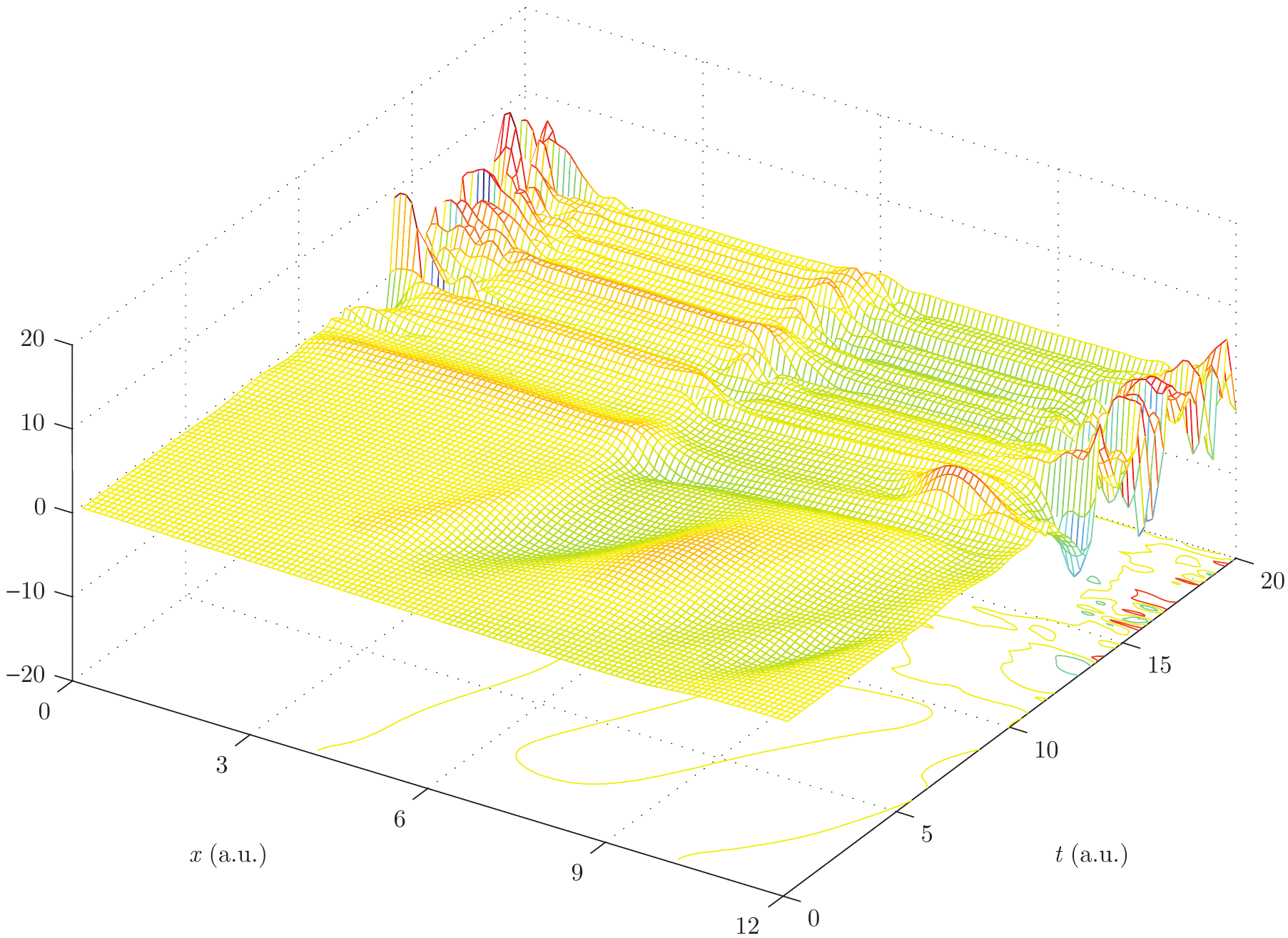}
\caption{The external potential $v_{\mathrm{ext}} = v[\Psi_0,n] - v_0$ that generates the prescribed density profile in the interacting system.}
\label{fig:v_ext}
\end{figure}

The resulting external time-dependent potential (where we subtracted the static potential, i.e., $v_{\mathrm{ext}} = v[\Psi_0,n] - v_0$) that does this in the interacting system is shown in Fig.~\ref{fig:v_ext}. The peaks at the boundaries are not numerical artefacts. This complex structure is the same for different spatial and temporal grids and indeed is needed to enforce that the rapidly moving wave function does not change its form too fast.   

\begin{figure} [H]
\centering
\includegraphics[width=9.0cm]{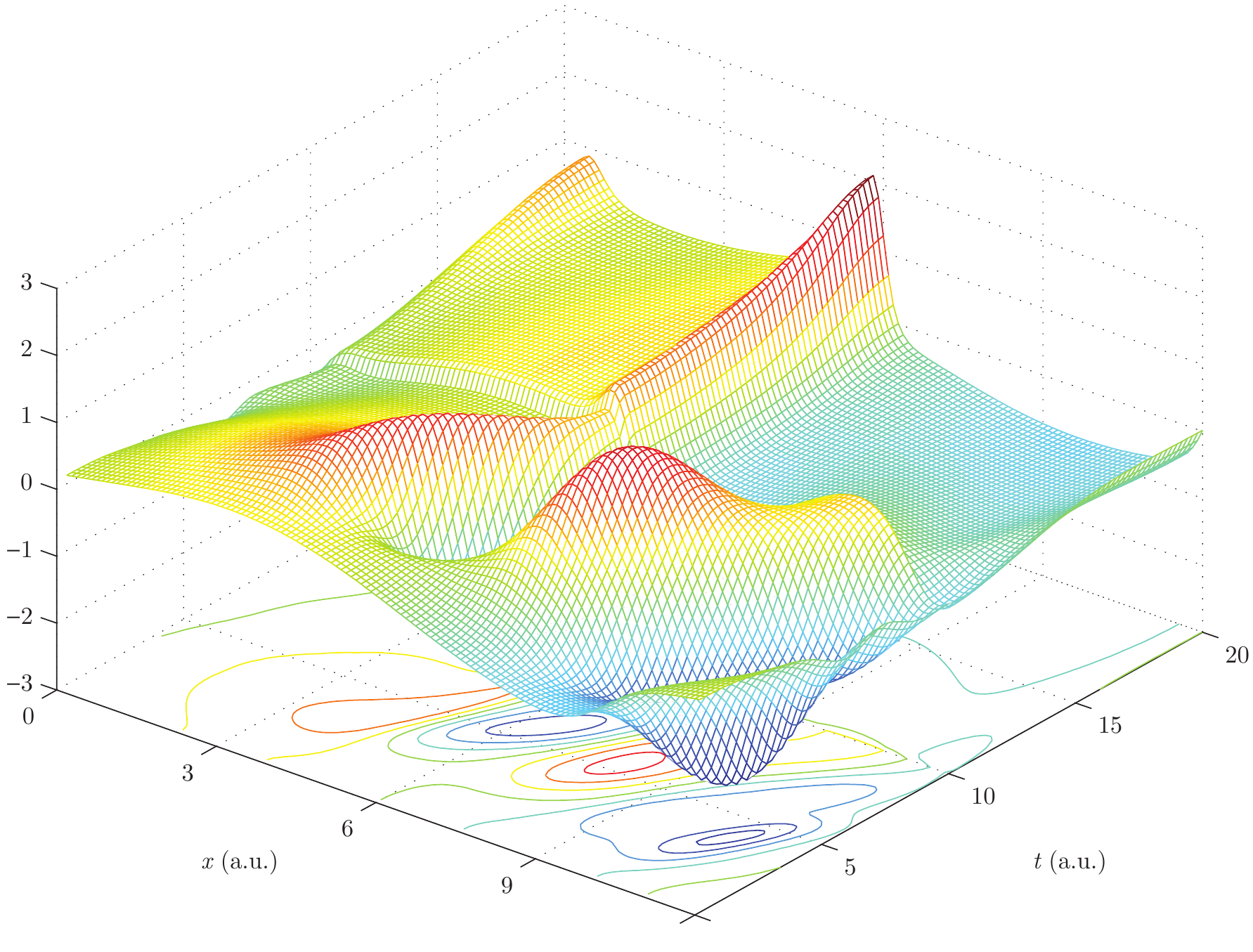}
\caption{The external potential $v_{\mathrm{ext}} = v[0,n] - v_0$ that generates the prescribed density profile in the non-interacting system.}
\label{fig:v_KSext}
\end{figure}

If we instead look at a non-interacting system with the same density profile (starting from the KS ground state with a single orbital of the form  $\varphi_0[0,n_0]$ as given in Sec.~\ref{sec:Example}.) we find that the external potential (again we have subtracted the static potential $v_0$) that enforces this charge-transfer behaviour does not have these extreme features (see Fig.~\ref{fig:v_KSext}).

These examples demonstrate the capability of this numerical realization of the $n \mapsto v$ mapping and also show that the basic ideas of the density-potential mapping can be used in practice also beyond TDDFT and the KS construction.

\section{Extensions to vector potentials and photons} 

\label{sec:extension}

The ideas of the density-potential mappings based on the Runge-Gross approach and its extension by van Leeuwen (see Sec.~\ref{sec:Uniqueness}) have been applied to a lot of different physical situations beyond the ones described by the standard Hamiltonian of (\ref{TDSEHamiltonian}), e.g., to superconducting systems \cite{wacker1994} and to open quantum systems \cite{diventra2007, appel2009, yuen2009, yuen2010} (for a list of references see Sec.~\ref{sec:Uniqueness}). In this section, we consider the application of these ideas to quantum systems driven by an external vector potential as an important example, which gives rise to (vector-)potential-current mappings. These mappings form the basis of time-dependent current-density-functional theory (TDCDFT) \cite{xu1985, vignale2004}. This density-functional approach can be easily extended to also include the interaction with photons \cite{rajagopal1994, ruggenthaler2011b, tokatly2013, ruggenthaler2014}. 
\\

In all our previous considerations we have neglected two important physical facts: relativity and photons. In principle we should use a kinetic-energy operator that is consistent with special relativity and the charged particles should interact via photons. The standard approach that takes these two requirements into account (and implies spin as well as the existence of positrons) is quantum electrodynamics (QED)\cite{ryder2006, engel2011}. While the predictions based on QED are extremely accurate, the theory has severe mathematical problems which express themselves, for instance, in divergent perturbative expressions \cite{greiner1996}. Despite these issues, the QED Hamiltonian (or equivalently its Lagrangian) is usually employed as a starting point to derive different approximate quantum theories which describe the properties of charged particles and photons in certain limits. For instance, if we assume that the energies of the charged particles are small compared to $mc^2$, then a non-relativistic treatment of the charged particles based on the Pauli-Fierz Hamiltonian \cite{pauli1938} is justified. If we further assume magnetic fields to be negligible and take the photons in Coulomb gauge (the polarization is restricted to the two transversal degrees of freedom and thus $\nabla \cdot \hat{\mathbf{A}} = 0$) \cite{greiner1996} we end up\footnote{Without further restrictions it cannot be guaranteed that the Pauli-Fierz Hamiltonian is well-defined \cite{hiroshima2002, hidaka2014}. Nevertheless, if we restrict our considerations to a box with periodic boundary conditions and introduce a highest allowed photon frequency, then the resulting Hamiltonian is self-adjoint. The following arguments, however, do not depend on this procedure and we therefore neglect these subtleties \cite{tokatly2013, ruggenthaler2014}.} with a Hamiltonian \cite{stefanucci2013, ruggenthaler2014} 
\begin{eqnarray}
\label{FullVectorPotentialHamiltonian}
\hspace{-1.5cm} \hat{H}(t) = & \hat{T} + \hat{H}_{\mathrm{EM}} + \hat{W} - \int \diff \bbr \, \hat{\mathbf{J}}(\bbr,t) \cdot \hat{\mathbf{A}}(\bbr) 
\\
&- \int \diff \bbr \, \left(  \hat{\mathbf{J}}(\bbr,t) \cdot \mathbf{a}_{\mathrm{ext}}(\bbr,t) +  \mathbf{j}_{\mathrm{ext}}(\bbr,t) \cdot \hat{\mathbf{A}}(\bbr)   \right) \nonumber
\\
&+ \int \diff \bbr \, \hat{n}(\bbr) \left( v_{\mathrm{tot}}(\bbr,t) - \frac{1}{2} \hat{\mathbf{A}}^2_{\mathrm{tot}}(\bbr,t) \right) \nonumber
\end{eqnarray}
that describes electrons subject to an external vector and scalar potential, i.e, $\mathbf{a}_{\mathrm{ext}}$ and $v$ respectively, and photons subject to an external charge current $\mathbf{j}_{\mathrm{ext}}$ and charge density $n_{\mathrm{ext}}$. The interaction between photons and electrons is described with the terms $\hat{W} + \int \hat{\mathbf{J}}\cdot \hat{\mathbf{A}}$, where
\begin{eqnarray}
\label{ChargeCurrent}
 \hat{\mathbf{J}}(\bbr,t) = \hat{\mathbf{j}}(\bbr) - \hat{n}(\bbr) \hat{\mathbf{A}}_{\mathrm{tot}}(\bbr,t)
\end{eqnarray}
is the charge current, the total vector potential is $\hat{\mathbf{A}}_{\mathrm{tot}} = \hat{\mathbf{A}} + \mathbf{a}_{\ext}$, and the vector-potential operator is given by
\begin{eqnarray}
 \hat{\mathbf{A}}(\bbr) = \int \frac{\diff \mathbf{k}}{\sqrt{2 \mathbf{k}^2 (2 \pi)^3}} \sum_{\lambda=1}^{2} \mathbf{\epsilon}(\mathbf{k}, \lambda) \left[ \hat{a}_{\mathbf{k} \lambda} \e^{ \imagi \mathbf{k} \cdot \bbr} +  \hat{a}^{\dagger}_{\mathbf{k} \lambda} \e^{- \imagi \mathbf{k} \cdot \bbr} \right],
\end{eqnarray}
where $ \mathbf{\epsilon}(\mathbf{k}, \lambda)$ are the two transversal polarization vectors \cite{greiner1996, ruggenthaler2014} and the creation and annihilation operators obey $[\hat{a}_{\mathbf{k}' \lambda'} , \hat{a}^{\dagger}_{\mathbf{k} \lambda}] = \delta(\mathbf{k}-\mathbf{k}') \delta_{\lambda \lambda'}$. The total scalar potential is given by $v_{\mathrm{tot}} = v + \int \diff \bbr' \, n_{\mathrm{ext}}(\mathbf{r'},t)/ 4 \pi |\bbr - \bbr'| $ and the free-photon energy operator is $\hat{H}_{\mathrm{EM}} = \int \diff \mathbf{k} \, \mathbf{k}^2 \, \sum_{\lambda=1}^{2} \hat{a}^{\dagger}_{\mathbf{k} \lambda} \hat{a}_{\mathbf{k} \lambda}$. The initial state $\Psi_0$ in this case is a combined initial state of electronic and photonic degrees of freedom.
\\

Now, we first consider situations where the coupling term $\int \hat{\mathbf{J}}\cdot \hat{\mathbf{A}}$ between the photons and the electrons is negligible. This is the case, if the initial state is separable into a purely electronic and photonic state and the transversal part of the internal charge current
\begin{eqnarray}
 \hat{\mathbf{J}}_{\perp}(\bbr,t) = \nabla \times \int \diff \bbr' \frac{\nabla \times \hat{\mathbf{J}}(\bbr',t)}{4 \pi |\bbr' - \bbr|}
\end{eqnarray}
can be discarded. In this situation the main contribution comes from the longitudinal internal charge current
\begin{eqnarray}
  \hat{\mathbf{J}}_{\parallel}(\bbr,t) = - \nabla \int \diff \bbr' \frac{\nabla \cdot \hat{\mathbf{J}}(\bbr',t)}{4 \pi |\bbr' - \bbr|}
\end{eqnarray}
for which by partial integration the coupling term is zero due to the Coulomb-gauge condition (the interaction between the electrons by the longitudinal charge current is taken into account fully by the Coulomb term $\hat{W}$). We therefore can decouple the electronic and the photonic degrees of freedom and have approximately \cite{stefanucci2013, vignale2004}
\begin{eqnarray}
\label{VectorPotentialHamiltonian}
\hspace{-1.5cm} \hat{H}(t) = & \hat{T} + \hat{W} - \int \diff  \bbr \,  \hat{\mathbf{J}}(\bbr,t) \cdot \mathbf{a}_{\mathrm{ext}}(\bbr,t) + \int \diff^3 r \, \hat{n}(\bbr) \left( v(\bbr,t) - \frac{1}{2} \mathbf{a}_{\mathrm{ext}}^2(\bbr,t) \right), \nonumber
\end{eqnarray}
where now also the internal charge current $\hat{\mathbf{J}} = \hat{\mathbf{j}} - \hat{n} \, \mathbf{a}_{\mathrm{ext}}$ does no longer depend on the photon field. If we only allow for scalar external potentials $v$ (and thus $\mathbf{a}_{\mathrm{ext}}=0$) we rederive our original Hamiltonian given in (\ref{TDSEHamiltonian}). On the other hand, if we keep the external vector potentials explicit we see that with respect to our previous considerations we have more freedom in the choice of our external fields to describe and control a quantum system. Consequently instead of a mapping from potentials to densities in this case we find a mapping of the form
\begin{eqnarray}
 \Psi: \mathcal{A}  \rightarrow  \mathcal{C}^{1}([0,T], \mathcal{H} ) 
\\
(v, \mathbf{a}_{\mathrm{ext}})  \; {\buildrel\rm \Psi_0 \over \mapsto }  \;  \Psi[v, \mathbf{a}_{\mathrm{ext}}], \nonumber 
\end{eqnarray}
where we assume $\Psi_0$ and the allowed external fields regular enough (for instance infinitely-often differentiable). The continuity equation in this case becomes then
\begin{eqnarray}
\label{ContinuityEq2}
 \partial_t n(\bbr,t) = -\nabla\cdot \mathbf{J}(\bbr,t),
\end{eqnarray}
and the charge current obeys a local-force equation of the form \cite{stefanucci2013, ruggenthaler2014} (suppressing all dependencies)
\begin{eqnarray}
\label{LocalForceEquation}
\hspace{-2.0cm} \partial_t J_k  = -n \left( \partial_k v -\partial_t a^{\mathrm{ext}}_k \right)  - Q_k +  J_l (\partial_k a^{\mathrm{ext}}_l - \partial_l a^{\mathrm{ext}}_k)
+ \partial_l \left( a^{\mathrm{ext}}_k J_l + a^{\mathrm{ext}}_l j_k  \right),
\end{eqnarray}
where $Q_k$ is defined as in Sec.~\ref{sec:observables} and summation over multiple indices is implied.

In the case of the density-potential mapping based on the Hamiltonian of (\ref{TDSEHamiltonian}) the densities $n$ and the potentials $v$ are functions with the same degrees of freedom. Now, however, we have much more freedom since we can choose $(v, \mathbf{a}_{\mathrm{ext}})$. A reasonable choice to set up a similar mapping would be $(n, \mathbf{J})$ since this pair would have the same degrees of freedom. If we consider the continuity equation (\ref{ContinuityEq2}) together with a prescribed initial state which implies an initial density $n_0$, then we see that the charge current determines the density uniquely. Therefore, we need to have a similar reduction of freedom in the external fields if we want to have an invertible mapping for the charge current $\mathbf{J}$. To find this restriction we first take a look at the case of (\ref{TDSEHamiltonian}), where we can add a time-dependent yet spatially constant function to $v$ and still have the same density. For the inversion of $v \mapsto n$ we have to restrict this freedom by fixing a gauge. Now we find that for any differentiable $\Lambda$ both the pair $(v, \mathbf{a}_{\mathrm{ext}})$ and 
\begin{eqnarray}
 v' &= v - \partial_t \Lambda
\\
 \mathbf{a}'_{\mathrm{ext}} & = \mathbf{a}_{\mathrm{ext}} + \nabla \Lambda
\end{eqnarray}
lead to the same charge current $\mathbf{J}$ and physical observables \cite{vignale2004, stefanucci2013, ruggenthaler2014}. By fixing this gauge freedom we find the desired restriction and thus only take into account physically inequivalent external fields. In our case we choose the \textit{radiation gauge} which fixes $v = 0$ and which leaves $\Psi_0$ unchanged by taking the initial condition $\Lambda(\bbr,0) = 0$ \cite{vignale2004}. This slightly simplifies the local-force equation (\ref{LocalForceEquation}). Following now the same steps as for the fundamental equation (\ref{TddftFundament}) in Sec.~\ref{sec:Uniqueness} and calculating all higher time-derivatives of $\mathbf{J}$ we find recursive equations for the Taylor coefficients of $\mathbf{J}$ in terms of the intial state $\Psi_0$ and the Taylor coefficients of $\mathbf{a}_{\mathrm{ext}}$. These equations were then used by Vignale to show (similar to Sec.~\ref{sec:Uniqueness}) the invertibility of a mapping
\begin{eqnarray}
 J: & \mathcal{A} &   \rightarrow  \mathcal{J}
\\
& \mathbf{a}_{\mathrm{ext}} &  \; {\buildrel\rm \Psi_0 \over \mapsto }  \;  \mathbf{J}[\mathbf{a}_{\mathrm{ext}}], \nonumber 
\end{eqnarray}
and to provide the construction of the time-analytic vector potential $\mathbf{a}_{\mathrm{ext}}$ for a given time-analytic charge current $\mathbf{J}$ \cite{vignale2004}. We point out that one could start the very same construction directly with the definition of $\mathbf{J} = \mathbf{j} - n \mathbf{a}_{\mathrm{ext}}$ since there the external vector potential appears already explicitly. That means that we suppress the time-derivatives of the complicated terms in (\ref{LocalForceEquation}) and collect them in time-derivatives of $\mathbf{j}$ and $n$ \cite{tokatly2011, ruggenthaler2014}
\begin{eqnarray}
\label{VectorTaylor}
 \mathbf{J}^{(k)} = \mathbf{j}^{(k)} - \sum_{l=0}^{k} \left( \begin{array}{c} k \\ l \end{array}
\right) n^{(l)} \mathbf{a}_{\mathrm{ext}}^{(k-l)}, 
\end{eqnarray}
where again the superindex $(k)$ referes to the $k$-th time-derivative at $t=0$. And consequently we have for vector potentials $\mathbf{a}_{\mathrm{ext}} \neq \mathbf{a}'_{\mathrm{ext}}$ (by more than a gauge) which first differ in the $k$-th order of their respective Taylor expansions that
\begin{eqnarray}
 \mathbf{J}^{(k)} - \mathbf{J}'^{(k)} = n^{(0)} \left( \mathbf{a}'_{\mathrm{ext}} - \mathbf{a}_{\mathrm{ext}} \right) \neq 0,
\end{eqnarray}
provided $n_{0}$ is non-zero everywhere (except maybe at the boundaries) and thus $\mathbf{J}[\mathbf{a}_{\mathrm{ext}}] \neq \mathbf{J}[\mathbf{a}'_{\mathrm{ext}}]$. This forms the basis of TDCDFT and allows us perform a self-consistent calculation in terms of the charge current $\mathbf{J}$ instead of considering the full TDSE with the Hamiltonian of (\ref{VectorPotentialHamiltonian}) \cite{ullrich2012, marques2012}. If we discretize the Hamiltonian~(\ref{VectorPotentialHamiltonian}) \cite{kurth2011} a rigorous iterative approach to the current-potential mapping similar to the one presented in Sec.~\ref{sec:Lattice} has been established by Tokatly \cite{tokatly2011}.
\\

If we do not assume that the photons are negligible, we have further degrees of freedom in the external variables since we can also choose different external charge densities and currents, i.e., $n_{\mathrm{ext}}$ and $\mathbf{j}_{\mathrm{ext}}$ respectively. From the purely photonic limit of (\ref{FullVectorPotentialHamiltonian}), we see that $\mathbf{j}_{\mathrm{ext}}$ couples to $\mathbf{A}$, and thus an invertible mapping $\mathbf{j}_{\mathrm{ext}} \mapsto \mathbf{A}$ seems possible. However, we have restricted the freedom of $\mathbf{A}$ by the Coulomb-gauge condition to only transversal degrees of freedom and thus we need to find a similar restriction also for the external current $\mathbf{j}_{\mathrm{ext}}$. If we determine the equation of motion for the vector-potential operator governed by the full Hamiltonian~(\ref{FullVectorPotentialHamiltonian}) we find \cite{ruggenthaler2014}
\begin{eqnarray}
 \Box \mathbf{A} + \nabla \int \diff \bbr' \frac{\partial_t n_{\mathrm{ext}}(\bbr',t) + \partial_t n(\bbr',t)}{4 \pi |\bbr -\bbr'|} =  \mathbf{j}_{\mathrm{ext}} + \mathbf{J} .
\end{eqnarray} 
which only guarantees the Coulomb gauge for $\mathbf{A}$ if we impose a continuity equation on the external charge current and density, i.e.,
\begin{eqnarray}
 \partial_t n_{\mathrm{ext}} = - \nabla \cdot \mathbf{j}_{\mathrm{ext}}.
\end{eqnarray}
Then the equation of above can be rewritten as 
\begin{eqnarray}
\label{MaxwellEquation}
 \Box \mathbf{A} - \nabla \int \diff \bbr' \frac{\nabla'j_{\mathrm{ext}}(\bbr',t) + \nabla'J (\bbr',t)}{4 \pi |\bbr -\bbr'|} =  \mathbf{j}_{\mathrm{ext}} + \mathbf{J},
\end{eqnarray}
where the second term on the left-hand side cancels explicitly any longitudinal component of $\mathbf{A}$ which would arise due to longitudinal components of $\mathbf{j}_{\mathrm{ext}} + \mathbf{J}$ \cite{ruggenthaler2014}. Consequently, any two inhomogeneities that differ only by a longitudinal function lead to the same internal vector potential $\mathbf{A}$, and hence $\mathbf{A}$ and $\mathbf{j}_{\mathrm{ext}}$ have the same degrees of freedom. Therefore, similar to the radiation-gauge condition on the external vector potentials, we only take into account physically inequivalent external charge currents and restrict to those $\mathbf{j}_{\mathrm{ext}}$ that differ by more than a longitudinal current.

As a consequence of this condition, for a given internal pair $(\mathbf{J}, \mathbf{A})$ there exists a unique external charge current $\mathbf{j}_{\mathrm{ext}}$ determined via Maxwell's Eq.(\ref{MaxwellEquation}). And since we can by a Taylor expansion similar to (\ref{VectorTaylor}) uniquely determine all the Taylor-coefficients of the (radiation-gauged) external vector potential from $(\mathbf{J}, \mathbf{A})$ we have an invertible mapping
\begin{eqnarray}
 \left(\mathbf{a}_{\mathrm{ext}}, \mathbf{j}_{\mathrm{ext}}  \right) \; {\buildrel\rm \Psi_0 \over \mapsto }  \; \left(\mathbf{J}, \mathbf{A}  \right)
\end{eqnarray}
from the set of Taylor-expandable external currents and potentials to the corresponding internal currents and potentials. This allows us to perform a density-functional treatment of a system of charged particles coupled to photons. In a similar manner also other Hamiltonians describing particle-photon systems give rise to an invertible mapping from external fields to internal fields \cite{rajagopal1994,ruggenthaler2011b,tokatly2013,ruggenthaler2014}. If we discretize the Hamiltonian of the charged particles and keep finitely many photonic modes then a rigorous iterative formulation similar to Sec.~\ref{sec:Lattice} can be provided \cite{farzanehpour2014}.

\section{Summary, open questions and outlook }

\label{sec:summary}


In this topical review we discussed in detail various aspects of the existence, uniqueness, and construction of the density-potential
mapping in TDDFT. This problem splits into a number of important subproblems.
The most basic of these subproblems is to determine the class of potentials and initial states for which the TDSE has a
unique solution. We identified an important class of potentials for 
which a solution can be guaranteed for any normalizable initial state $\Psi_0$. 
These are the potentials in the Kato class for which also the time-derivative is in the Kato class. We denoted this set of potentials by $\mathcal{V}$. If the initial state is in the domain $D (\hat{T})$ of the kinetic energy operator such that $\hT \Psi_0$ is normalizable then 
we can also guarantee that the solution stays in this domain and thus has finite total energy.  Therefore it is natural to
consider this class $\mathcal{V}$ as the class of physical potentials.
For any potential in $\mathcal{V}$ and any initial state in the domain of $\hat{T}$ 
we can find a time-dependent wave-function $\Psi (t)$ and subsequently a density $n(t)$.
This defines a mapping from potentials to densities.  Our next problem was to know whether this
map is invertible or whether it is possible that two potentials map to the same density.
An important role in this discussion was played by the local-force equation which gives a
direct relation between densities and potentials. We found that for this equation to be well-defined
we need to put extra conditions on the potentials. The resulting potentials are similar
to those that are generated by charge distributions as calculated from the Poisson equation.
Although point charges
are not allowed this includes the important physical case of finite atomic nuclei (softened
Coulomb potentials).
Then we discussed how the local-force equation played an important role in several
of the proofs of the density-potential mapping.  
The original proof by Runge and Gross had to require that the potential was a real-analytic function in time and that
the initial state $\Psi_0$ was infinitely differentiable with respect to spatial
coordinates. To remove these conditions we considered
an iterative solution of the local-force equation. This lead us to consider a number of
other issues, such as the invertibility of a certain Sturm-Liouville equation and
the linear response of the $q$-operator. We showed how we could use the
iterative scheme to prove the existence and uniqueness of the density-potential
mapping under certain conditions on the densities and initial states.
We further showed how we could define the fixed-point procedure on a lattice
and presented a numerical implementation of the fixed-point scheme
to construct the density-potential mapping and gave several
examples. In the discussion of the numerical implementation we saw how rather
abstract mathematical concepts like the self-adjoint domain of an operator becomes
important in practice. We finally discussed a TDDFT extension to vector potentials and photons.
As is clear from this summary the question whether a time-dependent density
can be obtained from some TDSE has many aspects and therefore it is
natural that there are still several open issues.

The main issue for a rigorous approach to the density-potential mapping in terms
of a fixed-point procedure, or equivalently in terms of a non-linear TDSE, are the properties
of the $q$-operator. We do not yet know sufficient conditions for the differentiability
and boundedness of its response functions. However, recent results for the wave function
\cite{penz2014} indicate that such conditions should be possible. Closely connected with these
issues is also the question of determining the most general set of external potentials for which
the fundamental equation of TDDFT (\ref{TddftFundament}) is well-defined. This implies the problem
of ensuring that an initial state which is four-times differentiable keeps this property through time.
For now we only know this to be rigorously true in the case of periodic systems with infinitely
differentiable potentials and interactions \cite{delort2010}. The next open problem is to guarantee
that the iterative procedure to determine the potential for a fixed initial state and density
really does reproduce this density as a solution of the TDSE. This holds true if the solution to
(\ref{RhoEquation}) for zero initial conditions and a general potential indeed is the zero
function. Up to now we only know this to hold for analytic potentials. And finally there is
the question whether one can extend the invertibility of the Sturm-Liouville equation to
all of $\mathbb{R}^3$. This would be desirable since then one can treat the standard
setting of quantum mechanics.

While we face a lot of mathematical challenges if we want to treat the density-potential
mapping in the most general setting, in practice we are usually safe. First of all, the restrictions
we had to impose to make the density-potential mapping rigorous, e.g., to only allow for
infinitely differentiable initial states in the Runge-Gross theorem, do not really matter
when actually solving the TDSE. The time-propagation of a smooth approximation to an initial
state with a cusp (as for ground states of Coulomb systems) can be made arbitrarily close
to the exact propagation (below any numerical accuracy). Further, since we need to put the
TDSE on a grid (or finitely many basis functions) to perform a calculation we are considering
indeed an approximation to the original problem in terms of a lattice. In this case we can rely
on the results of \cite{farzanehpour2012} to guarantee a well-defined density-potential
mapping. However, to guarantee that the discretised formulation represents the original
TDSE in the continuum limit for finer and finer grids certain analytic conditions have to be 
fulfilled, as becomes obvious from the examples in Sec.~\ref{sec:wavepacket} and the discussion 
of the numerical approximation to the 
propagator in Sec~\ref{sec:numerics}. Hence in practice usually the only real obstacle is to find 
better and more reliable approximation to the xc potential. Besides going beyond the usual time-local
approximations and also include previous times \cite{ullrich2012}, new and promising routes are
currently being developed by also employing different initial states and new auxiliary systems
such as strictly-correlated electrons \cite{seidl2007,gori2009,malet2013}.

However, answering the important open questions in the density-potential mapping will lead to
new insights into the fundamentals of TDDFT and the KS construction and hopefully
will also inspire more accurate approximations to the xc potential and
other time-dependent density functionals. 

\ack
The authors are grateful to N.T.~Maitra for fruitful discussions and her hospitality during their stay at the City University of New York, as well as to P.~Gori-Giorgi and K.~Giesbertz during their stay at the VU Amsterdam. We further thank S.E.B.~Nielsen for clarifying comments and help with the numerical examples.   
M.R.~acknowledges financial support by the FWF (Austrian Science Fund) through the project P 25739-N27. R.v.L.~acknowledges the Academy of Finland for support. 

\appendix

\section{Lebesgue and Sobolev spaces}

\label{app:Lp}

The function spaces $L^p(\Omega)$, $1 \leq p \leq \infty$, form not only the basic space for wave functions, where the norm defined on them is the used for an interpretation in terms of probabilities, but become relevant in this work also as the domains of the density-potential mapping. They consist of all Lebesgue-measurable functions $f : \Omega \rightarrow \mathbb{R}$ or $\mathbb{C}$ with finite $L^p$-norm, i.e.,
\[
\|f\|_p = \left(\int_\Omega \diff x \, |f(x)|^p \right)^{1/p} < \infty.
\]

The special case $p=\infty$ represents all functions that are bounded up to a set of measure zero and the associated norm $\|f\|_\infty$ is given by the smallest such bound. If we take $f$ roughly with amplitude $A$ and non-zero on a volume $V$ then the $L^p$-norm measures the quantity $AV^{1/p}$. This means that lower $L^p$-spaces allow more singularity while higher ones are more forgiving towards spreading, also expressed by the (continuous) embedding $L^p(\Omega) \subset L^q(\Omega)$ if $p > q$ on bounded domains $\Omega$ where the spread is not an issue. As all those spaces are normed vector spaces with always converging Cauchy sequences (completeness) they form proper Banach spaces. In the case $p=2$ the norm is directly linked to the usual inner product by $\langle f|f \rangle = \|f\|^2$ (note that we typically omitted the index $2$ in the norm in this important case) and $L^2(\Omega)$ has all the structure of a Hilbert space that has risen to eminent prominence within quantum theory.\\

The related class of Sobolev spaces includes the weak derivatives of several orders into its definition. Thus not only amplitude $A$ and volume $V$ are measured but also frequency $N$ with the sensitivity controlled by a parameter $m \geq 0$ defining up to what order derivatives get included into the Sobolev $W^{m,p}$-norm that will be concerned with the quantity $AV^{1/p}N^m$.\footnote{This idea is taken from an answer of Terence Tao on the collaborative website \textit{MathOverflow}.} This is already a strong indication towards the important Sobolev embedding theorems relating such spaces. For the definition of the $W^{m,p}$-norm we use a multi-index notation for the weak $\alpha$-th partial derivative and note that other equivalent definitions are possible.
\[
\|f\|_{m,p} = \sum_{|\alpha|\leq m} \|D^\alpha f\|_p
\]

Again the case $p=2$ yields Hilbert spaces denoted as $H^m(\Omega) = W^{m,2}(\Omega)$. For bounded intervals $I \subseteq \mathbb{R}$ of the real line the relation between absolute continuity and the Lebesgue integration leads to the identification $W^{1,1}(I) = \mathcal{AC}(I)$. The possibility of unique continuous continuation to the boundary points means we can give meaningful boundary conditions. For $m>1$ it holds that $H^m(I) \subset H^1(I) \subset W^{1,1}(I)$ and we used the notation $H_0^m(I)$ for functions in $H^m(I)$ with zero-boundary conditions up to the $(m-1)$-th derivative, deviating here from standard notation for closed intervals $I$. In a multi-dimensional setting one defines $W_0^{m,p}(\Omega)$ as the closure of the test functions under the $W^{m,p}$-norm. The definitive resource on almost all topics relating to Sobolev spaces is \cite{adams2003}.

\section{Generalization and functional variation of Schr\"{o}dinger solutions}

\label{app:existence}

For a different kind of generalization of solutions to the Schr\"{o}dinger equation following largely \cite{yajima1987} we employ the physical structure of the Hamiltonian
\begin{eqnarray}
\label{HamiltonianStructure}
 \hat{H}(t)=\hat{T} +f(t),
\end{eqnarray}
and try to get rid of the unbounded kinetic part $\hat{T}$ by unitary transformation. Here $f(t) = f(\bbr_1,\ldots,\bbr_N,t)$ is a scalar potential acting as a multiplication operator in spatial representation, including all interactions (possibly also of more than two particles) as well as external potentials. We define the unitary free evolution operator $\hat{U}_0(t)=\exp(-\imagi \hat{T} t)$ that solves the corresponding Cauchy problem $\imagi \partial_t \Psi(t) = \hat{T}\Psi(t)$ for any initial state $\Psi_0 \in \mathcal{H}$ (even if the initial state has infinite energy). With the help of $\hat{U}_0(t)$ we then perform a unitary transformation $\Psi(t)= \hat{U}_0(t) \tilde{\Psi}(t)$ of the original problem (\ref{Cauchy}) to $\imagi \partial_t \tilde{\Psi}(t) = \tilde{f}(t) \tilde{\Psi}(t)$ with $\tilde{f}(t) = \hat{U}_0^{\dagger}(t) f(t) \hat{U}_0(t)$ (this is one possible form of the so-called \textit{interaction picture}). Integrating this problem over time and transforming it back we find the \textit{mild form of the time-dependent Schr\"odinger equation} (mild TDSE)
\begin{eqnarray}
\label{MildSchroedinger}
 \Psi(t) = \hat{U}_0(t)\Psi_0 -\imagi \int_0^t \diff s \, \hat{U}_0(t-s)f(s)\Psi(s). 
\end{eqnarray}
This form generalizes the notion of a solution of the TDSE to $\Psi \in \mathcal{C}^{0}([0,T],\mathcal{H})$, i.e., functions that are (only) continuous in time as square-integrable spatial functions. We call solutions to (\ref{MildSchroedinger}) \textit{mild solutions}. Because those solutions include all initial states $\Psi_0 \in \mathcal{H}$ they are equivalent to the generalized solutions from Sec.~\ref{sec:Cauchy}.\footnote{We point out that an even more general definition of a solution to the TDSE would be possible if we defined the time-derivative in a weak sense. However, since such \textit{weak solutions} are not defined at every instance in time (and thus violate the usual notion of a physical wave function), they are commonly disregarded in physics literature. \cite{reed1975,lions1961}}

The idea is now to guarantee unique solvability of (\ref{MildSchroedinger}) by recursively putting $\Psi$ into the integral (thus generating all possible \textit{paths} of interaction) and showing that this mapping is a contraction and therefore has a unique fixed point. In practice it is enough to show boundedness of the mapping with an estimate involving $T$ then taking the time interval $[0,T]$ short enough and finally extending to arbitrary time intervals with a continuation procedure like in \cite{dancona2005}. To do so, this demands for a purpose-built Banach space of \textit{trajectories}, i.e., a wave function for all times in $[0,T]$, that is a subspace of $\mathcal{C}^{0}([0,T],\mathcal{H})$. An important stepping-stone towards such spaces is the Strichartz estimate for solutions to the free Schr\"{o}dinger equation using the spacetime norm of $L^\theta([0,T], L^q(\mathbb{R}^{3N},\mathbb{C}))$ with $(\theta,q)$ fulfilling a certain relation called \textit{Schr\"{o}dinger-admissible} \cite{dancona2005}.
\[
\|U_0 \Psi_0\|_{q,\theta} \leq \mathrm{const} \cdot \|\Psi_0\|_2
\]

This estimate exhibits a certain smoothing property of the free evolution. An equivalent result for non-free evolution is readily achieved by the fixed-point procedure described above. With the trajectory confined by the given inequality we have $\mathcal{C}^{0}([0,T],\mathcal{H}) \cap L^\theta([0,T], L^q(\mathbb{R}^{3N},\mathbb{C}))$ as the Banach space of quantum trajectories. Note however that the spatial domain here is $\mathbb{R}^{3N}$, Strichartz estimates for bounded domains are available, although not in this general form.

The set of allowed potentials for the mild TDSE to hold is then a complementary Banach space chosen in a way that $f \cdot \Psi$ is in the topological dual of the trajectory space. A physical consequence to that is $\langle \Psi(t)|f(t)\Psi(t) \rangle < \infty$, i.e., finite energy from the potential.

One drawback of this approach is that it does not include singular Coulombic potentials (as an interaction term or external potential) if more than two electrons in $\mathbb{R}^3$ configuration space are involved \cite{penz2014}. Yet it is general enough on the other hand to include sudden switch-on processes.

The mild TDSE (\ref{MildSchroedinger}) is also the starting point for the study of functional variations of trajectories. These are formed by varying the potential $f$ within its Banach space mentioned above. To fix notation $\Psi[f]$ is the solution of (\ref{MildSchroedinger}) for a fixed initial state and potential (internal and external) $f$. We form the directional derivative at $f$ in direction $g$ by
\[
\delta\Psi[f,g] = \lim_{\varepsilon \rightarrow 0} \frac{1}{\varepsilon} \big(\Psi[f+\varepsilon g]-\Psi[f]\big).
\]
The limit is taken in the Banach space of trajectories. In a different version of the interaction picture where the transformation is carried out with the unitary evolution system $\hat{U}([f],t,s)$ involving the Hamiltonian $\hat{T}+f(t)$ instead of only $\hat{T}$ this yields
\begin{eqnarray}
\label{Frechet}
\delta\Psi([f,g],t) = -\imagi \int_0^t \diff s \, \hat{U}([f],t,s) g(s) \Psi([f],s).
\end{eqnarray}

This variational derivative can be shown to be continuous in $f$ as a linear and bounded mapping from its second argument $g$ to variations of trajectories in the trajectory space and is thus a proper Fr\'echet derivative \cite{penz2014}. Application of this formalism can be carried over to observables and quantities such as the one-particle density and leads to the well-known non-equilibrium version of Kubo's formula. It further gives important justifications for non-equilibrium density-response theory (for an introduction see \cite{stefanucci2013}) and facilitates the apparatus of variational calculus in the TDDFT context.

\section*{References}

\bibliography{Topical_Review6}{} 

\end{document}